\documentclass[aps,amsfonts,nofootinbib,preprintnumbers,nobalancelastpage]{revtex4}
\usepackage[english]{babel}

\usepackage{amsmath,amssymb}
\usepackage[dvips]{graphicx}
\usepackage[sort&compress]{natbib}
\usepackage{bbm}
\usepackage[usenames,dvipsnames]{color}
\usepackage{footnote}
\usepackage{slashed} 
\usepackage{pdflscape} 
\usepackage{multirow}

\makeatletter
\def\endfmffile{%
  \fmfcmd{\p@rcent\space the end.^^J%
          end.^^J%
          endinput;}%
  \if@fmfio
    \immediate\closeout\@outfmf
  \fi
  \IfFileExists{\thefmffile.mp}{\immediate\write18{mpost \thefmffile}}{}
  \let\thefmffile\relax
}
\makeatother
\usepackage[normalem]{ulem}

\newcommand {\beq} {\begin{equation}}
\newcommand {\eeq} {\end{equation}}
\newcommand {\bea} {\begin{eqnarray}}
\newcommand {\eea} {\end{eqnarray}}

\newcommand{\cf}{{\it c.f. }}

\newcommand{\Xx}{X_{5/3} / B}
\newcommand{\MX}{ M_{X_{5/3}/ B}}
\newcommand{\mX}{ m_{X_{5/3}/ B}}

\definecolor{red1}{cmyk}{0,1,1,0.1}
\definecolor{blue1}{cmyk}{1,0,0,0}
\definecolor{ao}{rgb}{0.0, 0.0, 1.0}
\newcommand{\redc}[1]{{\color{red1} #1}}

\newcommand{\bluebold}[1]{{\textbf{\color{ao} #1}}}

\newcommand{\ignore}[1]{}
\newcommand{\nn}{\nonumber} \renewcommand{\bf}{\textbf}

\newcommand{\MET}{\sl{E_T}\,\,\,}

\newcommand{\GeV}{{\rm\ GeV}}

\newcommand{\TeV}{{\rm\ TeV}}
\newcommand{\fb}{{\rm\ fb}}

\def\sl#1{#1 \!\!\! \!\!   \!\! \slash}

\allowdisplaybreaks

\begin{document}
\title{LHC Top Partner Searches Beyond the 2 TeV Mass Region}
\date{\today}

\author{Mihailo Backovi\'{c}} 
\email{mihailo.backovic@weizmann.ac.il} 
\affiliation{Department of Particle Physics and Astrophysics, \\ Weizmann Institute of Science, Rehovot 76100, Israel} 
\author{Thomas Flacke}
\email{flacke@kaist.ac.kr}
\affiliation{Department of Physics, Korea Advanced Institute of Science and Technology, \\
335 Gwahak-ro, Yuseong-gu, Daejeon 305-701, Korea} 
\author{Seung J. Lee} 
\email{sjjlee@kaist.ac.kr} 
\affiliation{Department of Physics, Korea Advanced Institute of Science and Technology, \\
335 Gwahak-ro, Yuseong-gu, Daejeon 305-701, Korea} 
\affiliation{School of Physics, Korea Institute for Advanced Study, Seoul 130-722, Korea} 
\author{Gilad Perez} 
\email{gilad.perez@weizmann.ac.il} 
\affiliation{Department of Particle Physics and Astrophysics, \\ Weizmann Institute of Science, Rehovot 76100, Israel} 

\begin{abstract}
 We propose a new search strategy for heavy top partners at the early stages of the LHC run-II, based on lepton-jet final states. Our results show that final states containing a boosted massive jet and a hard lepton, in addition to a top quark and possibly a forward jet, offer a new window to both detecting and measuring top partners of mass $\sim 2 \TeV$. Our resulting signal significance is comparable or superior to the same sign di-lepton channels for top partner masses  heavier than roughly 1 TeV. Unlike the di-lepton channel, the selection criteria we propose are sensitive both to $5/3$ and $1/3$ charge top partners and allow for full reconstruction of the resonance mass peak. 
 Our search strategy utilizes a simplified $b$-tagging procedure and the Template Overlap Method to tag the massive boosted objects and reject the corresponding backgrounds. In addition, we propose a new, pileup insensitive method, to tag forward jets which characterize our signal events.  
 We consider full effects of pileup contamination at 50 interactions per bunch crossing. We demonstrate that even in the most pessimistic pileup scenarios, the significance we obtain is sufficient to claim a discovery over a wide range of top partner parameters. While we focus on the minimal natural composite Higgs model, the results of this paper can be easily translated into bounds on any heavy partner with a $t\bar{t}Wj$ final state topology. 
\end{abstract}


\maketitle

\section{Introduction}\label{sec:intro}
The discovery of the Higgs boson at the Large Hadron Collider (LHC) is a great victory for the Standard Model (SM) of particle physics. With its minimal scalar sector of electroweak symmetry breaking,
the SM at short distances is a complete weakly coupled theory up to very large energy scales. 
Furthermore, the SM admits a set of accidental symmetries that eliminate proton decay and suppress custodial, flavor and CP violating processes.
Even though the SM cannot explain several experimental observations such as the neutrino masses, the baryon asymmetry of the universe and the origin of dark matter one cannot deduce with any certainty the energy scale at which the extensions of the SM would be relevant, with the exceptions of the Planck scale and the scale of the Landau pole of the hyper charge interactions.
The only fuzzy scale, potentially accessible to the LHC, is related to the recently discovered Higgs boson.
As a fundamental scalar the Higgs mass is ultra-violet (UV) sensitive. Hence,  we expect that on the quantum level the Higgs mass will pick up large contributions from high energy scales, resulting in a very large mass of the Higgs boson. This, of course, is in direct contradiction with our direct and indirect knowledge of the Higgs boson dynamics. 

A simple possibility to stabilize the Higgs mass and the electroweak scale in a controlled manner is to add new fields to SM, with the same gauge quantum numbers as the SM fields, such that the contributions of the new fields to the Higgs mass eliminate the UV sensitivity. 
In the absence of interactions the Higgs will loose its quantum sensitivity (setting quantum gravity aside), and hence the most severe known sensitivity of the Higgs to quantum corrections arise as a result its large coupling to the top quark. To ensure the stabilization of the electroweak scale, the virtual contributions of some of the new particles to the Higgs mass should cancel the contributions coming from the SM top quarks. These new states are collectively denoted as top partners. In known examples the partners might be scalars as in the case of supersymmetry or fermions as in the case of composite Higgs models (CHMs). 
Current bounds on the top partner masses are roughly $\gtrsim 700 \GeV$ for supersymmetric scalar states and $\gtrsim 800 \GeV$ for composite-Higgs fermionic states (see {\it e.g.} Refs.~\cite{Chatrchyan:2013uxa,ATLAS-CONF-2014-036} for recent results).

While the bounds on the top partner masses are fairly strong they are not bullet proof, and they also only result in moderate pressure on naturalness (here we are not concerned with various definitions of fine tuning). Probably the most relevant question amidst the "LHC battle for naturalness" is how are we going to discover top partners (if any exists) or improve the bounds on the top partners both in terms of mass reach and in terms of robustness. The two criteria can be used to guide the focus of theoretical, phenomenological and experimental effort. 

One can define two "mini-frontiers" for the battle for naturalness at the LHC~\cite{Perez:2013oaa}:
\begin{enumerate} 
\item  The mini energy frontier, where the effort is directed towards searching for ultra massive top partners. The experimental focus of the energy frontier searches is defined by the highest center-of-mass energies that can be reached by the LHC. 
\item The mini intensity frontier, where the effort is focused on searching for partners with mass below or near the current bounds. The mini intensity frontier focuses the searches for top partners to possibilities that partners are elusive (i.e. when for some reason the current searches are not sensitive enough to their presence). 
\end{enumerate}
The physics describing the above frontiers is qualitatively different both in terms of the phenomenology describing them and in terms of the necessary experimental effort.
 It is important to notice that prior to the start of the LHC the starting points of the framework of supersymmetry and pseudo-Nambu-Goldstone boson (pNGB) composite Higgs models were different in the context of naturalness. If we were to remove our LHC-based knowledge (the results of the ATLAS and CMS  direct searches) then supersymmetric models are not subject to any substantial pressure from naturalness. For instance, stop (as well as most of the other superpartners) masses close to that of the top quark are not in conflict with existing data. This is not the case when  pNGB composite Higgs models are considered as the combination of LEP and Tevatron data is already constraining the model's decay constant $f$ to lie above the $f>{\cal O} (800\,\rm GeV)$ scale~\cite{Grojean:2013qca,Ciuchini:2013pca}. Beyond the mere fact that this rather strong constraint on the value of $f$ forces some amount of fine tuning, it also suggests that we should have expected that the composite fermion resonances would be somewhat heavy with masses probably larger than $f\,.$   Even at the centre of mass energy of 8 TeV, the typical fermonic top partner production cross sections and the collected luminosity were simply not enough to produce the heavy partners. Thus, there is very little surprise that the first run of the LHC, which was limited in centre of mass energy, did not observe them. In order to make experimental progress on fermionic top partner searches at the LHC, it is hence necessary to focus on the region of parameter space where the top partner masses are larger than $f$. So far, the parameter space region of heavy fermionic top partners has not been explored, providing the main motivation for our current study of heavy top partners at the mini energy frontier.
  
 The main focus of this work is  to study the reach of the LHC in the next run to discover and measure (or exclude) the presence of top partners in a regions of model parameter space which results in large top partner masses. When searching for top partners one needs to distinguish between event topologies of pair produced and singly produced top partners~\cite{DeSimone:2012fs,Azatov:2013hya}. While the former is more robust as the partners are produced via SM QCD processes it suffers from a severe ``large $x$ suppression'' from the parton distribution function (PDF) for large top partner masses. As two heavy particles are produced, the quarks and gluons in the proton have to carry a high $x$ in order to achieve a heavy final state. The expected reach of searches for doubly produced top partners is rather limited even when considering high luminosities ~\cite{SalamWeiler}. 
 Single production processes, on the other hand,  are model dependent but are subject to much lower level of PDF suppression and thus can potentially lead to a much better experimental reach. 
 
 Following the original papers that have emphasized the importance of the same sign lepton signal~\cite{Dennis:2007tv, Contino:2008hi, Mrazek:2009yu,} of most of the fermionic top partner studies so far focused on final. Standard Model processes are highly unlikely to produce final states with two same sign leptons, deeming such signals a ``clean'' signature of BSM physics.\footnote{It is important to note that since the dominant backgrounds to the same sign di-lepton processes come from detector effects such as photon conversions, accurate estimates of background channels are challenging.}
 However, the same sign lepton searches are only applicable for the exotic $5/3$ charged top partners. Furthermore, the di-lepton final states suffer from low branching ratios and from the fact that the resonance masses are smeared due to the missing energy having at least two hard neutrino components.

 In this paper consider the case where the heavy partners decay to hadronic-leptonic final states. For other studies involving hadronic final state see~\cite{DeSimone:2012fs, Azatov:2013hya, Ortiz:2014iza, Gripaios:2014pqa}. We provide a strategy and a detailed phenomenological study which shows that in preferred regions of pNGB composite Higgs models one can discover top partners (at 5 sigma CL) with mass as high as 2 TeV at the 14 TeV LHC run, and with integrated luminosity of roughly 35\,fb$^{-1}$.
 Furthermore, in the absence of signal one can exclude the presence of 2 TeV partners (at 2 sigma CL) with as little as 
10\,fb$^{-1}$. 
 
 Our study adopts the Template Overlap Method (TOM) ~\cite{Almeida:2011aa,Almeida:2010pa, Backovic:2012jj, Backovic:2013bga} to tag the highly boosted decay product of the partners and in part reject the corresponding SM backgrounds. The final state of our signal events is characterized by multiple $b$-jets, which we employ through a semi-realistic $b$-tagging procedure. We take into account the contamination from pileup, assuming average of 50 interactions per bunch crossing and show where the effects of pileup on our selection criteria can be mitigated and where additional improvement might be necessary. Finally, our study of singly produced top partners employs the presence of a high energy forward jet in the signal events, which is in principle susceptible to contamination from pileup. We propose a modification of forward-jet tagging, whereby we cluster the jets int the forward region using a small cone ($e.g.$ r = 0.2). We show that the signal distributions are hardly affected by reduction in forward jet cone size,  while the background is significantly suppressed upon requiring a forward jet tag. As this is the first time that such a technique is proposed we present the results with and without the use of this new forward-jet tagger. 
 
 At large top partner masses the mass splitting between the partners due to electroweak symmetry breaking is subdominant. Hence, as we are not confined to the same sign di-lepton final states, our event selection strategy is adequate for searches for all partners that decay to tops and $W$s and not only the $5/3$ charged ones. For the sake of concreteness and simplicity our current study focuses only on the relatively simple final state of $t\bar{t}Wj$. Note, however,  that it is straight forward to generalize our study to include other final states as well.  
 
 In Section \ref{sec:model} we provide a brief introduction to our benchmark composite Higgs model. We include only the bare minimum of information directly relevant for the phenomenology of top partners and postpone a detailed discussion  of the composite Higgs models and derivations of the equations until the Appendix. Section \ref{sec:model} also contains a discussion of dominant production and decay modes of fermionic top partners. The main results of the paper are discussed in detail in Section \ref{sec:results}. We include a detailed overview of our forward jet tagging proposal in Section \ref{sec:fwdjet}, as well as discuss our simplified $b$-tagging algorithm and the boosted jet tagger in Sections \ref{sec:tagger} and \ref{sec:btag}. We present the results on the sensitivity of Run-II LHC searches for 14 TeV to heavy top partner masses  in Section \ref{sec:sensitivity}. Finally, in Section \ref{sec:ifsignal} we comment on the use of various final states to extract additional information about top partners, if a signal is ever observed. A highly detailed description of our benchmark composite Higgs model, top partner production mechanisms and decays can be found in the Appendix.

\section{Partially composite Top partners}\label{sec:model}
\subsection{Brief Description of the Benchmark Model} \label{sec:description}
In this articles, we use the Minimal Composite Higgs Model (MCHM)~\cite{Agashe:2004rs} as a benchmark for illustrating the performance of our event selection searches for top partners. Here we give a brief overview of the model features important for our phenomenology study. For a detailed description of the model see Appendix~\ref{sec:app3}. 

The Higgs doublet in MCHM is a Goldstone boson multiplet which arises from the breaking of a global $SO(5)\times U(1)_X$ down to $SO(4)\times U(1)_X\simeq SU(2)_{R}\times SU(2)_{L}\times U(1)_X$ of a strongly coupled theory. The $SU(2)_{L}$ and  a $U(1)$ subgroup of $SU(2)_R\times U(1)_X$ are gauged in order to provide the electroweak gauge bosons.\footnote{The inclusion of the  $U(1)_{X}$ factor is required  in order to assign different hypercharges $Y = T^3_{R} + X$  to the up-type and down-type quarks.} 

The low energy description of the strongly coupled sector with ``weakly coupled'' deformations is expected to contain additional scalar, fermionic and vector resonances, typically at a scale $g_{*}f$, where $f$ is the scale of compositeness and $g_{*}$ is a strong coupling, $\mathcal{O}(1) \le g_{*} \le 4\pi $.
Electroweak precision measurements tend to push the mass bounds on  scalar and vector resonances towards the multi-TeV range while light ($\sim$ TeV) fermionic resonances are required
in order to accommodate an effective potential for the Higgs which induces Electroweak Symmetry Breaking (EWSB) and the Higgs mass. 

We use a bottom-up approach and only include a minimal set of light fermionic resonances: a top partner multiplet in the $\bf 5$ of $SO(5)$. The partner multiplet contains a partner with electric charge $5/3$ $(X_{5/3})$, a partner with charge $-1/3$ ($B$), and three partners with charge $2/3$ ($T_{f1,2}$ and $T_s$), where $T_s$ is a singlet while the other four states form a $\bf 4$  under the $SO(4)$.  A generic feature of composite Higgs models is that the $5/3$ charge partner ($X_{5/3}$) is the lightest state amongst the partners in the $\bf 4$. Furthermore, if one neglects the electrical charge sign of the decay products, the phenomenological signatures of $X_{5/3}$ and the $B$ are identical. We will hence focus our effort on searches for $X_{5/3} / B$ states and postpone the searches for other top partners until future studies.

Upon the diagonalization of the mass matrix (see the Appendix for more detail) the masses of the top, and the partners, are given by:
\begin{eqnarray}
	m_t \,\,\,\,\,\,\,&=&\frac{v}{\sqrt{2}}\frac{| M_1-e^{-i\phi} M_4|}{f} \frac{y_L f}{ \sqrt{M_4+y^2_L f^2}}\frac{y_R f }{ \sqrt{|M_1|^2+y^2_R f^2}}+\mathcal{O}(\epsilon^3), \label{eq:topmass} \nn \\
	M_B\,\,\,\,\, &=& \sqrt{M_4^2+ y_L^2 f^2},  \nn\\
	M_{X_{5/3}} &=& M_4,\nn\\
	M_{Tf1}&=& M_4 +\mathcal{O}(\epsilon^2),\nn\\
M_{Tf2}&=& \sqrt{M^2_4+y^2_L f^2} +\mathcal{O}(\epsilon^2),\nn\\
M_{Ts}&=& \sqrt{|M_1|^2+y^2_R f^2} +\mathcal{O}(\epsilon^2), 		
 \label{eq:massesm}
\end{eqnarray}
where $M_1$ and $M_4$ are the singlet and fourplet mass scales, $\phi$ is a relative phase between them (see~\cite{Azatov:2014lha} for a detail discussion on the model's flavor parameters), $f$ is the compositeness scale, $y_{L,R}$ are the left handed/right handed pre-yukawa couplings, and $\epsilon \equiv v/f$. 
Eq. \eqref{eq:massesm} reveals an important point which we will employ in the following sections. The mass splitting between the $M_{5/3}$ and $B$ goes as $f/M_4$, implying that the heavier the $X_{5/3}$ partner is, the more mass degenerate it becomes with the $B$ state, provided $y_L$ is not too big. 

Our current study will focus only on the $tW$ decays of the top partners, since this is the only mode $X_{5/3}$ can decay to due to charge conservation. The dominant couplings of  $X_{5/3}$ and $B$ states are of strength

\bea
|g^{R}_{XWt}|&=& \left|\frac{g}{\sqrt{2}}\frac{\epsilon}{\sqrt{2}}\left(\frac{ y_R f |M_1|}{M_4 M_{Ts}}-\sqrt{2} c_R \frac{e^{-i \phi } y_R f}{M_{Ts}}\right)+\mathcal{O}(\epsilon^2)\right|,\nn\\
\left|g^{R}_{BWt}\right|
&=& \left| \frac{g}{\sqrt{2}}\frac{\epsilon}{\sqrt{2}}\left(\frac{ y_R f \left(|M_1| M_4+e^{-i \phi}  y_L^2  f^2 \right)}{ M^2_{Tf2} M_{Ts} }-\sqrt{2} c_R  \frac{e^{-i \phi } y_R f}{ M_{Ts} }\right)+\mathcal{O}(\epsilon^2)\right|,
\label{eq:XWtcouplm}
\eea
where $c_R$ is a right-handed strong sector coupling between the partners in the $\bf 1$ and $\bf 4$.\footnote{Notice that these couplings are chiral, where the partner couplings to left handed tops are suppressed by $O(\epsilon^2)$. The dominance of  right hand couplings to tops result in characteristic features in the angular and $p_T$ distributions of the top decay products \cite{Agashe:2006hk,Almeida:2008tp} and could  help reveal the structure of top partner couplings (in case a signal is observed at the future LHC runs). } 

\subsection{Production of Top Partners}\label{sec:model.prod}
\begin{figure}[t]
\begin{center}
\begin{tabular}{cc}
\includegraphics[width=.5\textwidth]{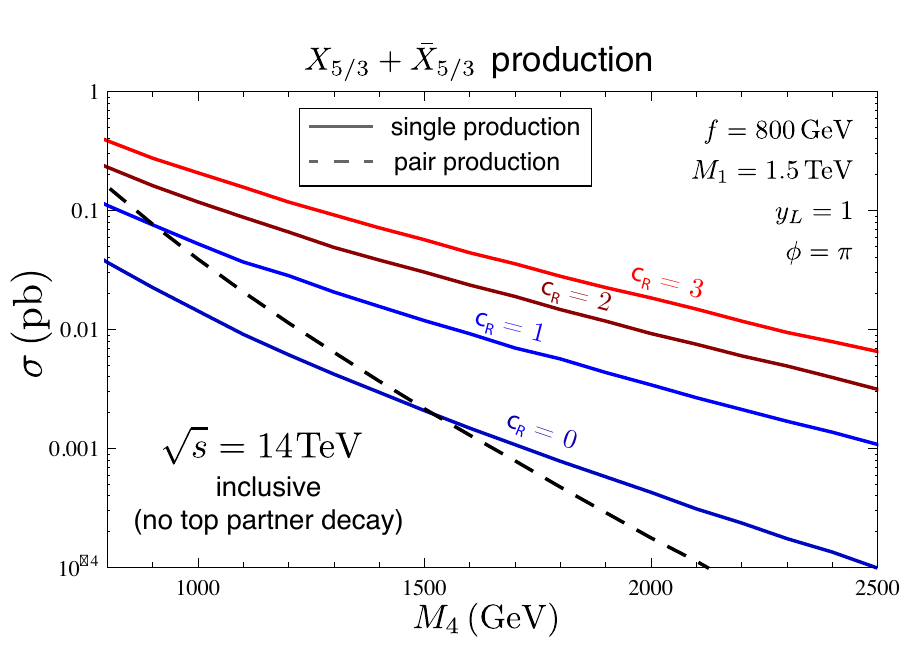} &
\includegraphics[width=.5\textwidth]{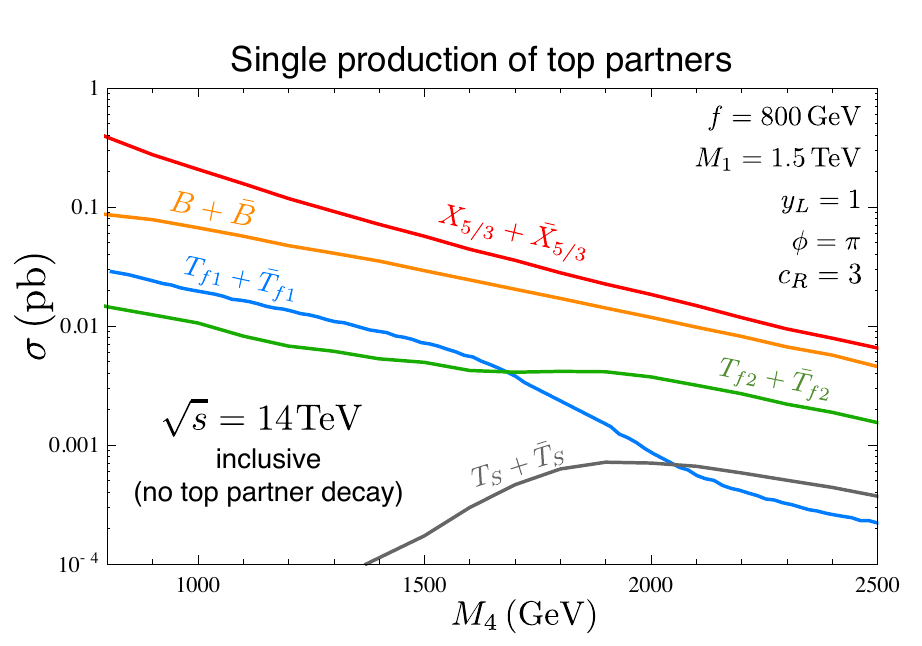} 
\end{tabular}
\end{center}
\caption{Production cross sections of partially composite top partners:\\
Left: Pair-production cross section for $X_{5/3}$ and single-production cross section for $X_{5/3}$ or $\bar{X}_{5/3}$ as a function of $M_4$ for different values of $c_R$. Other parameters are fixed to $f=800$ GeV, $M_1=1.5$ TeV, $\phi= \pi$,  $y_L = 1$\,. \\
Right: Single-production cross section for $X_{5/3}$ or $\bar{X}_{5/3}$, as compared to single-production cross section of other top-partners and their antiparticles as a function of $M_4\,$.  Other parameters are fixed to $f=800$ GeV, $M_1=1.5$ TeV, $\phi= \pi$, $c_L=c_R=3$, $y_L = 1\,$. } 
 \label{fig:prodX53}
 \end{figure}

The top partners are colored and can therefore be pair-produced via QCD interactions, where the production cross section only depends on the mass of the respective top partner. The top partners can also be single-produced via the interactions of Eq.\eqref{eq:XWtcouplm}. For low top partner masses, pair production dominates, but for higher top partner masses, single-production becomes kinematically favorable, as can be seen in Fig.~\ref{fig:prodX53}.\footnote{The single production cross section depends on the model parameters beyond the mass of the top partners as can be seen already from the couplings in Eq.\eqref{eq:XWtcouplm}. Hence, the top partner mass scale at which single production becomes dominant depends on the model parameters. We will return to this point momentarily.} Since here we are interested in \textit{heavy} top partners, we will focus our attention on single production only.

\begin{figure}[t]
\begin{center}
\includegraphics[width=0.8\textwidth]{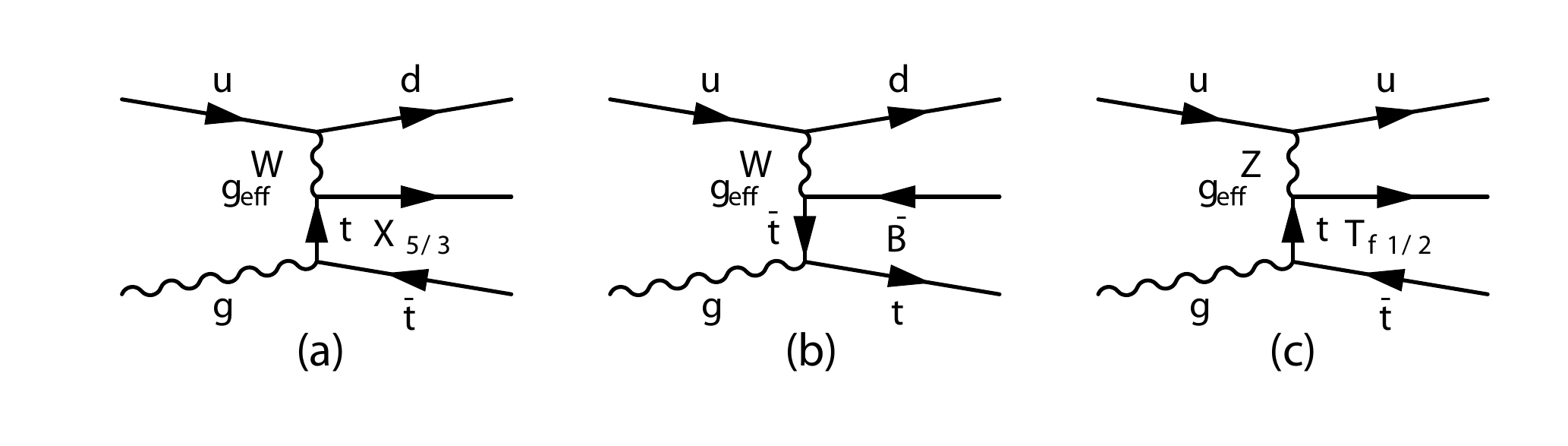}
\end{center}
\caption{Dominant single-production channels for the top partners $X_{5/3}$, $B$, $T_{f1}$ and $T_{f2}$ (from left to right) at a proton-proton collider.} 
 \label{fig:prodch}
 \end{figure}

Fig.~\ref{fig:prodch} shows the dominant production channels for the respective top partners.
The $X_{5/3}$ partner is produced together with a jet and an anti-top, where the dominant effective coupling is right-handed. Due to the larger up quark PDF in the proton, $X_{5/3}$ production is preferred as compared to $\bar{X}_{5/3}$ production, which requires a $d$ or a $\bar{u}$ in the initial state. The $\bar{B}$ is produced together with a jet and a top via a right-handed coupling with preference of $\bar{B}$ over $B$ production, again due to the larger up quark PDF. The fourplet top partners $T_{f1}$ and $T_{f2}$ are produced together with a jet and a top via a right-handed coupling. Analogously, their anti-particles are produced together with a jet and an anti-top. As their production arises from a $Z$ which is radiated off an initial state $u$, the production rates for them and their antiparticles are comparable. Finally, the singlet top partner $T_s$ dominantly couples to $Wb$ via a left-handed coupling. It can thus be produced together with only a jet, but requires a (PDF suppressed) $b$ quark in the initial state. Due to the larger up-quark PDF, $T_s$ production is preferred over $\bar{T}_s$ production at a proton-proton collider. 

 The effective couplings Eq.~(\ref{eq:XWtcouplm}) relevant for single-production\footnote{The analogous couplings for the charge $2/3$ partners are given in Eq.~(\ref{eq:XWtcoupl}).} depend not only on the mass of the top partners but also directly on the other model parameters -- in particular on $c_R$ (for $X_{5/3}, T_{f1}, T_{f2}$ and $B$ single production) and $c_L$ (for $T_s$ single production) -- but also on the pre-yukawa couplings and the relative phase $\phi$. In the large $c_{L,R}$ limit, the production cross sections of the fourplet (singlet) states scale with $|c_R|^2$ ($|c_L|^2$). For $c_R\sim 1$, the first ($c$-independent) and second ($c$-dependent) term contributions to the effective couplings in Eq.~(\ref{eq:XWtcouplm})  become comparable in magnitude, and can cancel or enhance each other depending on the phase of $c_R$ and $\phi$. As an illustration, the left panel of Fig.\ref{fig:prodX53} shows  the single production cross section of  $X_{5/3}$ and $\bar{X}_{5/3}$ for different values of $c_R$ as a function of $M_4$, where we fixed the other model parameters to $f=800$ GeV, $M_1=1.5$ TeV, $\phi= \pi$,  $y_L = 1$, and $y_R\sim O(1)$, while making sure to reproduce the top mass. For comparison we also show the pair production cross section for $X_{5/3}+ \bar{X}_{5/3}\,$.
 
In the limit of large $c_{R}$, the production cross section of the $\bar{B}$ is marginally lower than the one for $X_{5/3}$ because the  $B$ is slightly heavier. The $T_{f1,2}$ production cross sections are lower because the dominant production channel involves two couplings to the $Z$ rather than to the $W$ which yields a suppression of $(\frac{g/2 c_w}{g/\sqrt{2}})^4\sim 0.4\,.$\footnote{For $c_R\sim\mathcal{O}(1)$ the situation becomes more involved because the dependence of $g^{R}_{XWt}$, $g^{R}_{BWt}$, and $g^{R}_{Tf1,2Wt}$ on the phase of $c_R$ and $\phi$ differ.}  As an illustration, Fig.~\ref{fig:prodX53}, right panel, shows  the single production cross section of  $X_{5/3}$, $B$, $T_{f1}$, $T_{f2}$, $T_s$ and their antiparticles as a function of $M_4$ where we fixed the other model parameters to $f=800$ GeV, $M_1=1.5$ TeV, $\phi= \pi$, $c_L = c_R = 3$, $\lambda_L = 1$, and $y_R$  by the requirement to reproduce the top mass.

\subsection{Top Partner Decays}

For all top-partners, the dominant couplings to $W,Z,h$ and an SM quark are chiral (either left- or right-handed coupling dominates). In this case, the partial widths for a decay of a fermion $F$ into a fermion $f$ and a gauge boson or Higgs are
\bea
\Gamma(F\rightarrow W f) &=&M_F\frac{M^2_F}{m^2_W}\frac{|g|^2_{\rm eff}}{32 \pi}\Gamma_W\,,\\
\Gamma(F\rightarrow Z f)  &=&M_F\frac{M^2_F}{m^2_W}\frac{|g|^2_{\rm eff}}{32 \pi}\Gamma_Z\,,\\
\Gamma(F\rightarrow h f)  &=&M_F\frac{|\lambda|^2_{\rm eff}}{32 \pi}\Gamma_h\,,
\eea
where $\Gamma_{W,Z,h}= 1+\mathcal{O}\left(\frac{m^2_{W/Z/h}}{M^2_F}\right)$ are kinematic functions, and $M_F$ is the mass of the fermion. Using these relations we can estimate the partial widths and BRs of the different top partners, using the effective couplings Eqs.~(\ref{eq:XWtcoupl} - \ref{eq:Tshtcoupl}).

\bigskip
For the $X_{5/3}$ partner one obtains\\
\beq
\Gamma(X_{5/3}\rightarrow W t) \approx M_4 \frac{M^2_4}{m^2_W}\frac{|g|^2_{\rm eff}}{32 \pi} \\
= M_4 \, \frac{y^2_R}{32 \pi} \left|\frac{|M_1|-\sqrt{2}c_R e^{-i\phi} M_4}{M_{Ts}}\right|^2\,.
\eeq
There are several interesting features of the $X_{5/3}$ decay width to $tW$. First, note that although the effective coupling is $\mathcal{O}(\epsilon)$, the partial width is not $\epsilon$ suppressed. For large $c_R$ (and $M_1$ and $M_4$ of similar size), it is proportional to $|y_R^2 c^2_R|$. For $|y_R^2 c^2_R|\lesssim 5$, this still yields a narrow resonance $(\Gamma/M\lesssim 25\%)$, but for larger values of $y_R c_R$ the resonances become broad. Resonances of ultra-large widths are difficult to measure since they tend to ``blend'' into the continuum spectra of differential cross sections. Hence, sections of parameter space which can be probed by the future LHC runs are limited by the width/mass resolution.

Since $X_{5/3}$ is the lightest partner state in the $\bf 4$ (fourplet) such that decays into $B,T_{f1},T_{f2}$ and SM particles are kinematically forbidden, hence $X_{5/3}$ always decays into $Wt$.\footnote{The singlet partner $T_s$ can be lighter than the $X_{5/3}$ if $M_1 < M_4$, but even then, the ``cascade'' decay $X_{5/3}\rightarrow W T_s$  is  kinematically suppressed by a factor $\sim(1-(M_{Ts}/M_4)^2)^2$, so that this it only plays a role when we have substantial mass splitting. We do not consider this extreme case further as this scenario, in which the fourplet partners are at mass scale substantially above the singlet partner  is much better tested by searching for $T_s$, directly.} 

For the $B$ decay width and its branching ratios, the analogous discussion applies. The total $B$ width is of similar size as the $X_{5/3}$ width (\cf Appendix \ref{sec:app3} for the explicit expression). The decay $B\rightarrow W t$ dominates over $B\rightarrow Z b$ and $B\rightarrow h b$ because effective couplings for the latter decays are of higher order in $\epsilon$. ``Cascade'' decays $B\rightarrow W T_{f1,2}$ are kinematically suppressed (if not forbidden) due to the small mass splitting between $B$ and $T_{f1,2}$. 

For more details on top partner decays see Appendix \ref{sec:app3}. 
\bigskip
\bigskip
\subsection{Single Production Cross Section - Same Sign Di-leptons vs. Lepton-Jet Final States}

In addition to very interesting event topology, the single $\Xx$ production is also interesting because at high enough $\MX$ it becomes the dominant production mode.
The kinematics of singly produced $\Xx$ events are mostly determined by two parameters: $\MX$ and $\Gamma_{\Xx}$ (modulo effects of spin correlations), while the production cross section is subject to many other model parameters. Here we are not interested in details of models but in general features of $t\bar{t}Wj$ event topologies and will hence leave the production cross section as a free parameter. We consider a range of $\MX$, while keeping the width $\Gamma(\Xx) \sim 15-20 \%$ of $\MX$. Keeping the cross section a free parameter has an additional benefit of presenting the analysis in a model independent fashion and being able to apply our results to other new physics searches in the $t\bar{t}Wj$ channel. 

In order to determine the ``reasonable range'' of cross sections, we consider several combinations of model parameters in a general partially composite model. We do not make any assumptions about the mass hierarchy in the model ({\it e.g.} we do not only consider the decoupling limit of $M_1 \gg M_4$), while we make sure that each model parameter point reproduces the correct $m_t$.

\begin{figure}[htb]
\begin{center}
\includegraphics[width=3.5in]{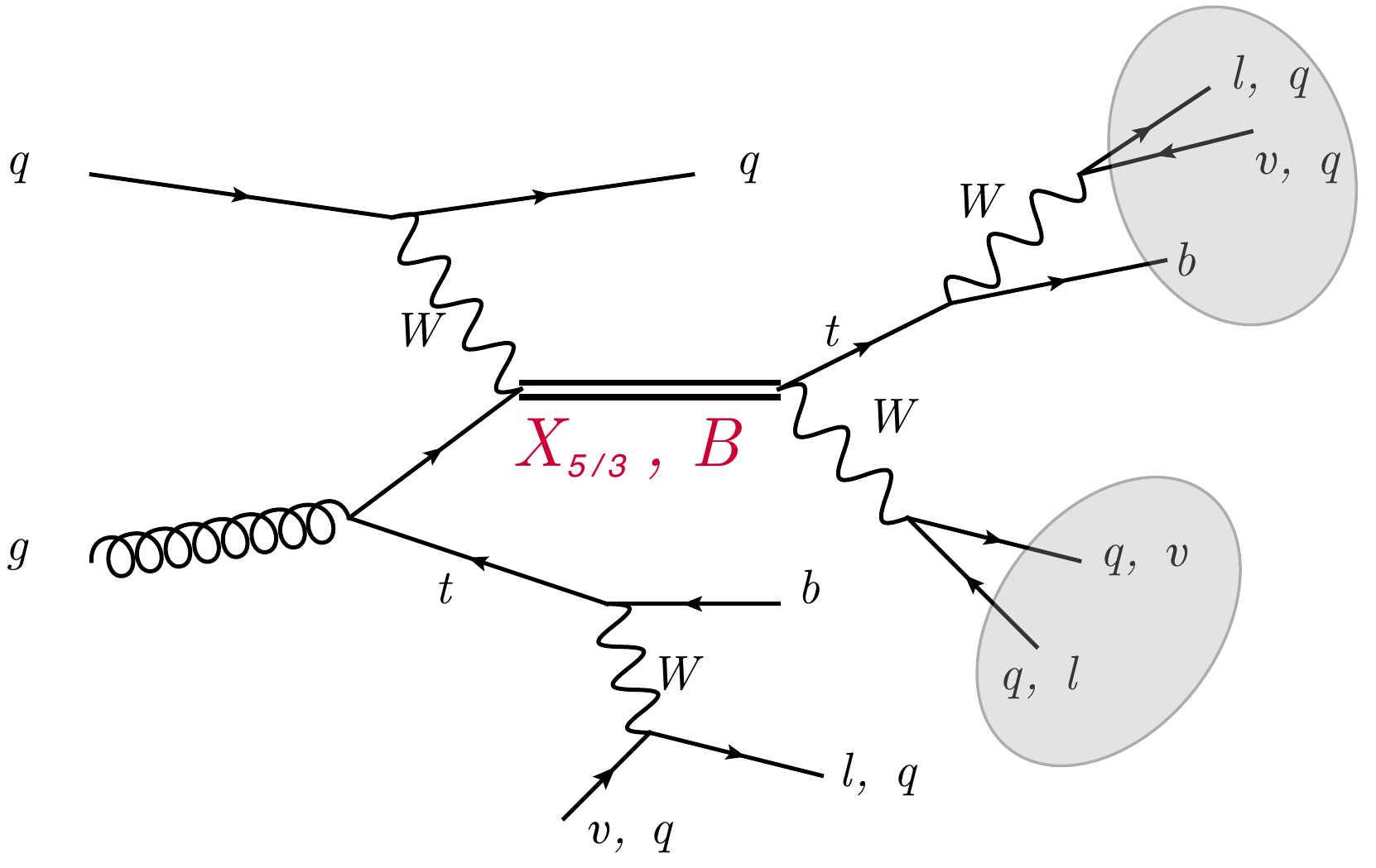}
\caption{Single production of top partners with decay channels. We consider events characterised by a boosted $tW$ system in the case of $\Xx$, as denoted by the ovals, in addition to a high energy forward jet and a top. Notice that the only difference in the $X_{5/3}$ production and $B$ production is the sign of the decay products' charges. We consider inclusive decays of the initial state radiated top. }
\label{fig:prod_diag}
\end{center}
\end{figure}

The current limits of $\Xx$ partners place $\MX \gtrsim 1 \TeV$. Hence, if $\Xx$ is to be found during the future runs of the LHC, it will be found almost exclusively in the events containing at least one boosted top quark and one boosted $W$. Previous searches for $\Xx$ partners focused mostly on the same sign di-lepton searches, due to the extremely clean signal, but at a cost of the signal rate. Compared to the inclusive single $\Xx$ production, the signal rate is diminished by the branching ratio of $W$ decays to leptons, resulting in $$\sigma_{2l} = \sigma_{\rm tot} \times {\rm Br}(W \rightarrow l \nu)^2 = \sigma_{\rm tot} \times (2/9)^2 \sim  0.05 \, \sigma_{\rm tot}\,,$$
where $\sigma_{\rm tot}$ is the inclusive $\Xx$ single production cross section. In addition, we checked that the geometric acceptance ($i.e.\, |\eta_l| < 2.5$) for two leptons in a same sign di-lepton final state is $~50\%$, implying that the total same sign di-lepton cross section is at least a factor of 2 smaller after the event selections.
Instead, here we propose to search for top partners in channels which contain at least one lepton and a fat jet. Fig.~\ref{fig:prod_diag} shows an example diagram of singly produced $\Xx$, including the decay modes, where we take the initial state radiated top to decay inclusively. Compared to the same sign di-lepton searches, the starting signal cross section in our search strategy is $$\sigma^* = \sigma_{\rm tot} \times2\times {\rm Br}(W \rightarrow l \nu) \times {\rm Br}(W\rightarrow j j) = \sigma_{\rm tot} \times 2 \times (2/9) \times (2/3) =  6 \,\sigma_{2l}\,,$$
if we consider both the top and the W decaying hadronically (but not simultaneously). Note that the signal cross section is increased roughly by an additional factor of two for high $\MX$, where we expect $X_{5/3}$ and $B$ to be nearly mass degenerate. The same sign di-lepton cross section, however,  remains the same at high $\MX$, as the top and the $W$ from the $B$ decay are of the opposite charge

\bigskip
\section{Results}\label{sec:results}
\begin{figure}[htb]
\begin{center}
\includegraphics[width=3.5in]{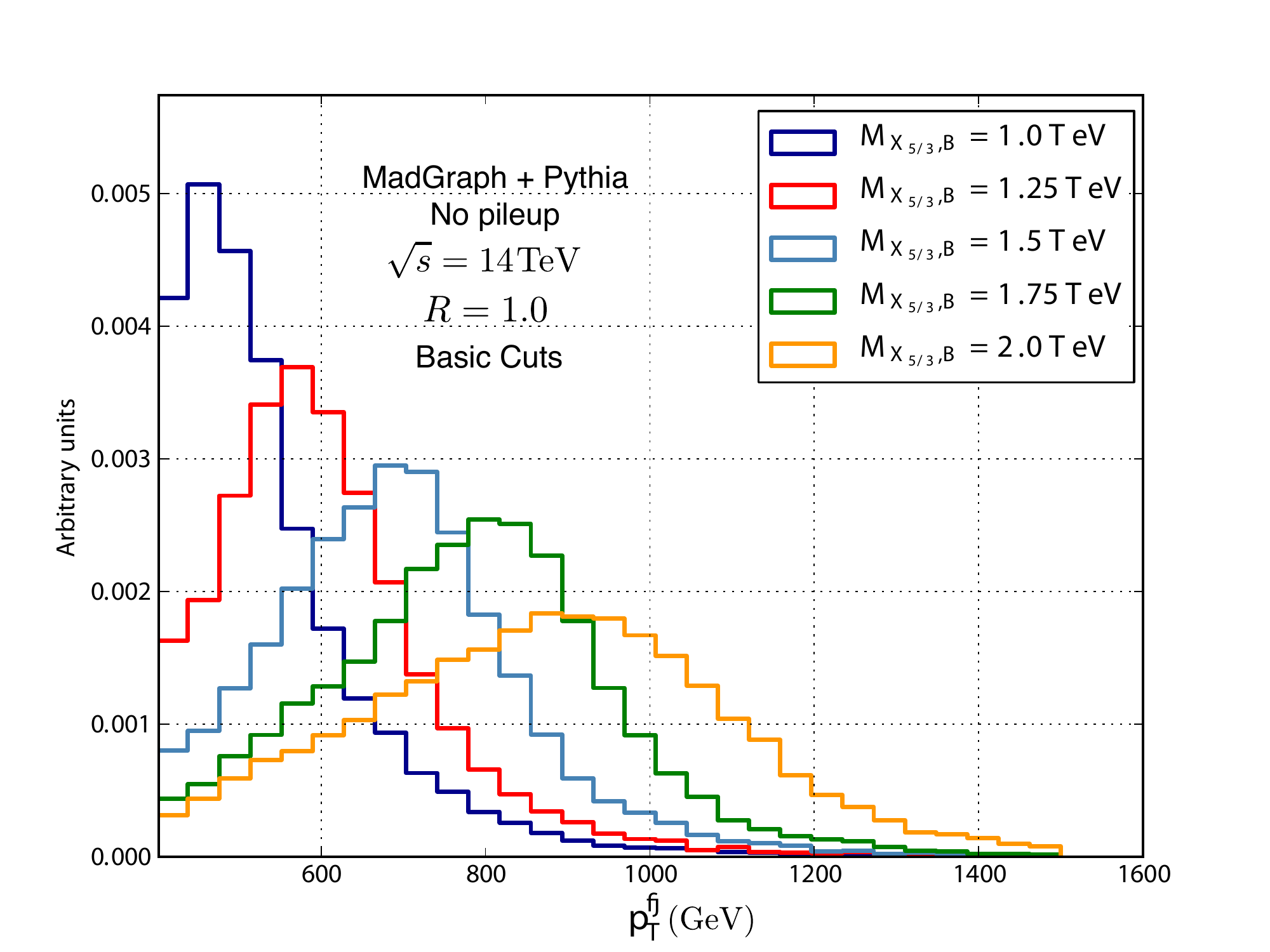}
\includegraphics[width=3.5in]{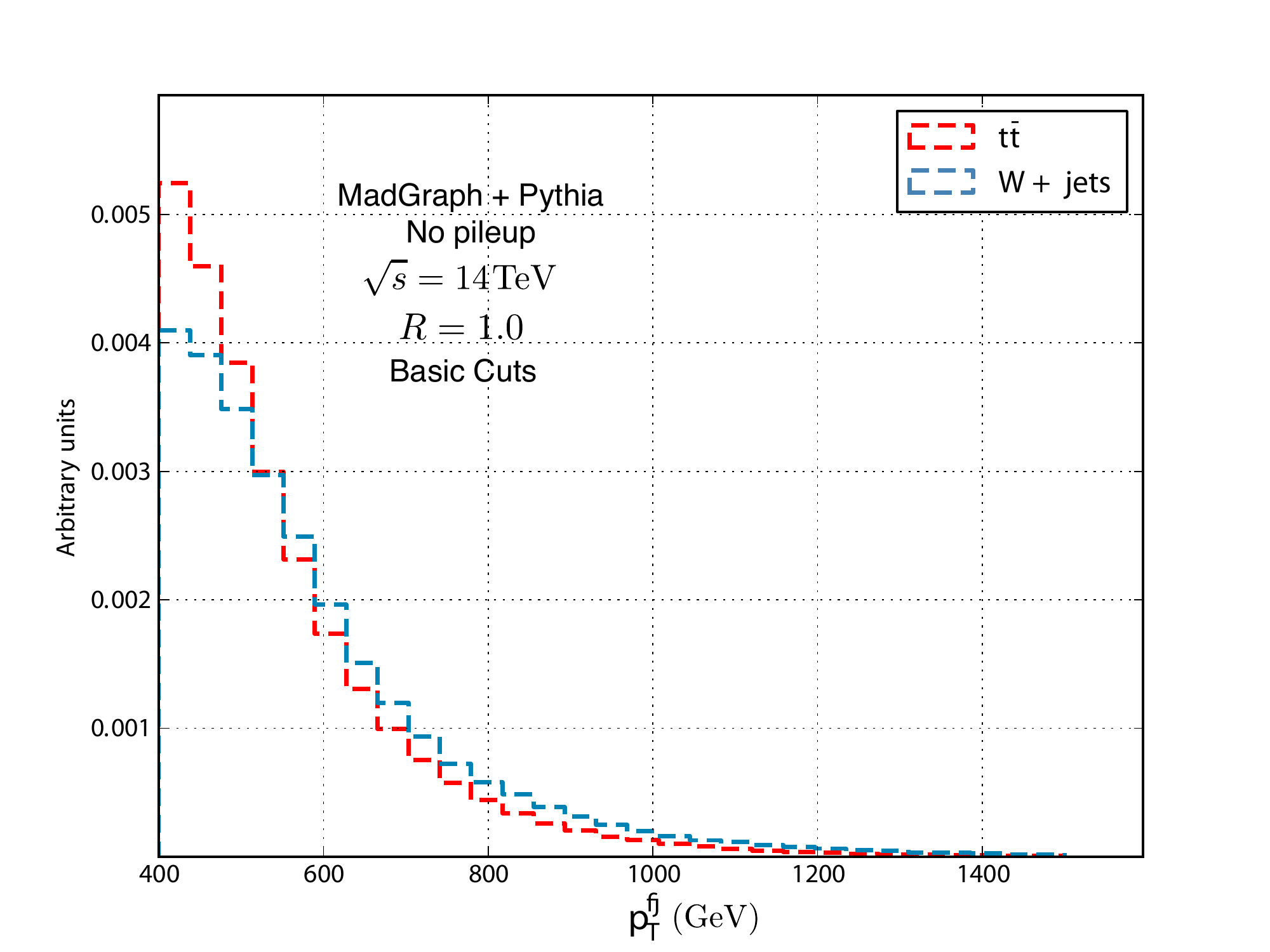}
\caption{Distribution of the hardest fat jet $p_T$. Left panel shows the signal distributions for various masses of $\MX$, while we show the backgrounds on the right panel. All plots are normalised to unit area.}
\label{fig:signal_pt}
\end{center}
\end{figure}

We proceed to discuss the main results of the paper. The signal events at a $\sqrt{s}~=~14~\TeV$ $pp$ collider are characterised by four distinctive features:
\begin{enumerate}
\item A single, high energy forward jet. 
\item One boosted $t$ or one boosted $W$ ($\MX \gtrsim 1 \TeV$), as can be seen in Fig.~\ref{fig:signal_pt}\,. 
\item One hard ($p_T > 100 \GeV$) lepton, resulting from a top or $W$ decay.
\item Two $b$ jets, one of which can be a part of a top fat jet.  
\end{enumerate}

Fig.~\ref{fig:signal_pt} shows the features of the signal and background fat jet $p_T$ spectrum. The $p_T$ distribution of background events is characterised by a steep decline as a function of transverse momentum. Conversely, the signal distributions tend to peak at roughly $\sim \MX / 2,$ with the PDF broadning effects becoming significant at high $\MX$, as the partner becomes more likely to be produced off-shell. 

As we will demonstrate in the following sections, our event selection based on the unique single $\Xx$ event topology, combined with boosted jet techniques, $b$-tagging and forward jet tagging can achieve sensitivity to $\Xx$ top partners over a wide range of model parameters at the $14 \TeV$ run of the LHC. We further argue that our results are comparable and in some cases superior to the same sign di-lepton searches, with an additional advantage that our method allows for the reconstruction of the resonance. 

In Section~\ref{sec:description} we pointed out that at large $\MX$ we expect the $X_{5/3}$ top partner and the $B$ to be nearly mass degenerate if the left hand yukawa coupling is not too large, a fact which has significant implications on the phenomenology of the heavy top partners and highlights a key advantage of our method over the same sign di-lepton searches. Since we do not consider the charge of the leptons as a part of the selection, the fact that the mass splitting between $X_{5/3}$ and $B$ is small means that our search is sensitive to both channels, effectively doubling the signal cross section. Conversely, requiring a presence of two same sign leptons would essentially veto the $B$ production, as the $B$ partner decays to a top and $W$ of the opposite charge. In the following sections we will consider the production of top partners both individually and under the assumption they are mass degenerate where relevant. 

\newpage
\subsection{Data Simulation and Event Pre-selection}

We generate all our simulated events at $\sqrt{s} = 14 \TeV$ $pp$ collider, using leading order \verb|MadGraph 5| \cite{Maltoni:2002qb} and shower them with \verb|Pythia 6| \cite{Sjostrand:2006za}, with a fixed renormalisation and factorisation scale and assuming the \verb|CTEQ6L| \cite{Pumplin:2002vw} parton distribution functions. In order to improve the statistics in the background channels, we impose a generation level cut of $H_T >  600 \GeV$ on the background events, where $H_T$ is the sum of all hard parton $p_T$ values in the event. We require that all final state hard level patrons are generated with $p_T > 15 \GeV, $ and impose a rapidity acceptance of $\eta_j < 5.0$ for the quarks and $\eta_l < 2.5$ for the leptons. The background $t\bar{t}$ events are matched to 1 extra jet while $W$+jets is matched up to four extra jets, using the MLM matching scheme \cite{Mangano:2006rw} with the matching parameters  $Q_{\rm min} = 20$ and \verb|xqcut = 30|.

For the purpose of pileup studies, we generate a large sample of minimum bias events using \verb|Pythia 6| with default tunes. We simulate the effects of pileup contamination on signal/background events by adding to each event a random number of pileup events drawn from a Poisson distribution centered around $\langle N_{\rm vtx} \rangle = 50$.

Next, we cluster the showered events using the \verb|FastJet| \cite{Cacciari:2011ma} implementation of the anti-$k_{T}$ algorithm \cite{Cacciari:2008gp}, where we use $R=1.0$ for the fat jets and $r= 0.4$ for the light and $b$-jets. For the purpose of pileup mitigation, it is useful to consider a smaller $R$ for higher $p_T$ fat jets. The pileup contamination scales as the jet area, hence a $R=0.6$ cone will experience only $\sim 40 \%$ of the pileup effects on a $R=1.0$ cone. However, decreasing the fat jet cone involves an elaborate procedure of calibrating jet energy scales and other systematics which is beyond the scope of our current work. For simplicity, here we will cluster all fat jets with $R=1.0$ with the caveat that pileup effects can further be mitigated by reducing the fat jet cone.

\begin{table}[!]
\begin{center}
\begin{tabular}{|c|c|c|}
\hline
Channel & $\sigma (H_T > 600 \GeV) [{\rm fb}]$ & $\sigma (H_T > 1.2 \TeV) [{\rm fb}]$ \\
\hline
$t\bar{t}+jet$ &  $14.4 \times 10^3$  & $1.2 \times 10^{3}$ \\
$W$+jets & $30.6 \times 10^3$ & N/A\\
\hline
\end{tabular}
\end{center}
\caption{Background channels to single $\Xx$ production, before Basic Cuts. We will only consider $t\bar{t}$ and $W+$ jets in our analysis. Branching ratio of $2/9$ for leptonic decays of the $W$ in $W$+jets is included in the cross section, as well as the branching ratio of $(2/9)\times(2/3)$, for the semi-leptonic $t\bar{t}$ decays. For improved statistics at high $\MX,$ we consider $t\bar{t}$ samples with two $H_T$ cuts, while we only take $W$+jets sample with $H_T > 600 \GeV$ since at the end of the analysis it is a sub-leading background. \label{tab:bgd0}}
\end{table}

We consider signal events in which both the top and the $W$ daughters of $\Xx$ decay leptonically (but not simultaneously), while we take the other, non-boosted top to decay inclusively. 

Table~\ref{tab:bgd0} shows a list of possible backgrounds and the corresponding cross sections. The main background channel in our search strategy is SM $t\bar{t}$ production and $W$ +jets, while even at generation level the other SM backgrounds are subleasing. Since we require at least one hard lepton in our analysis, we will only consider the background channels in which one of the tops or $W$ bosons decays leptonically.  We normalize the $t\bar{t}$ cross section to the NNLO result from Ref. \cite{Czakon:2013goa}, while the NLO corrections in $W+$ jets is not expected to be large. Here we will consider a conservative estimate of $K_{W+ {\rm jets}} = 2.0$ for the NLO $K$-factor.  In the following sections, we will show that our results are not strongly affected by the $W+$ jets $K$-factor. 

All events are subject to \textbf{Basic Cuts}:

\bea
	|\eta_{{\rm fj}, l, j} |< 2.5\,, \,\,\,\,\,\,\,\, p_T^{\rm fj} > 400 \GeV\,, \,\,\,\,\, \nn \\
	p_T^{l, j} > 25 \GeV\,, \,\,\,\,\,\,\, \MET > 40 \GeV, \nn \\
	N_{j}(\Delta R_{{\rm fj}, j} > r + R) \ge 1\,, \,\,\,\,\,\,\,\,\,\,\,
\eea

where $l$ represent the hardest lepton with mini-ISO $> 0.8$ \cite{Kaplan:2008ie},  ``fj'' stands for the fat jet, ``$j$'' stands for light jets, and $N_{j}(\Delta R_{{\rm fj}, j} > r + R) $ is the multiplicity of $r=0.4$ jets with $p_T > 25 \GeV$ and $|\eta_j| < 2.5$ which are isolated from the $R=1.0$ fat jet by $\Delta R > r + R$.

In addition to Basic Cuts, we consider a series of additional selections designed to further suppress the background channels while maintaining as much of the signal as possible. In order to suppress the $t\bar{t}$ background further we require
\begin{equation}
	m_{j^\prime l} > 200 \GeV,
\end{equation}
where $j^{\prime}$ is the hardest jet isolated from the fat jet by $\Delta R = R + r = 1.4$, and $l$ is the hardest mini-isolated lepton in the event. 
The rest of the cuts we employ in this analysis deserve more attention and are described in detail in the following sections.

\newpage
\subsection{Tagging of Boosted Objects} \label{sec:tagger} 

Events which pass the Basic Cuts are subject to jet-substructure analysis. Many available methods for boosted top tagging exist in the literature (see for instance Refs. \cite{Ellis:2007ib,Abdesselam:2010pt,Salam:2009jx,Nath:2010zj,Almeida:2011ud,Plehn:2011tg,Altheimer:2012mn,Soper:2011cr, Soper:2012pb, Jankowiak:2011qa, Krohn:2009th,Ellis:2009me, Backovic:2012jk, Backovic:2013bga, Hook:2011cq, Thaler:2010tr,Thaler:2011gf, Thaler:2008ju,Almeida:2008yp, Almeida:2011aa,Almeida:2010pa} and references therein). In addition, several interesting proposals for boosted $W$ tagging appeared recetly in Refs. \cite{Rentala:2014bxa, Cogan:2014oua, Larkoski:2014wba}). Here, we use the \verb|TemplateTagger v.1.0|~\cite{Backovic:2012jk} implementation of the Template Overlap Method \cite{Almeida:2011aa,Almeida:2010pa, Backovic:2012jj, Backovic:2013bga} as our boosted jet tagger,  by virtue of the weak susceptibility of the method to pileup contamination. 

TOM approach to  jet substructure aims to match the energy distribution of a jet to a parton-like configuration of heavy particle decays. The output of the method is the overlap score $Ov$, a measure of likelihood that a jet is, say, a top quark, a Higgs or a $W$ boson,  as well as the partonic configuration ($i.e.$ peak template) which maximized the $Ov$ score. The latter is of much importance, as one can in principle approximate the fat jet with the peak template. We will utilize this possibility in the following sections when considering effects of pileup on the measurements of the top partner mass.

Our analysis of jet substructure follows the prescription of Ref. \cite{Backovic:2013bga}, with the main difference that we divide events into hadronic top and hadronic $W$ candidates before analyzing the fat jets. Note that the our work in this paper represents the first use of TOM as a boosted $W$ tagger. We begin by selecting the hardest mini-isolated lepton in the event and determining whether it originated from a top quark or a $W$. If there is a $p_T > 25 \GeV, $ $r=0.4$ jet within $\Delta R=1.0$ from the hardest lepton, such that $m_{l_1, \, \MET, \, \rm lj} > 90 \GeV$ \footnote{For simplicity, here we determine the $z$ component of the neutrino by assigning $\eta_{\nu} = \eta_{l}$. We find that this is a good approximation in the boosted regime. } we declare that the lepton is a part of a leptonically decaying top, and hence the hardest fat jet in the event is a $W$ candidate. Otherwise, we declare that the $W$ decayed leptonically, and that the hardest fat jet is a top candidate. An alternative method of determining the ``candidacy'' of a fat jet would be to simply use the fat jet invariant mass cut, but such a choice requires techniques to subtract or correct for pileup contamination of the jet mass. Here, instead,  we aim for pileup insensitive criteria for both jet substructure and event selection, whenever possible. 

The leptonically decaying $t/W$ also serves as a pileup insensitive estimator of the fat jet $p_T$ in the Template Overlap analysis, as the fat jet and the leptonic object recoil agains each other. We find that the scalar sum of the leptonic object constituent's $p_T$ ($i.e.$ the lepton, missing energy and ,if the leptonic object is a top, a light jet) is a good estimator of the fat jet transverse momentum \cite{Backovic:2013bga}. 

\begin{figure}[htb]
\begin{center}
\includegraphics[width=3.2in]{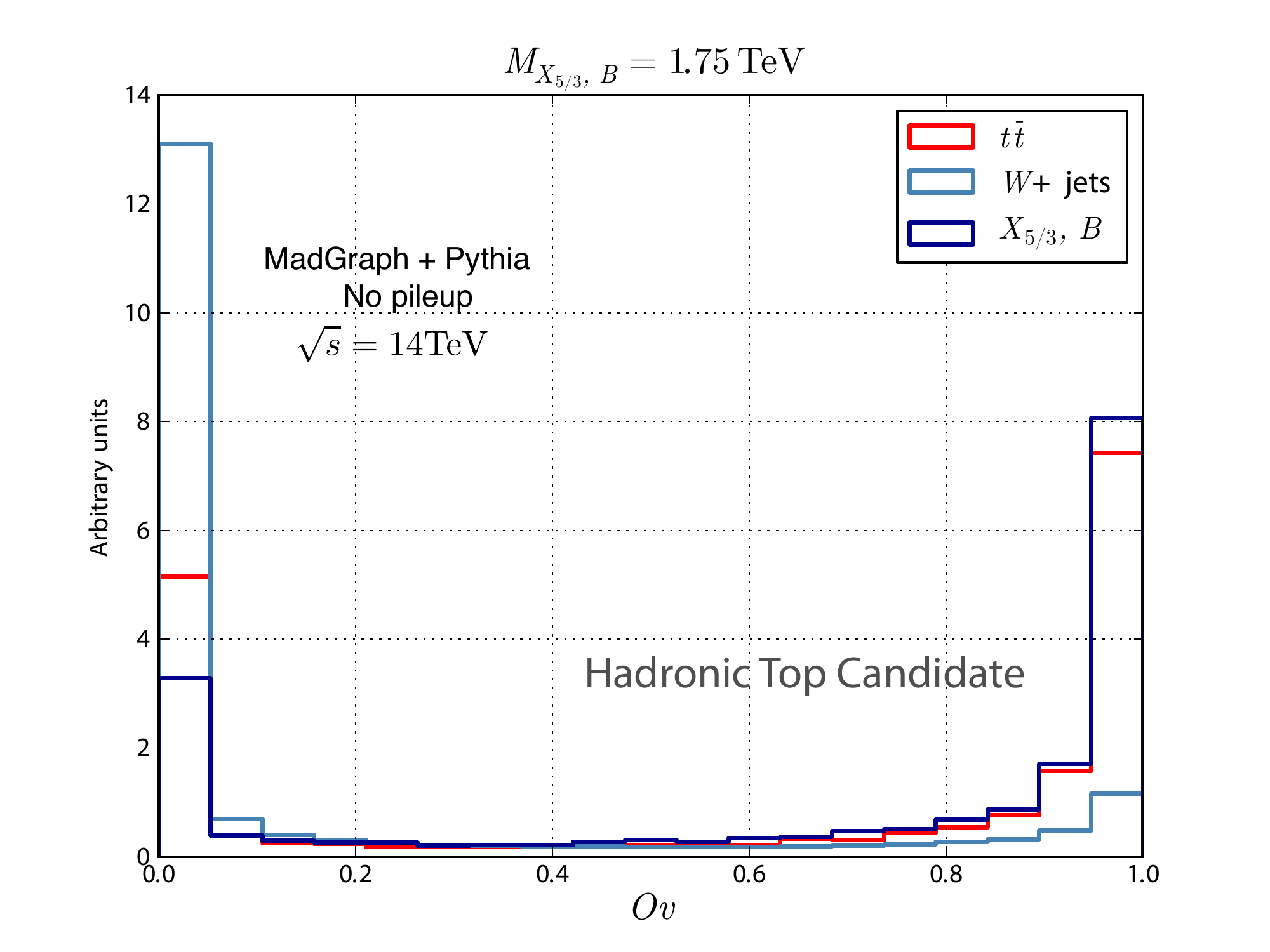}\includegraphics[width=3.2in]{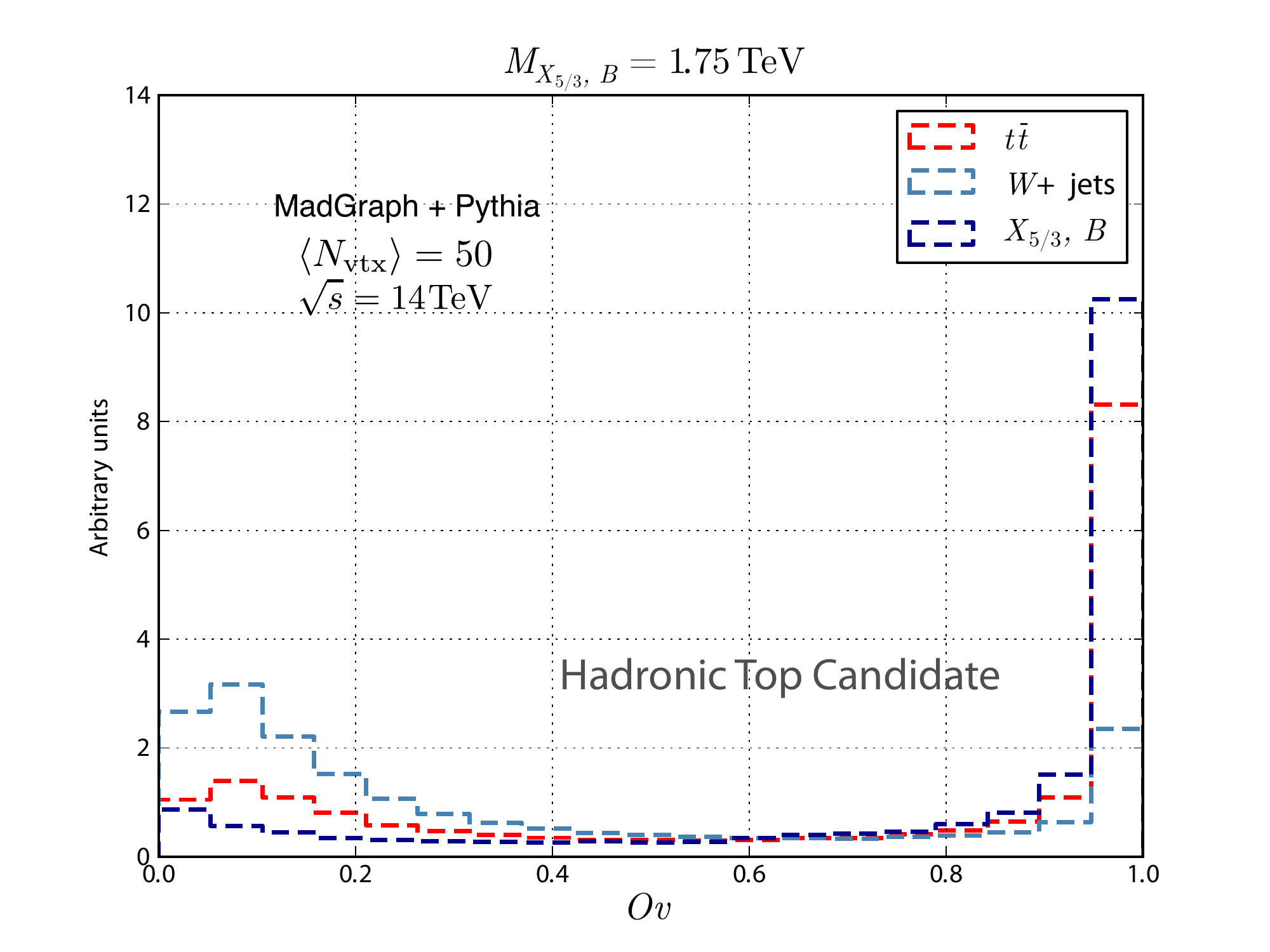}
\includegraphics[width=3.2in]{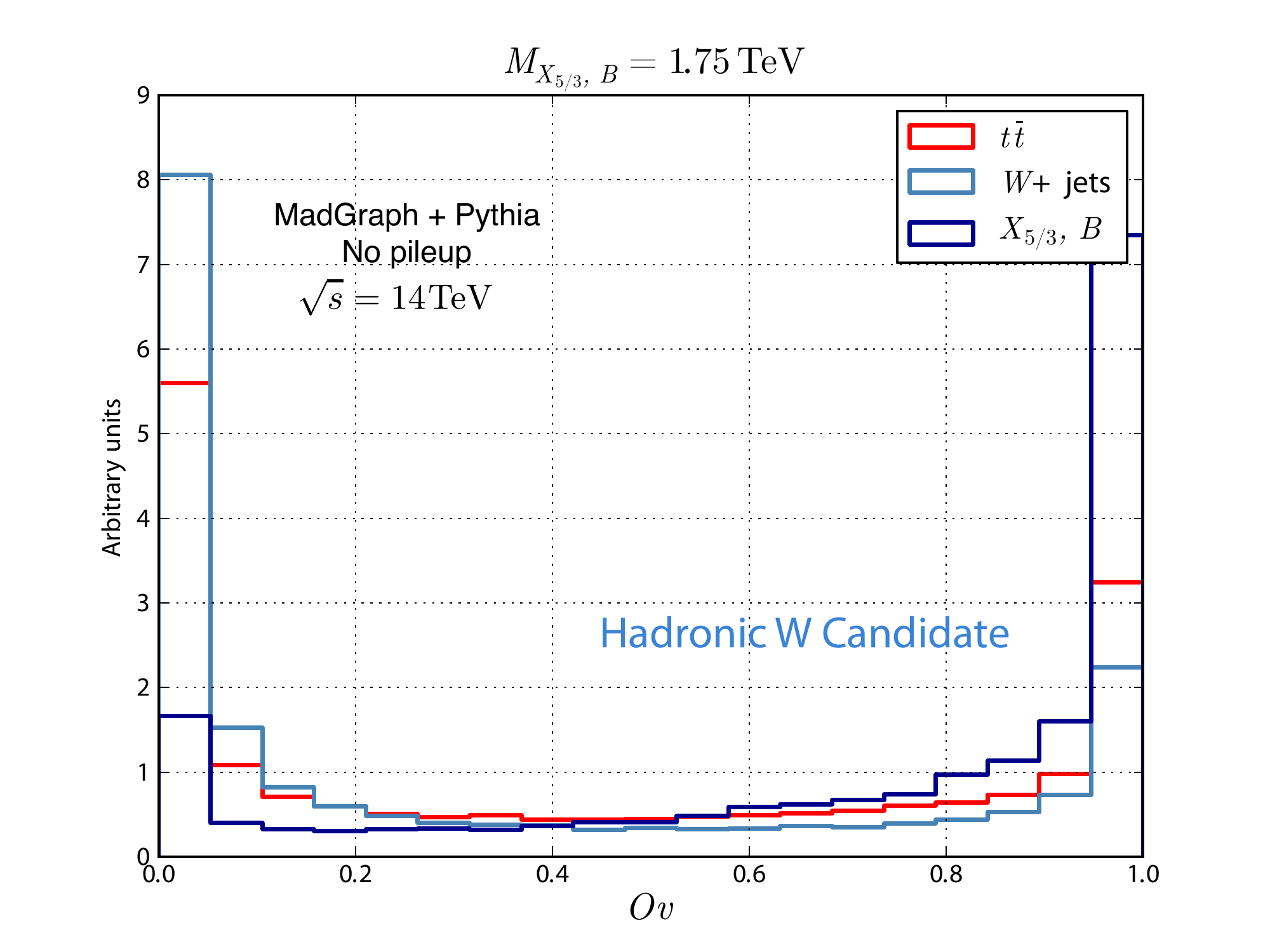}\includegraphics[width=3.2in]{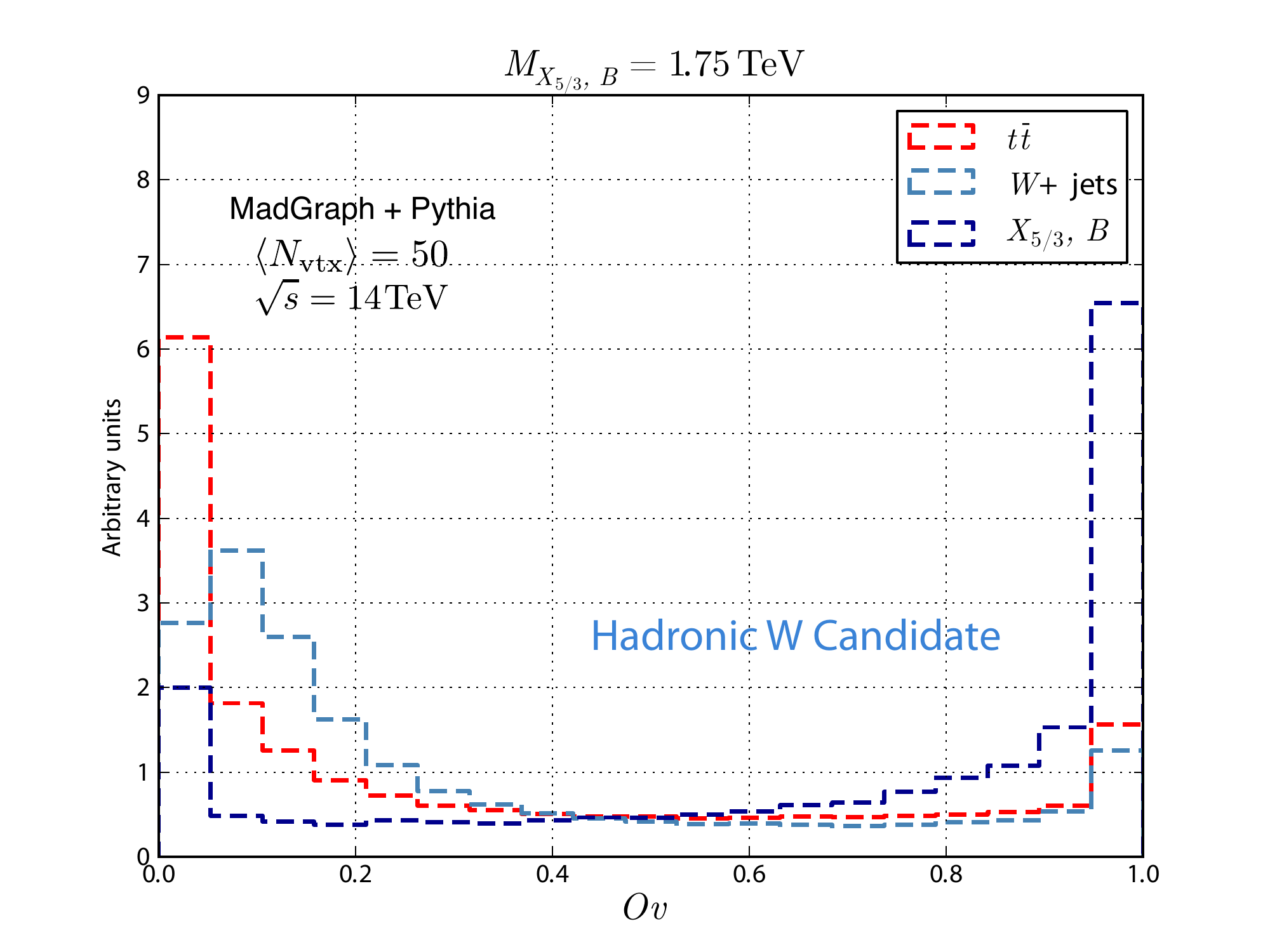}
\caption{Template overlap distributions for signal and background channels. The left panel shows the peak template distributions for hadronic $t/W$ (top panel / bottom panel) candidate events with no pileup (solid lines), while the right panel is the peak overlap for hadronic $t/W$ (top panel /bottom panel) candidate events in the presence of 50 average pileup events (dashed lines). The plots assume Basic Cuts and $p_T > 500 \GeV$ for the fat jet. Notice that the signal distribution is weakly affected by pileup contamination. }
\label{fig:overlap}
\end{center}
\end{figure}

In order to speed up the numerical calculations, we generate template states at fixed $p_T$, and use 10 bins of width $\delta p_T=100 \GeV$, starting from $p_T = 450 \GeV$. The templates are produced assuming $50$ steps in $\eta, \phi$, while we scale the template sub-cones using the $p_T$ scaling rule of Ref. \cite{Backovic:2012jj}. We produce two separate sets of templates, the three body template sets for top states and two body template sets for the $W$ states, where we use the appropriate set based on whether the fat jet is a top candidate of a $W$ candidate. Note that the use of the \textit{leptonic} top TemplateTagger does not add much to the analysis, as the background objects already contain a leptonically decaying top (in case $W$ is the fat jet), and the leptonic $W$ is too simple of an object to require a substructure analysis (in case that $t$ is the fat jet). 

Finally, for an event to pass our boosted object selection, we require that the fat jet has an overlap score:
\beq
	Ov > 0.5,
\eeq
for both the hadronic top and hadronic $W$ candidates.

Figure~\ref{fig:overlap} shows an example distribution of Template Overlap for signal and background events, after the Basic Cuts. The left panel shows only the events which were categorized as hadronic top candidates, while the right panel shows the corresponding plot for hadronic $W$ candidates. In both cases the $W$+jets events are rejected very well by TOM, as our lepton requirement deems that the $W$ decays leptonically and the fat jet is hence either a light jet or a combination of light jets which get clustered together. Semi leptonic $t\bar{t}$ events are more challenging to reject via Template Overlap, since the final state content in terms of jet substructure is more similar to signal events. If a $t\bar{t}$ is categorised as a hadronic top candidate, TOM will likely tag the event with a high overlap score, since the fat jet is indeed a hadronically decaying top. If the events is categorized as a hadronic $W$ candidate, the expected peak overlap score will likely be lower since TOM will try to match the substructure of a top to a decay of a W boson. 

It is important to note that when it comes both to $t\bar{t}$ and $W$+jets background, higher order effects on the \textit{shape} of the kinematic distributions become significant at high energies. Extra hard gluons are likely to appear in a highly energetic $t\bar{t}$ final state, causing the top-antitop system not to appear back to back in the transverse plane.  Such ``asymmetric'' events offer an additional handle to reject Standard Model di-top events. Proper treatment of the effect requires a full NLO event simulation, which is beyond the scope of our current study. It is impotent to note that since here we only consider a leading order $t\bar{t}$ sample matched to one extra jet, our estimates for the Template Overlap's ability to reject Standard Model $t\bar{t}$ events is likely underestimated. 

One of the most attractive features of TOM is its weak susceptibility to pileup contamination. Refs. \cite{Backovic:2012jj, Backovic:2013bga} showed that the effects of pileup are not significant on TOM  (up to 50 pileup events). The low susceptibility to pileup is a manifest of the fact that, by construction, TOM is sensitive mostly to the hard energy depositions within the fat jet and less so to the incoherent soft radiation. Here we find similar results both in the case of the top as well as the W, as shown in Figure~\ref{fig:overlap}. The signal distributions maintain a very similar shape upon the addition of pileup contamination, with the signal efficiency of the $Ov > 0.5$ cut remaining at $\sim 65 \%$ for both hadronic top and hadronic $W$ candidate events. The shape of the background distributions is affected more drastically in the presence of pileup, however, notice that the region of $Ov > 0.5$ remains weakly affected, resulting in a small effect on the background fake rate upon the overlap selection cut.

\newpage
\subsection{Forward Jet Tagging}\label{sec:fwdjet}

The event topology in Fig.~ \ref{fig:prod_diag} offers another interesting handle on background mitigation -- a high energy forward jet \cite{DeSimone:2012fs}. 
The question of how well forward jet tagging (FJT) will perform in the high pileup environment of the future LHC runs remains open \cite{ATLASPileupJets, CMS:2014ata}. Yet, there is much interesting physics one can do with forward jets. 
Single top production, tagging Higgs events which originate from vector boson fusion and understanding of the proton structure at high $x$ are just some of the examples. 
Here we are interested in forward jets only as event tags. The problem of forward jet tagging hence becomes simpler, as we are not concerned with precise measurements of forward jet energy and transverse momentum.

We propose a novel approach to forward jet tagging, which addresses the effects of pileup contamination (at $50$ interactions per bunch crossing). Pileup contribution  to jet $p_T$ goes as $\delta p_T \sim R^2$, where $R$ is the jet cone, resulting in a shift of the jet kinematic observables to higher values and a broadening of the kinematic distributions. In addition, larger jet cones are more likely to produce fake pileup jets, thus increasing the overall forward jet multiplicity.  In order to limit the pileup contamination in the forward region, here we propose to cluster the jets in the forward region with a cone smaller than the standard $r = 0.4$ ($i.e.$ $r=0.1, 0.2$). Notice that this approach does not require an elaborate re-calibration of jet observables as we do not propose to measure the forward jet, just tag it.

\begin{figure}[htb]
\begin{center}
\includegraphics[width=3.5in]{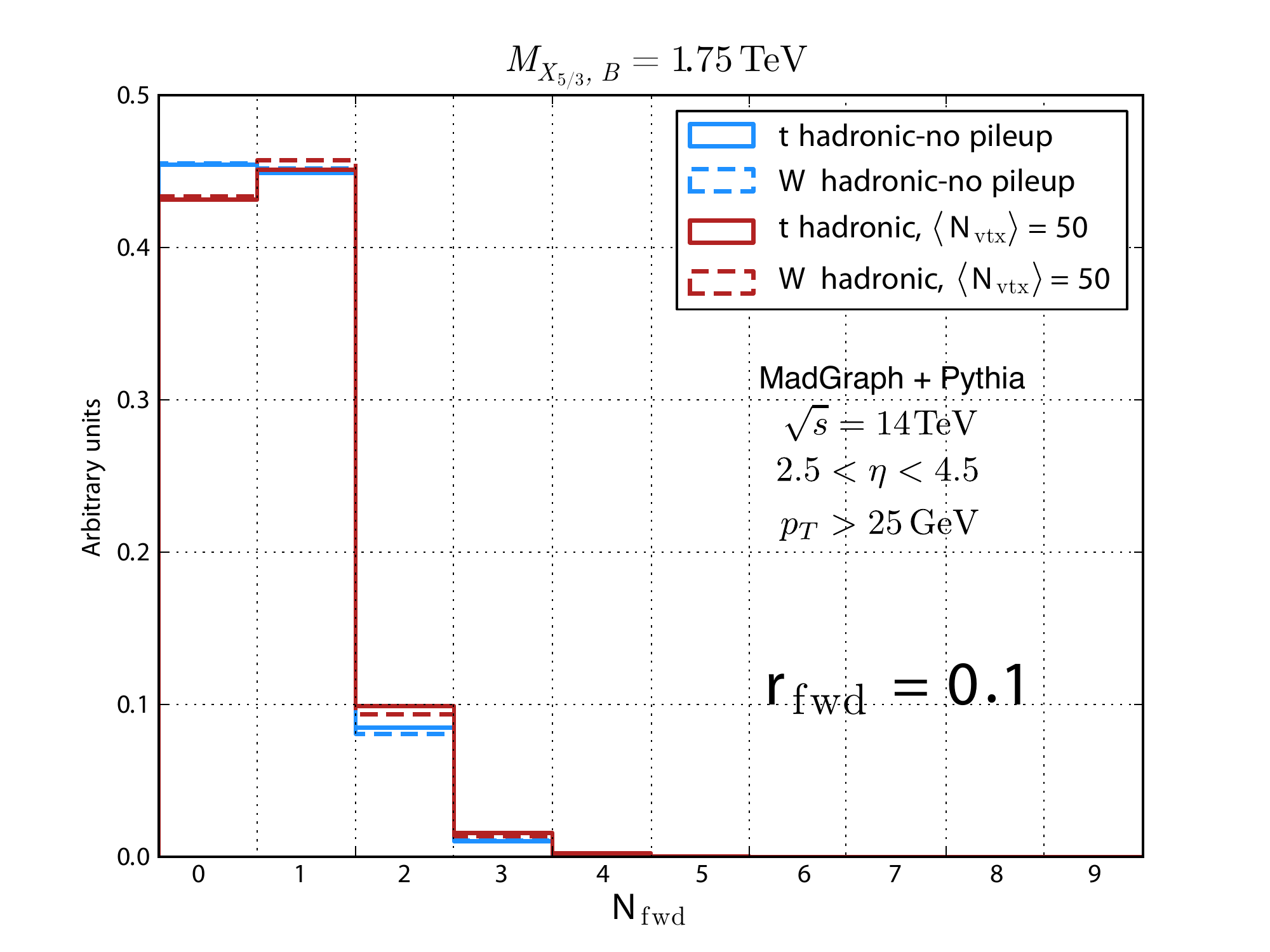}\includegraphics[width=3.5in]{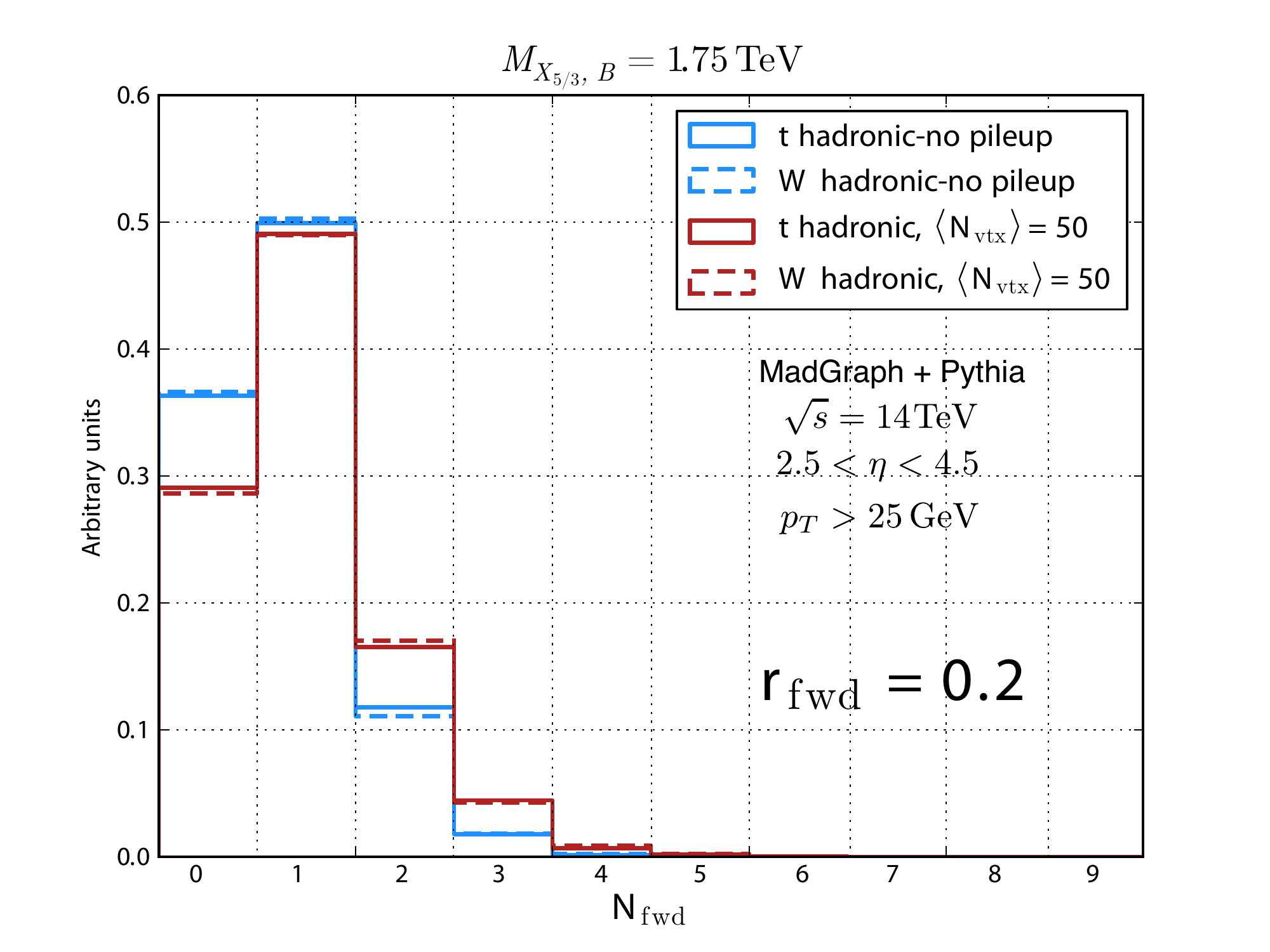}
\includegraphics[width=3.5in]{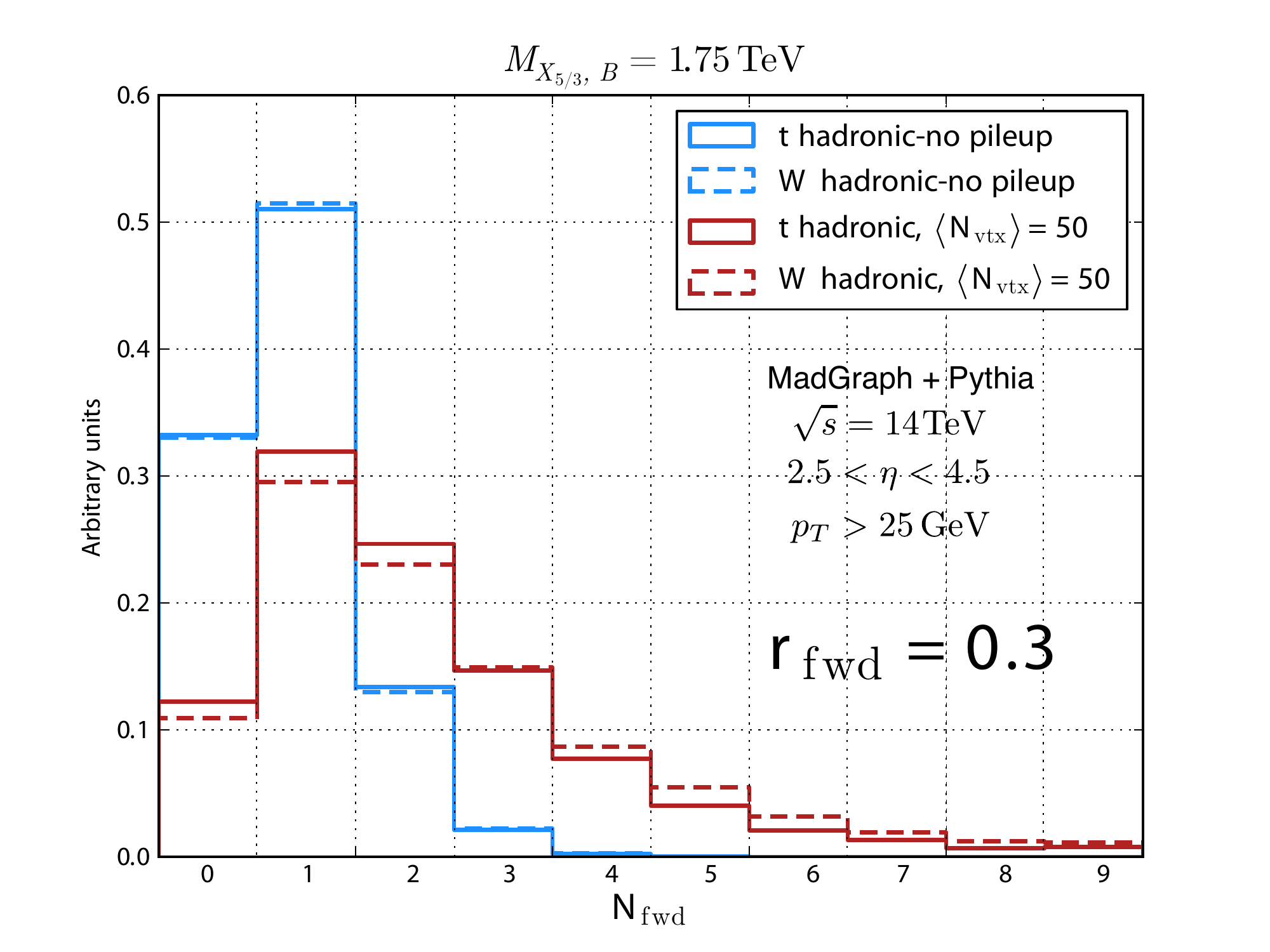}\includegraphics[width=3.5in]{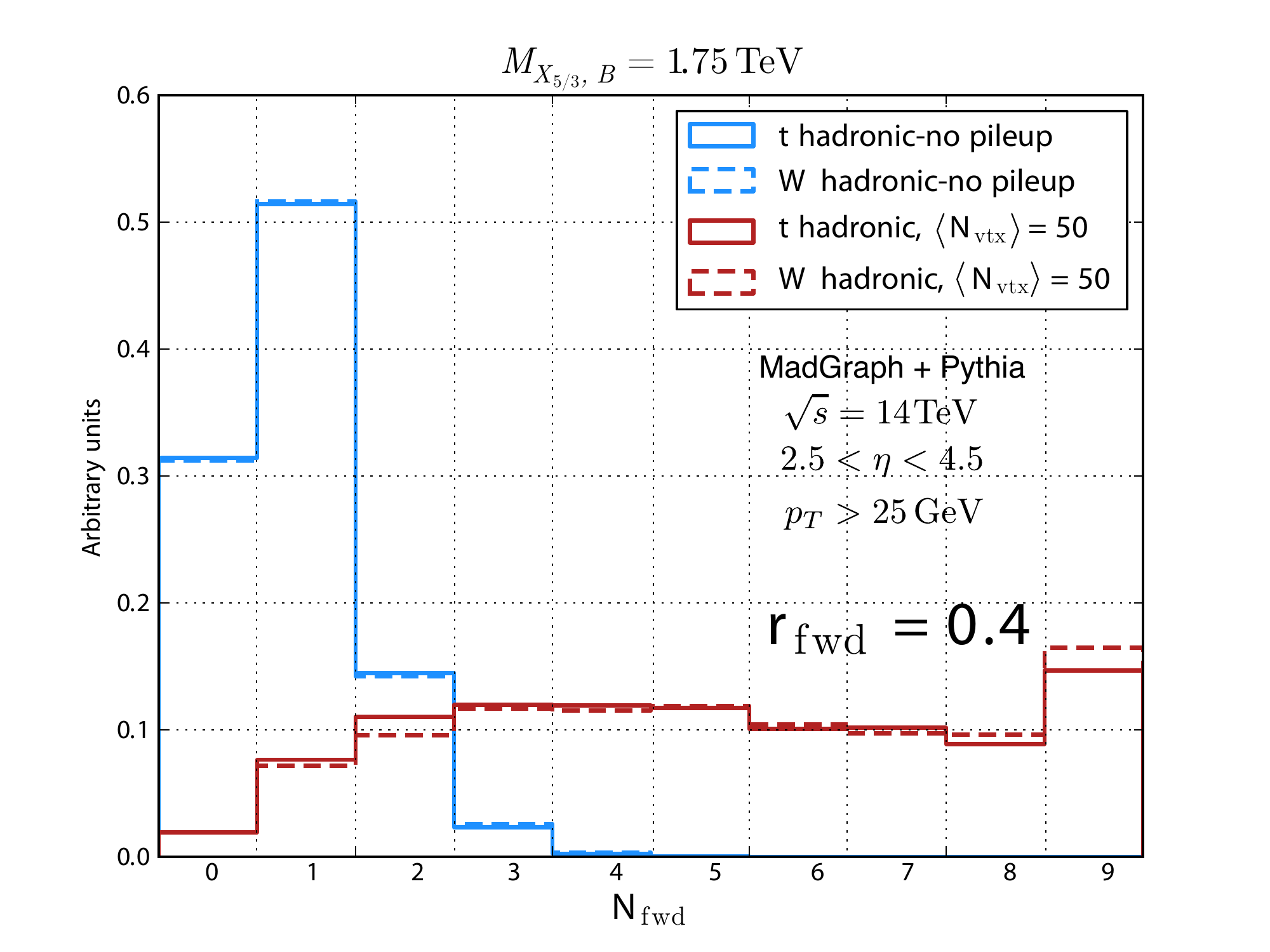}
\caption{Effects of pileup on the multiplicity of forward jets. For the purpose of illustration, we show signal events with $\MX = 1.75 \TeV$. The solid lines are hadronic top candidate events, while the dashed lines are for hadronic $W$ candidates, as defined in Section~\ref{sec:tagger}. We find that the effects of pileup are negligible for $r_{\rm fwd}=0.1$ at 50 interactions per bunch crossing, but at a cost of signal efficiency. $r_{\rm fwd}=0.2$ shows some effects of pileup, with the signal efficiency significantly improved. Notice the enormous effect pileup has on the forward jet multiplicity if standard ATLAS $r_{\rm fwd}=0.4$ jets are used. The first bin in each plot is $N_{\rm fwd}=0.$}
\label{fig:fwd_jets}
\end{center}
\end{figure}

We define forward jets by clustering the entire event using a cone of radius $r^{\rm fwd}$ and then selecting the jets in the event which satisfy the following criteria:
\begin{equation}
			p_T^{\rm fwd} > 25 \GeV, \,\,\,\,\,\,\,  2.5 < \eta^{\rm fwd} < 4.5\, , \label{eq:fwd_jet_def}
\end{equation}
where $p_T^{\rm fwd}$ and $\eta^{\rm fwd}$ are the transverse momentum and rapidity of the forward jet. We then define forward jet tagging by requiring  the number of forward jets in the event $N^{\rm fwd} \ge 1$. 

How is the forward jet multiplicity affected by pileup? Figure~\ref{fig:fwd_jets} provides the answer. Clustering the event with a standard ATLAS $r^{\rm fwd}=0.4$ cone results in a dramatic shift in the forward jet multiplicity distribution, with as many as 10 forward jets easily appearing in an event at 50 interactions per bunch crossing. Reducing the cone size to $r^{\rm fwd} =0.1$ almost extinguishes the effects of pileup, but at a cost to signal efficiency as only about $50 \%$ of the signal events pass the forward jet tagging requirement. We find that $r^{\rm fwd} = 0.2$ gives a good compromise between effects of pileup and signal efficiency, and throughout the rest of this paper we will adopt the term ``forward jet'' to mean a jet of radius $r^{\rm fwd} =0.2$ which passes the forward jet criteria of Eq.~\eqref{eq:fwd_jet_def}.

\subsection{$b$-tagging}\label{sec:btag}
Our analysis utilizes the presence of multiple $b$-jets in the signal, whereby we use information from the hard process to simulate the $b$-tagging procedure. We define every $r=0.4$ jet to be $b$-tagged if there is a hard process $b$ or $c$ quark within $\Delta R = 0.4$  from the jet axis. We consider the benchmark efficiency of $75 \%$ for every $b$ jet to be tagged as a $b$, with the fake rate of $18 \%$ and $1\%$ for $c$ and light jets respectively. We further consider a \textit{fat jet to be $b$-tagged} if there is a $b$-tagged $r=0.4$ jet within $\Delta R = 1.0$ from the fat jet axis. 

We apply different $b$-tagging criteria based on whether the fat jet is a hadronic top or hadronic $W$ candidate. Namely, we require:
\begin{itemize}
\item One $b$-tagged fat jet ($i.e.$ $\Delta R({\rm fj}, $b$) < 1.0$), and at least one $b$-tagged $r=0.4$ jet outside the fat jet ($i.e.$ $\Delta R({\rm fj},$b$) > 1.4$) if the fat jet is a \textbf{hadronic top candidate}. Note that the criteria for an event to be a hadronic top candidate also require the $r=0.4$ jet to be isolated from the hardest lepton ($i.e.$ $\Delta R(l, b) > 1.0$).
\item One fat jet without a $b$-tagged $r=0.4$ jet within $\Delta R=1.4$ from the fat jet axis ($e.g.$ anti-$b$-tagged) and at least one $b$-tagged $r=0.4$ jet outside the fat jet,  if the fat jet is a \textbf{hadronic $W$ candidate}.
\end{itemize}

How large of a $b$-tagging efficiency should we expect for the signal events? Naively, we would assume that the fraction of events which contain two true $b$-jets is $\sim 1.0$. When folded into the above mentioned $b$-tagging efficiencies, we would hence expect the overall signal $b$-tagging efficiency to be $\sim 0.5.$ 

\begin{figure}[htb]
\begin{center}
\includegraphics[width=3.5in]{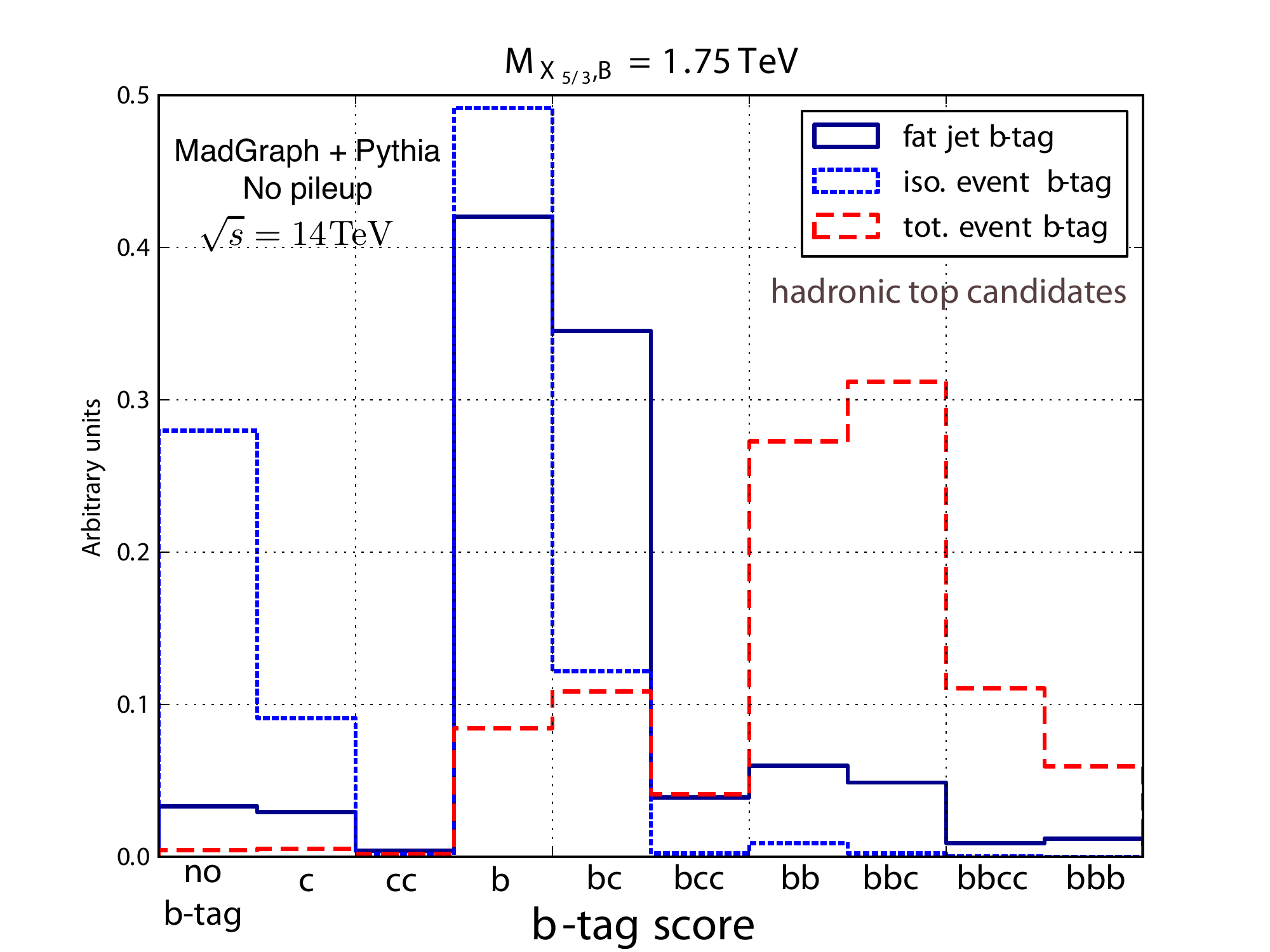}\includegraphics[width=3.5in]{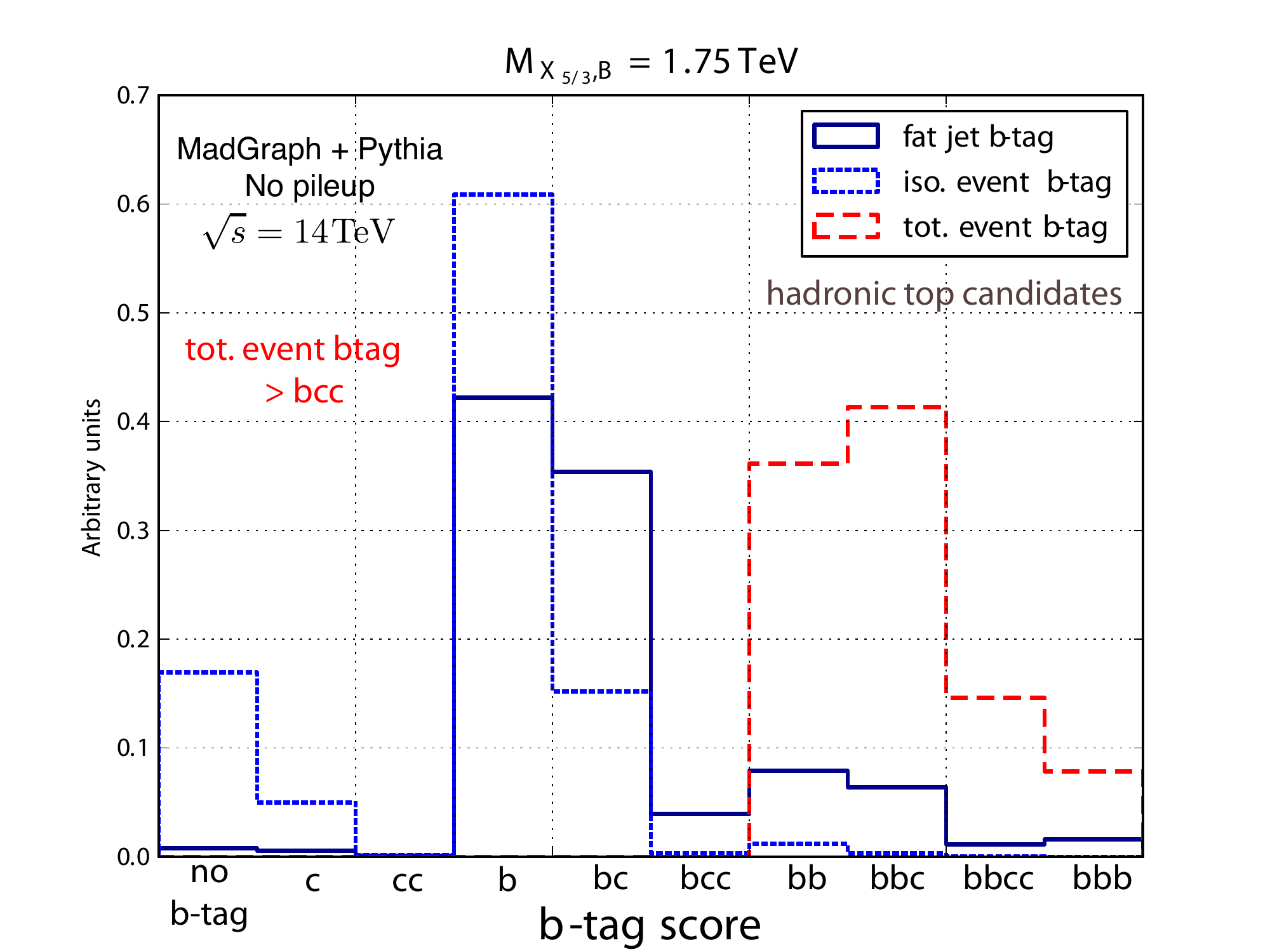}
\caption{Example of $b$-tag scores of signal events. The solid blue line represented the $b$-tag score of the hardest fat jet in the event, as prescribed in the bulleted list of this Section. The dotted blue line refers to the event $b$-tag score considering isolated $b$-jets only. The dashed red curve is the event $b$-tag score assuming only the acceptance requirements of $p_T^j > 25 \GeV$ and $| \eta_j | < 2.5$. We show all events which pass the Basic Cuts and have an overlap score of $Ov > 0.5$ on the left panel, while the right panel assumes only the events which pass the Basic Cuts, overlap cuts \textit{and} contain two proper $b$-jets.  No $b$-tagging efficiencies have been applied.}
\label{fig:btag}
\end{center}
\end{figure}

Figure~\ref{fig:btag} shows more precise and complete information on the $b$-tagging of signal events (for the purpose of illustration, here we show only hadronic top candidate events). From the left panel, we can see that the geometrical acceptance for events which contain two proper $b$-jets is $\sim 80 \%, $ as represented by dashed, red histogram area with a $b$-tag score $\ge bb$. The probability that the highest $p_T$ fat jet of a signal event will contain a proper $b$-tag is $\sim 90\% $, due to the large degree of collimation of the top decay products and the large fat jet clustering cone $R=1.0$. 

In addition, we find that the isolation criteria on the $b$-jet outside the fat jet reduce the signal efficiency by an additional $20-30 \%$, as seen in the right panel of Fig.~\ref{fig:btag}. 
The effect can be understood almost entirely from a simple geometrical argument.  Consider for instance the $b$-tagging criteria for hadronic top candidate events. Because anti-$k_T$ jets are roughly circular in $\eta, \phi$, the fraction of the available detector area in which a $b$-jet will be isolated both from the fat jet and the hardest lepton is given by:
\begin{equation}
	\epsilon (b {\rm -tag\, isolated}) \sim 1- \frac{\pi ((r+R)^2 + R^2)}{2 \pi \Delta \eta_a }\,,
\end{equation}
where $\Delta \eta_a$ is the detector acceptance in rapidity for the $r=0.4$ jets ($i.e.$ -2.5 to 2.5), $r$ is the radius of the $b$-tagged jets, and $R$ is the radius of the fat jet. The $(R+r)^2$ term serves to isolate the $b$-jet from the fat jet while the term proportional to $R^2$ isolates the jet from the lepton. 
Jet rapidity acceptance  is roughly $\Delta y \approx 5$, although this is an under-estimate since tracks with $|y| < 5$ are all taken into account during jet reconstruction. Next, for $b$-jets clustered with $r=0.4$ and fat jets with $R=1.0$ one obtains $\epsilon (b {\rm- tag \, isolation})\sim 0.7$, roughly the fraction of isolated $b$-tag events with a $b$-tag score greater than $b$ in the left panel of Fig.~\ref{fig:btag}.

We conclude that the expected $b$-tagging efficiency for the hadronic top candidate events (including the $75\%$ efficiency of $b$-tagging) will be of order
\begin{equation}
	\epsilon(b{\rm -tag}) \sim 0.8 \times 0.7 \times (0.75)^2 \sim 0.3\,.
\end{equation}

A full study of pileup effects on $b$-tagging requires detailed detector information, an endeavor which is beyond the scope of our current analysis. However, we would like to point out that the experimental studies of Ref. \cite{Capeans:1291633} 
suggest that $b$-tagging performance at the LHC will perform well at 50 interactions per bunch crossing. 

\subsection{Resonance Mass Reconstruction}

Reconstructing the mass of the top partner in our analysis involves one fat jet, a hard lepton, missing $E_T$ and possibly a $r=0.4$ jet (if the event topology is such that the top  decays leptonically). In such a situation, there are three issues which might arise:
\begin{itemize}
\item \textit{Combinatorics:}  The signal final state is characterized by no less than 5 small cone ($i.e.$ $r=0.4$) jets. Determining which jets resulted from a decay of a top or a $W$ is hence a challenge. 
\item \textit{Pileup contamination:} Pileup contamination not only shifts mass distributions to higher values and broadens them, but also creates ``pileup jets'' which could additionally complicate the combinatorial issues. 
\item \textit{Missing energy reconstruction:} Reconstructing the resonance mass in our case involves reconstructing the $z$ component of the neutrino coming from the resonance decay as well as a contribution from a possible additional neutrino from an initial state radiated top. 
\end{itemize}
The resonance mass reconstruction method we propose bypasses all of the above-mentioned issues. Since $\MX\sim~O(1 \TeV)$, the resonance decay products are boosted, with average $p_T \sim \MX /2$. As we will show shortly, simply selecting a hardest fat jet in the event, a lepton (with a possible $r=0.4$ jet in its vicinity) and missing energy suffices to reconstruct $\MX$, and hence eliminates many combinatorial issues. 
Boosted final states also allow us to make an approximation of $\eta_\nu = \eta_l, $ where $\nu$ is the total missing transverse momentum in the event and $l$ is the hardest lepton. For the purpose of $\MX$ reconstruction, we find this approximation to be adequate without loss of generality.
We avoid significant effects of pileup contamination by reconstructing $\MX$ using the pileup-insensitive peak template momenta instead of the fat jet momentum (as in Ref. \cite{Backovic:2013bga}),  while the effects of pileup on the lepton, missing energy and a possible $r=0.4$ jet in the vicinity of the lepton are manageably weak at $\langle N_{\rm vtx} \rangle = 50$ pileup events. Finally, we have checked that the possible additional neutrino coming from the initial state radiated top does not significantly contribute to the missing transverse energy. 

\begin{figure}[htb]
\begin{center}
\includegraphics[width=3.5in]{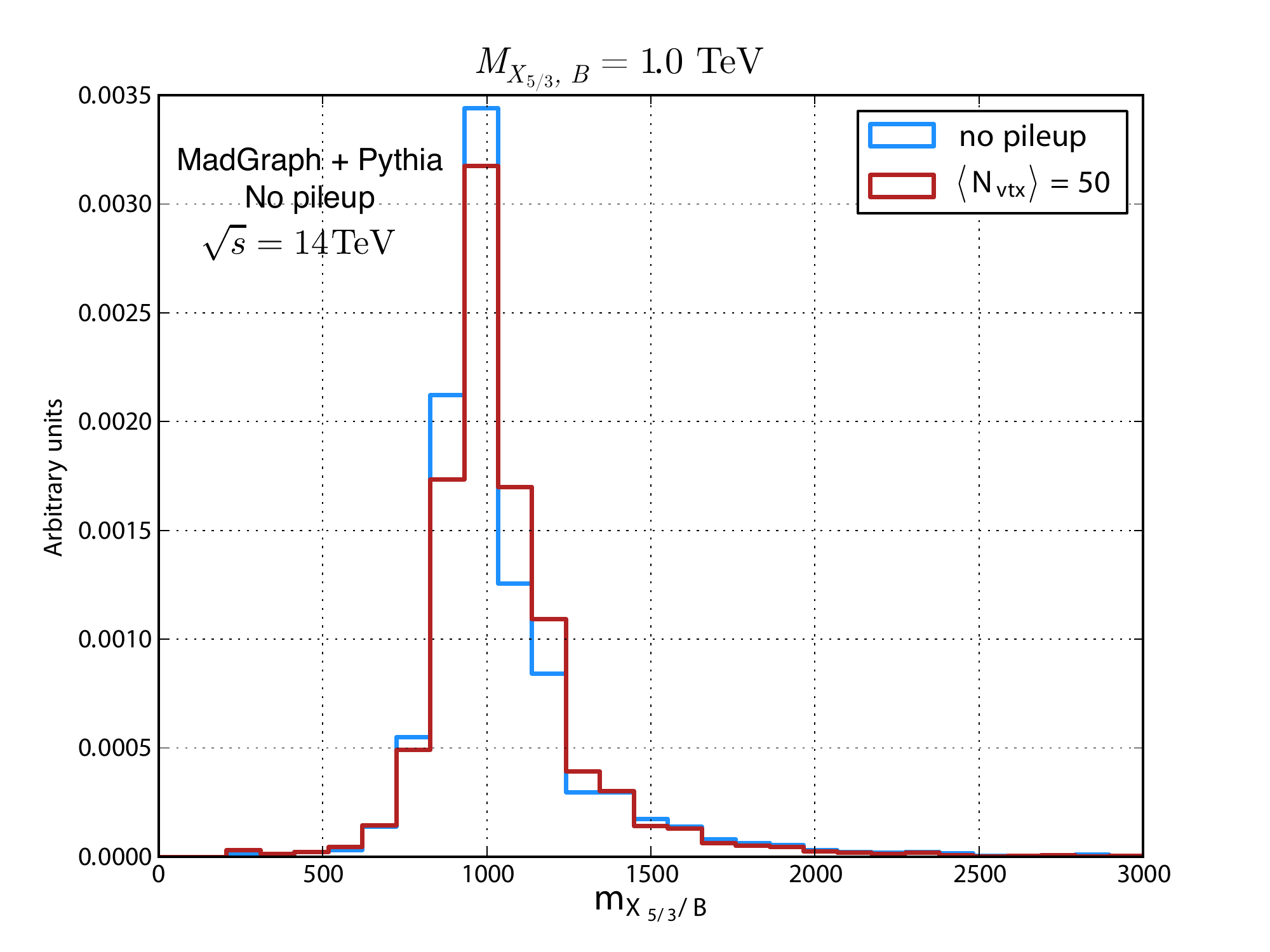}\includegraphics[width=3.5in]{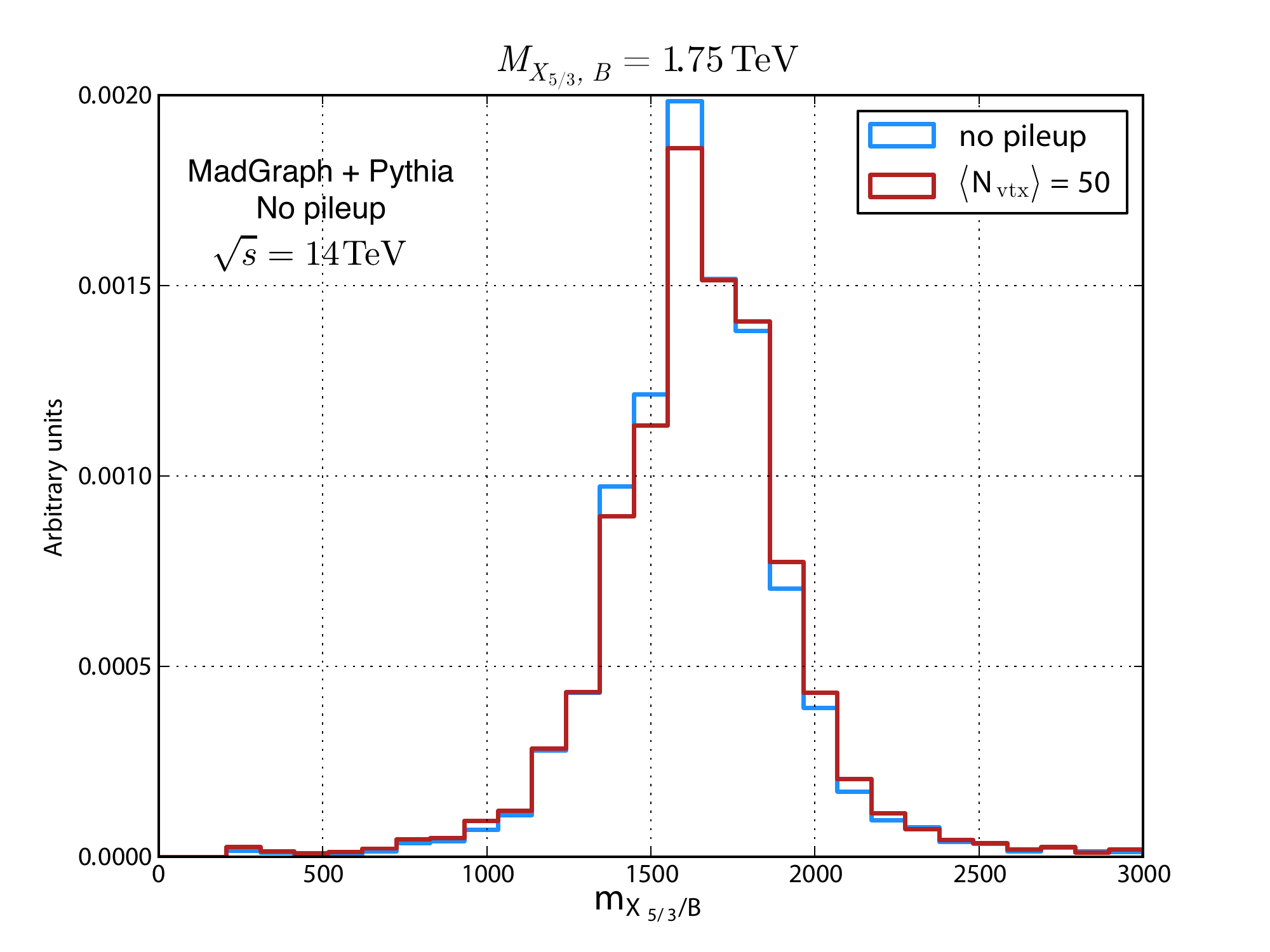}
\caption{Efficiency of $\MX$ reconstruction in a pileup environment . In each case, we construct $\mX $ from the missing energy, the hardest lepton and the peak template. The blue line assumes no pileup while the red line assumes $50$ average pileup events. Only hadronic top candidate events are shown.}
\label{fig:res_mass}
\end{center}
\end{figure}

Figure~\ref{fig:res_mass} shows two examples of mass reconstruction for signal events only, where we denote the true resonance mass as $\MX$ and the reconstructed mass as the lowercase $\mX$. The solid blue lines represent the $\mX$ distribution if no pileup was present, while the red lines show the corresponding distribution at $\langle N_{\rm vtx} \rangle = 50$ pileup events. In both cases, our mass reconstruction method is able to resolve the resonance peak to a very good degree, while effects of pileup on the mass peak resolution remain weak at average 50 pileup events.

\subsection{Projected $\MX$ Sensitivity} \label{sec:sensitivity}

The main question we would like to answer in this paper is how sensitive will the future LHC runs be to different $\MX$ in our analysis framework? For this purpose we analyzed  several signal event samples with $\MX~=~(1.0, 1.25, 1.5, 1.75, 2.0) \TeV$. Varying other model parameters will change the value of the single production cross section but will not significantly affect the event kinematics. Hence, we fix all other couplings and scales and leave the inclusive production cross section $\sigma_{\Xx}$ a free parameter. An additional benefit of considering $\sigma_{\Xx}$ as a free variable is that our results in this section can be applied to other searches for BSM physics in the same final state channel. In this section we assume no pileup contamination and postpone the discussion of 50 interactions per bunch crossing until the next section.

To illustrate the ability of our proposal to reject SM backgrounds, we begin with the example cutflow results in Table~\ref{tab:cutflow1a}. We chose the values for the inclusive cross section in each table to be roughly in the mid-range of the cross section values for model parameters which address the hierarchy problem and result in a correct mass of the top quark. 
Perhaps the most exciting result of our analysis is that the future LHC runs will be sensitive to $\MX \sim 2.0 \TeV$ top partners, where we find that $5 \sigma$ sensitivity should be achievable with $35 \,{\rm fb^{-1}}$ of data (assuming $b$-tagging and no forward jet tagging), while requiring a signal cross section large enough to give $\sim 10$ events, as shows in Table~\ref{tab:cutflow1a}. Note that because our event selection is sensitive to both $X_{5/3}$ and $B$ production, the final signal cross section we achieve for hadronic top candidate events alone is higher than the the naive estimate of the same sign di-lepton cross section (assuming a 50 \% geometric acceptance for the two leptons).

\begin{figure}[!]
\begin{center}
\includegraphics[width=3.0in]{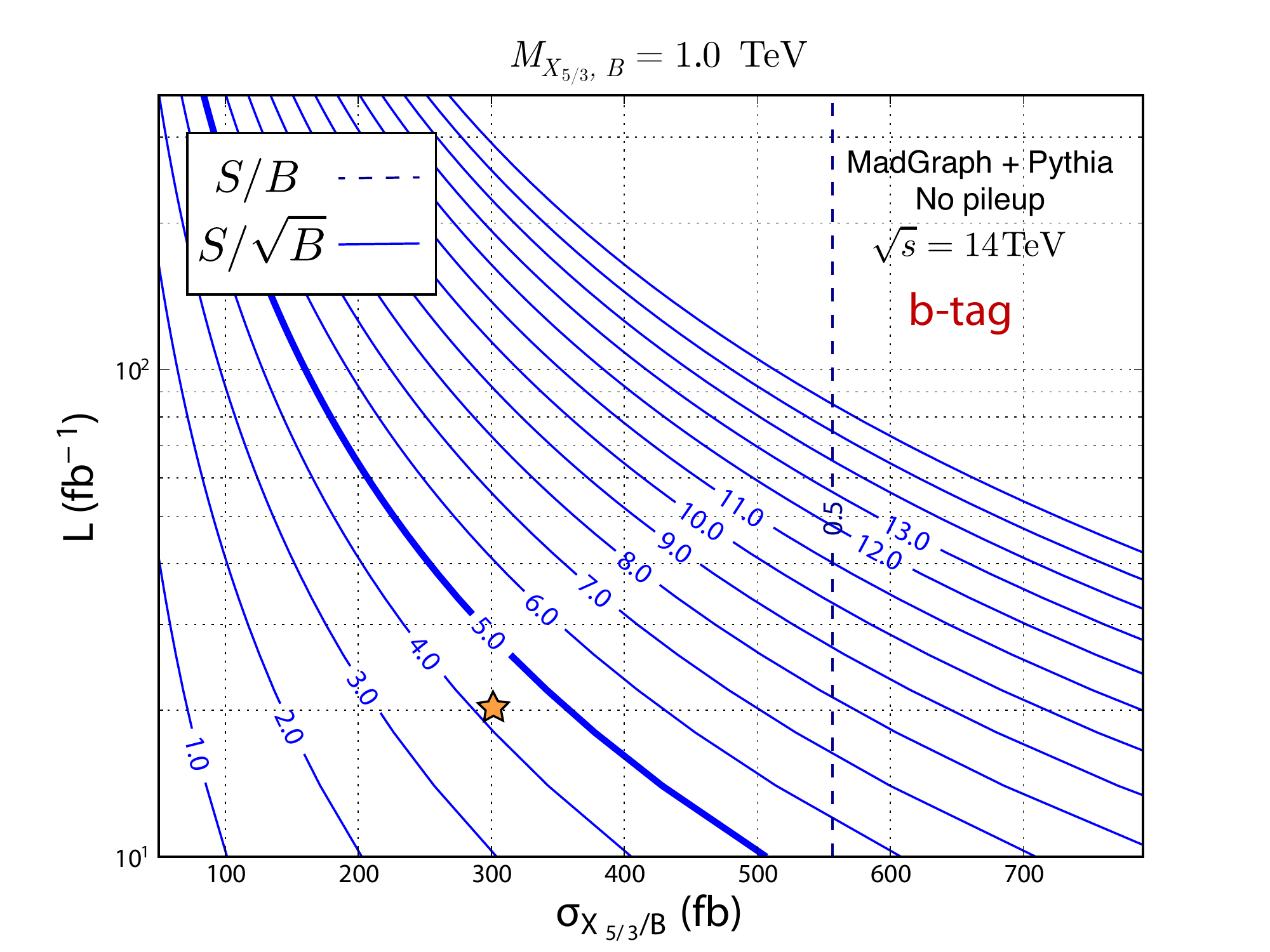}\includegraphics[width=3.0in]{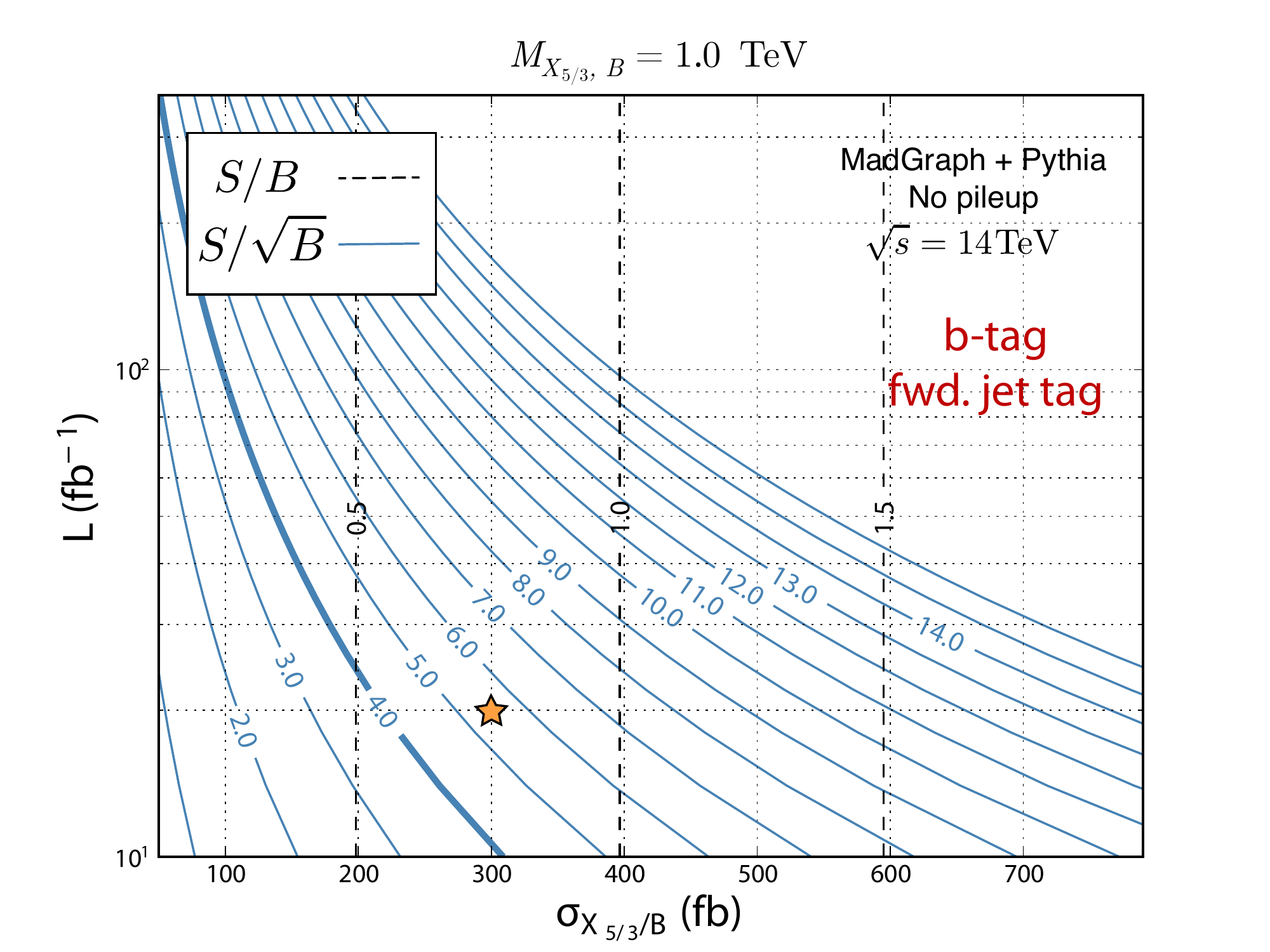}
\includegraphics[width=3.0in]{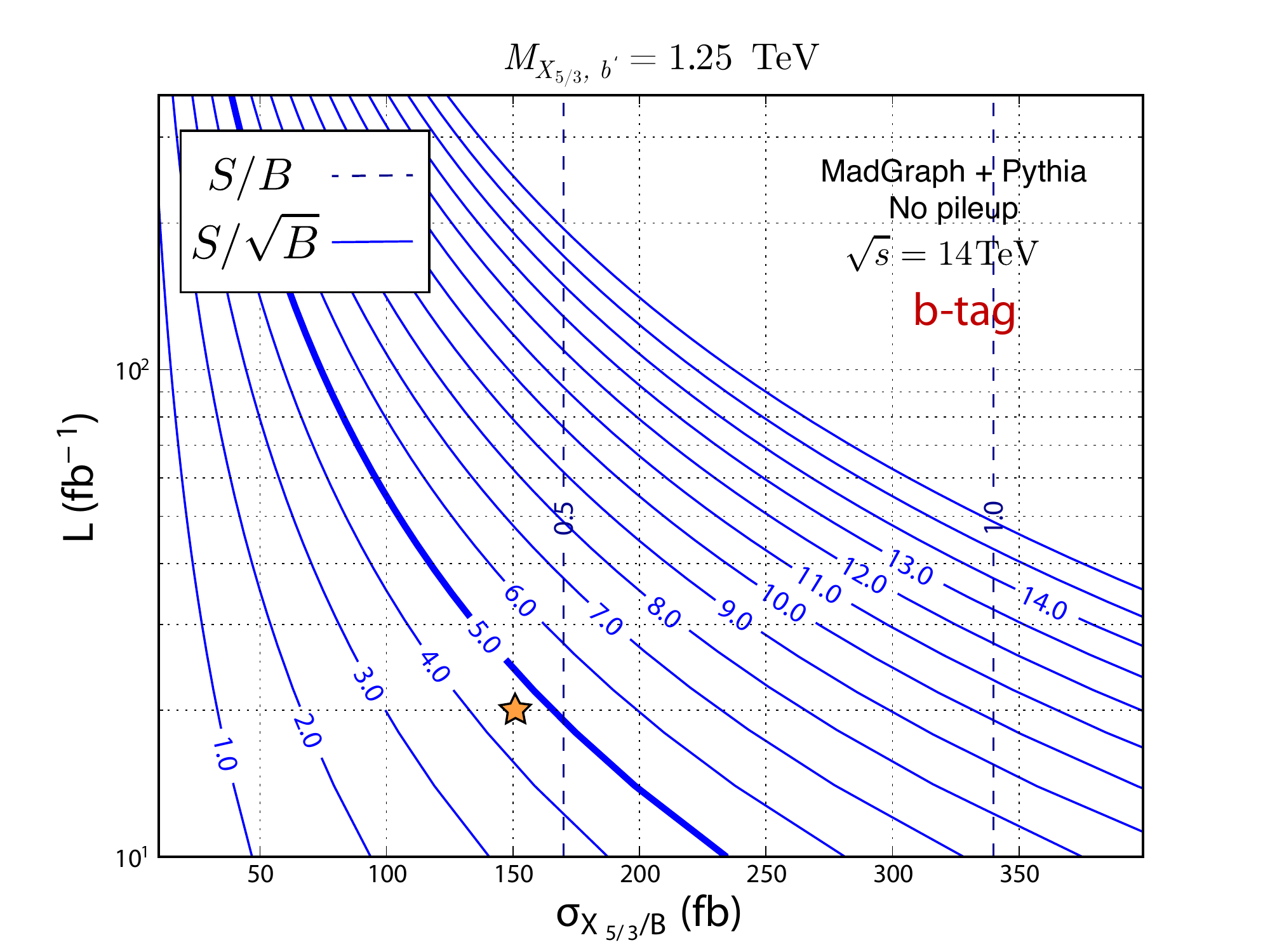}\includegraphics[width=3.0in]{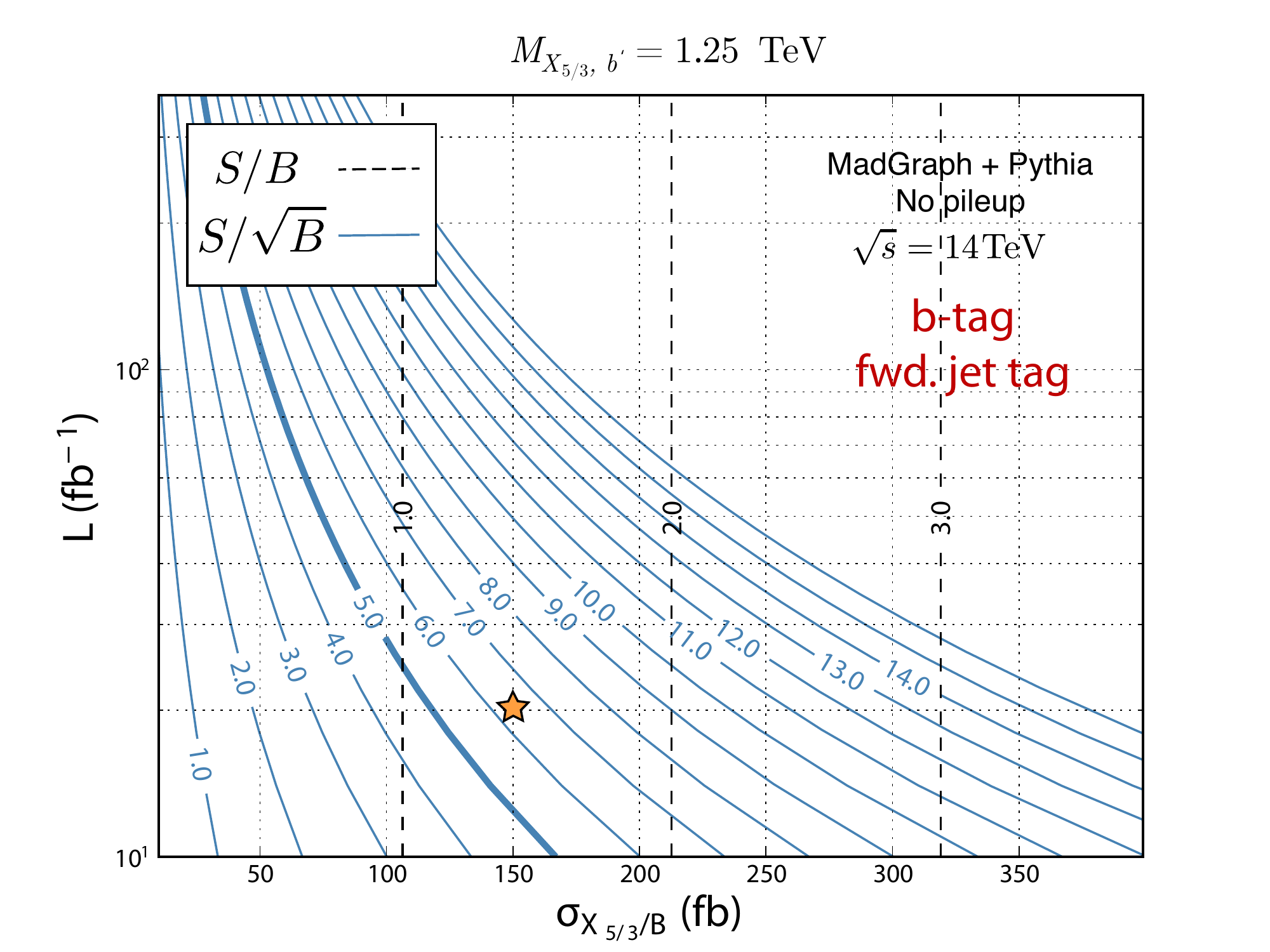}
\includegraphics[width=3.0in]{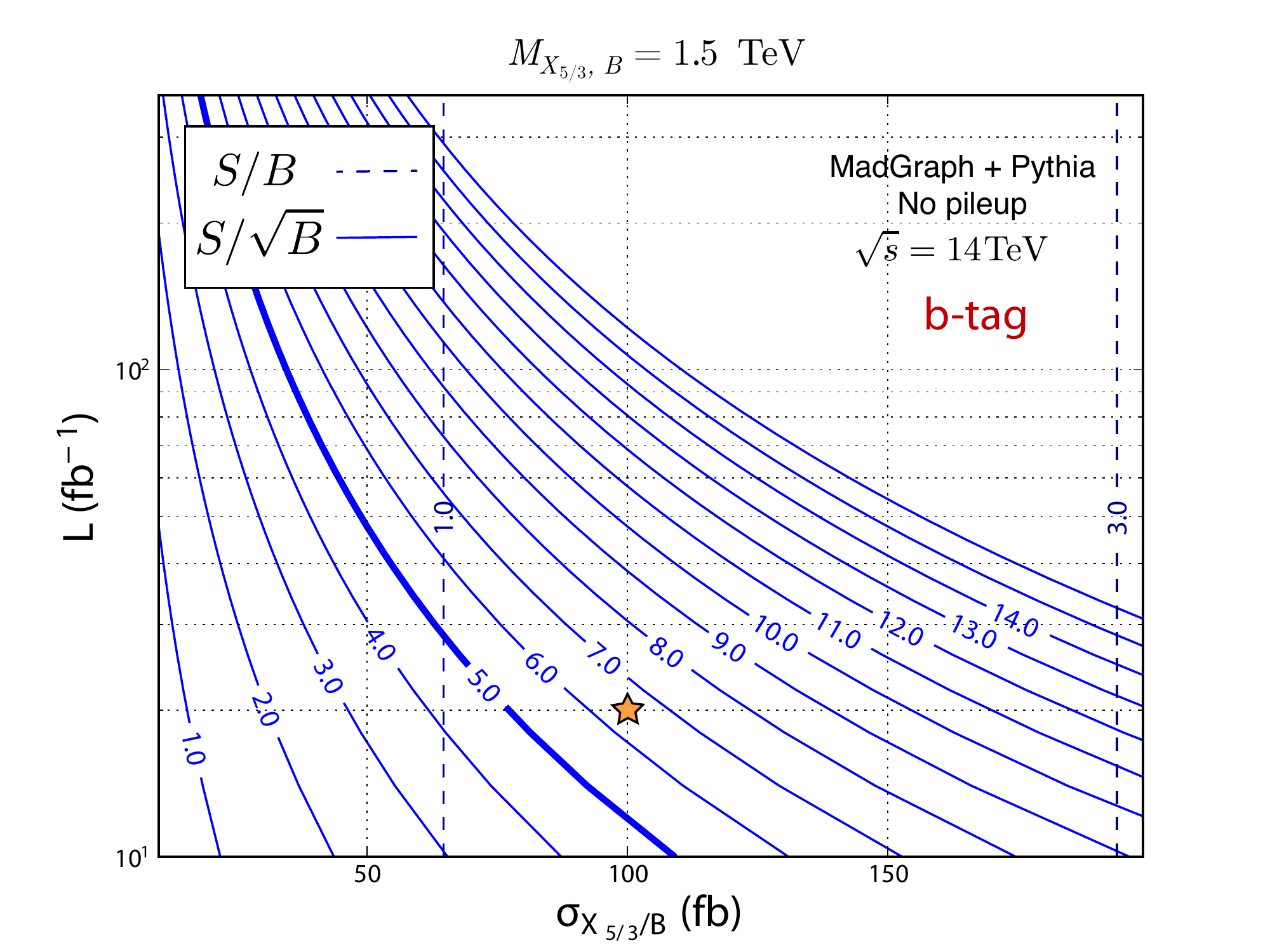}\includegraphics[width=3.0in]{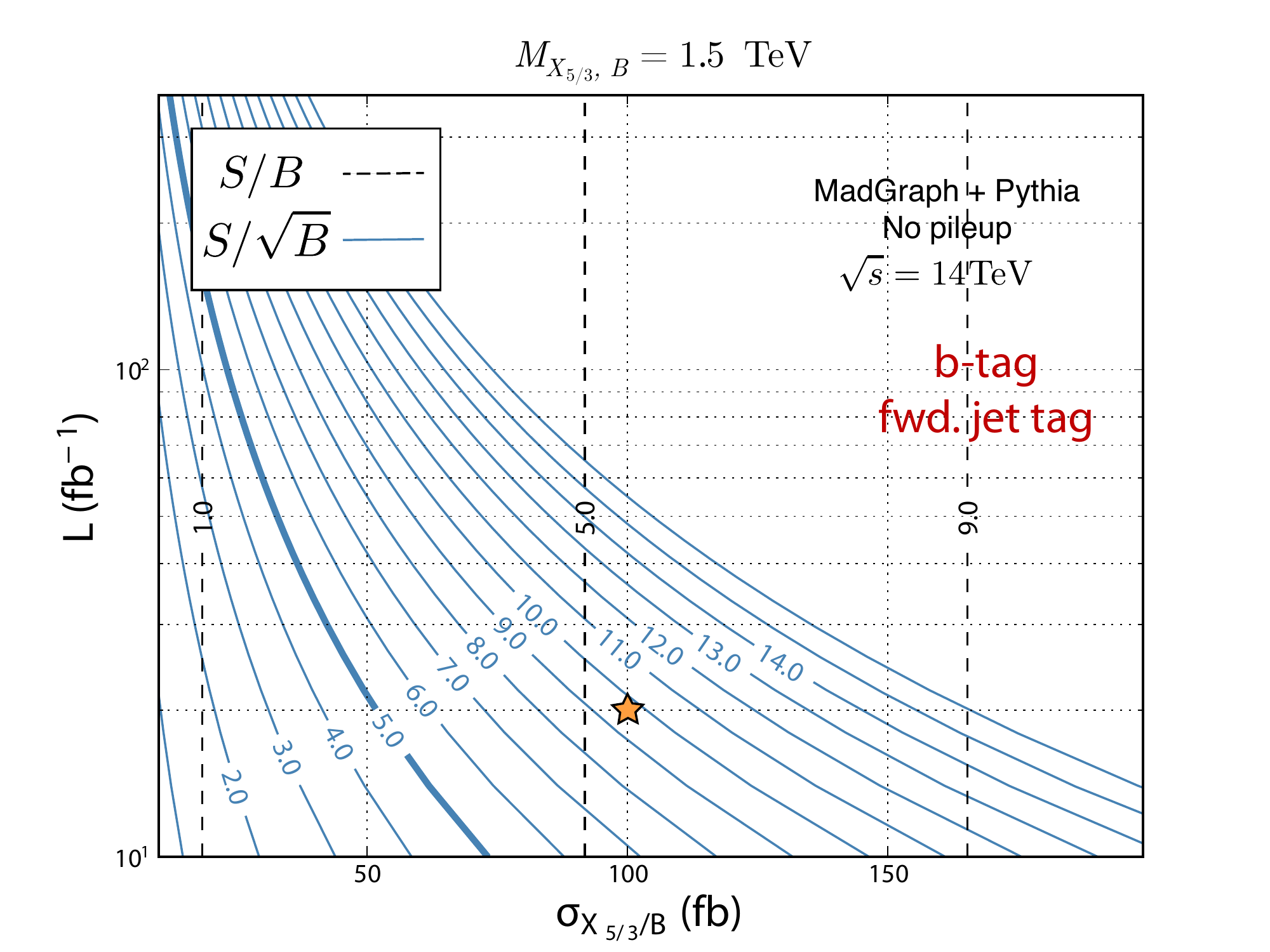}
\includegraphics[width=3.0in]{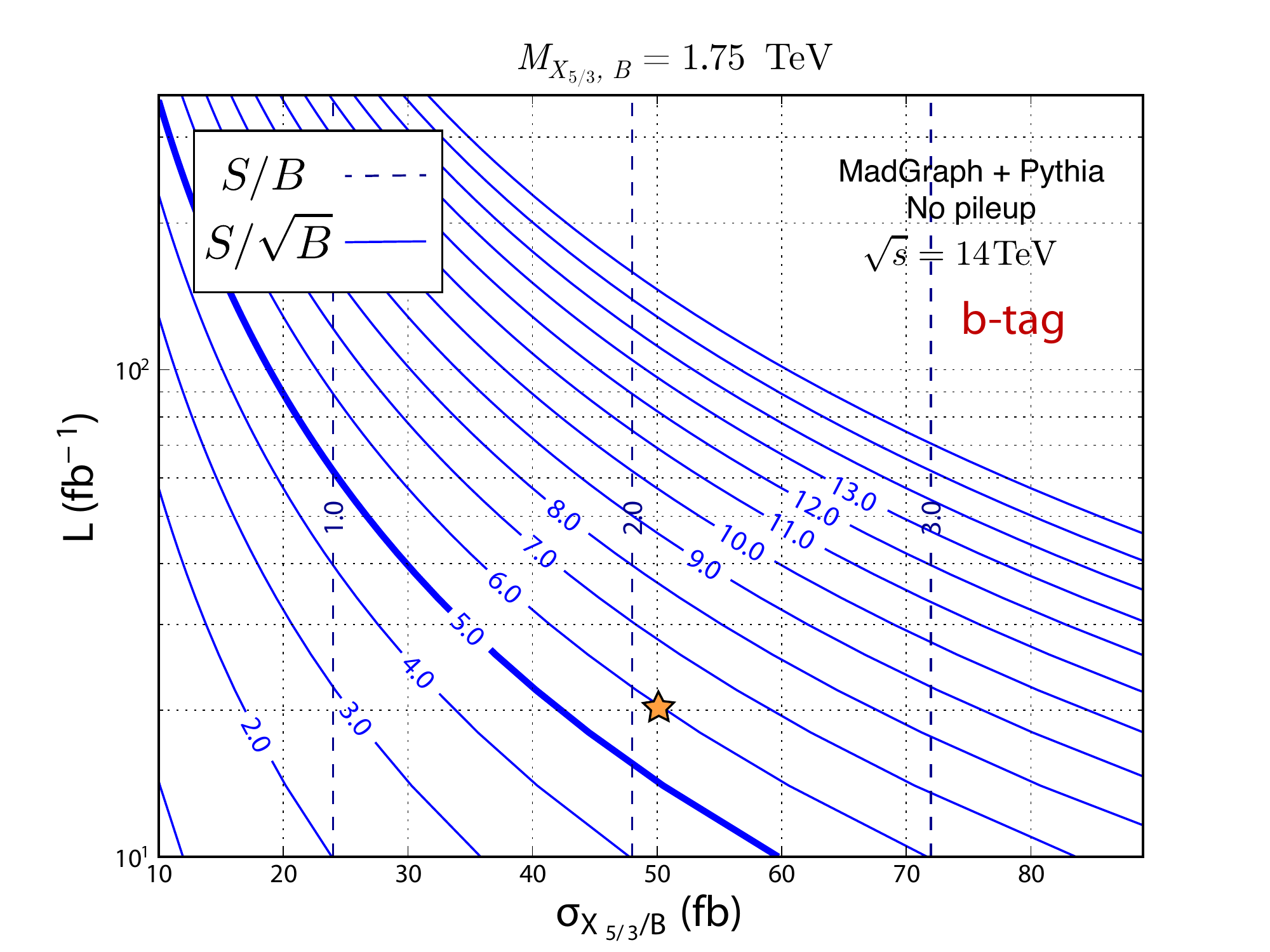}\includegraphics[width=3.0in]{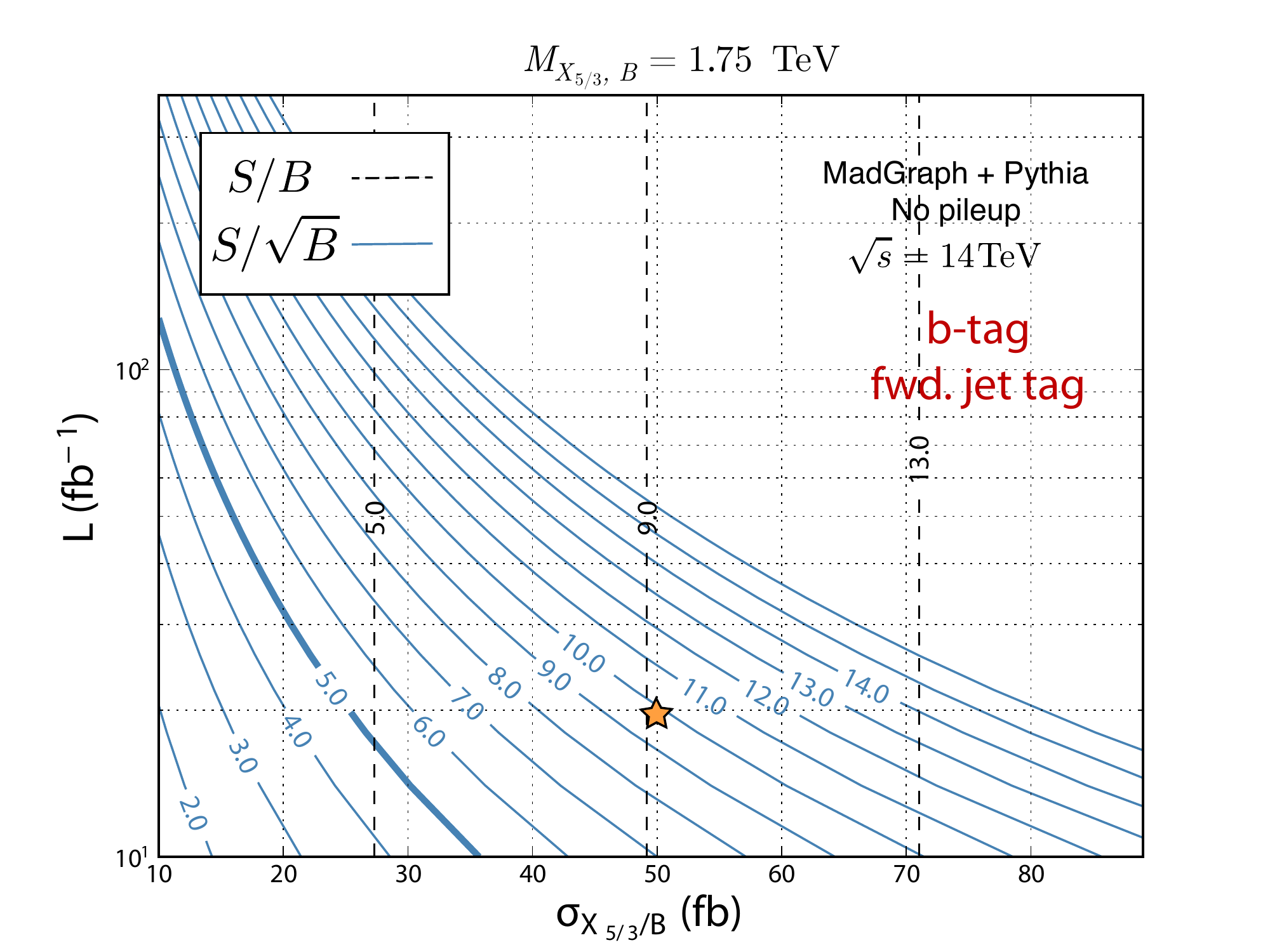}

\caption{Sensitivity to various $\MX$. The plots on the left \textbf{assume $b$-tagging but not forward jet tagging}, while plots on the right \textbf{assume both $b$-tagging and forward jet tagging}. The inclusive signal cross section and integrated luminosity are on the $x$ and $y$ axes respectively.  We only show hadronic top candidate events no pileup contamination. The blue solid lines represent contours of constant $S/\sqrt{B}$. The dashed lines are $S/B$. The selection cuts for each $\MX$ reflect the ones in Table~\ref{tab:cutflow1}, where we mark the point presented in the table by a star. }
\label{fig:SBresults}
\end{center}
\end{figure}

We present detailed information for masses lower than 2 TeV in Table~\ref{tab:cutflow1}. We find that the LHC run at $14 \TeV$ can achieve $S/B > 1$ for $\MX > 1 \TeV,$ with $\sim 5 \sigma$ significance using the $b$-tagging proposal of Section~\ref{sec:btag} alone, while the addition of a forward jet tag from Section~\ref{sec:fwdjet} results in an almost background free signal and a significant improvement in significance at an additional $25-30 \%$ signal loss. Forward jet tagging alone is not sufficient to produce desirable sensitivity to any of the $\MX$ we considered, except for very large signal cross sections and integrated luminosities. However, complemented by $b$-tagging, we find that forward jet tagging can significantly improve the $\MX$ sensitivity.
 The lower $\MX$ ($e.g.\, \MX \sim 1 \TeV$) cases benefit more from forward jet tagging, as the signal cross section is larger and hence allows for lower final signal efficiency. We find that an additional factor of $\sim 4-6$ in $S/B$ improvement is typically achieved by adding a forward jet tag. We show a detailed comparison of results with and without forward jet tagging in Fig.~\ref{fig:SBresults}, where the left panels assume the $b$-tagging criteria without the forward jet tag, while the right panels assume both $b$-tagging and a forward jet tag.

There are several interesting features of our analysis. First, we find that for $\MX >  1 \TeV$, the $S/\sqrt{B}$ we achieve in the hadronic top channel is significantly better than for hadronic $W$ candidate events, even though it results in a $50 \%$ lower signal efficiency, while the significance is comparable for $\MX \sim 1 \TeV$. The effect can be attributed to the asymmetry in the proportion of hadronic top vs. hadronic $W$ candidate events in the background as defined in Section~\ref{sec:tagger}.  For instance, the signal events contain hadronic top and hadronic $W$ candidate events in the equal proportion, while the background $t\bar{t}$ events we consider always contain a leptonic top. Hence, the amount of SM $t\bar{t}$ events which will be categorised as hadronic top events is smaller and will amount only to the events in which the $b$ quark from the leptonic top decay happens to land far enough from the lepton. Notice that the probability that a $b$ quark will land far from the fat jet axis increases with the decrease in the fat jet $p_T$, hence the comparable $S/\sqrt{B}$ at lower $\MX$. The second interesting feature of our results is that the sensitivity to signal events increases with $\MX$. One of the reasons for higher efficiency at higher $\MX$ is that the TOM reconstruction and tagging of boosted objects becomes more efficient at higher $p_T$. The hard parts of a boosted jet, which TOM is designed to tag, become more prominent features of a fat jet at high $p_T$, while a higher degree of collimation of signal fat jets at high $p_T$ make it less likely that radiation will ``leak'' out of the $R=1.0$ cone. In addition, the fact that the high $p_T$ tails of background distributions fall-off as several powers in $p_T$ and faster than the signal distributions, imply that at high $\MX$ we expect less background contamination. 

The final signal efficiencies for $\MX < 2.0 \TeV$, where we do not expect a large degree of mass degeneracy between the $X_{5/3}$ and $B$, are roughly at the level as the naive estimate of a background free same sign di-lepton analysis (assuming a detector acceptance of $50 \%$), with the additional benefit that our method allows for good reconstruction of the resonance in a pileup-insensitive manner. 

\begin{figure}[!]
\begin{center}

\includegraphics[width=3.5in]{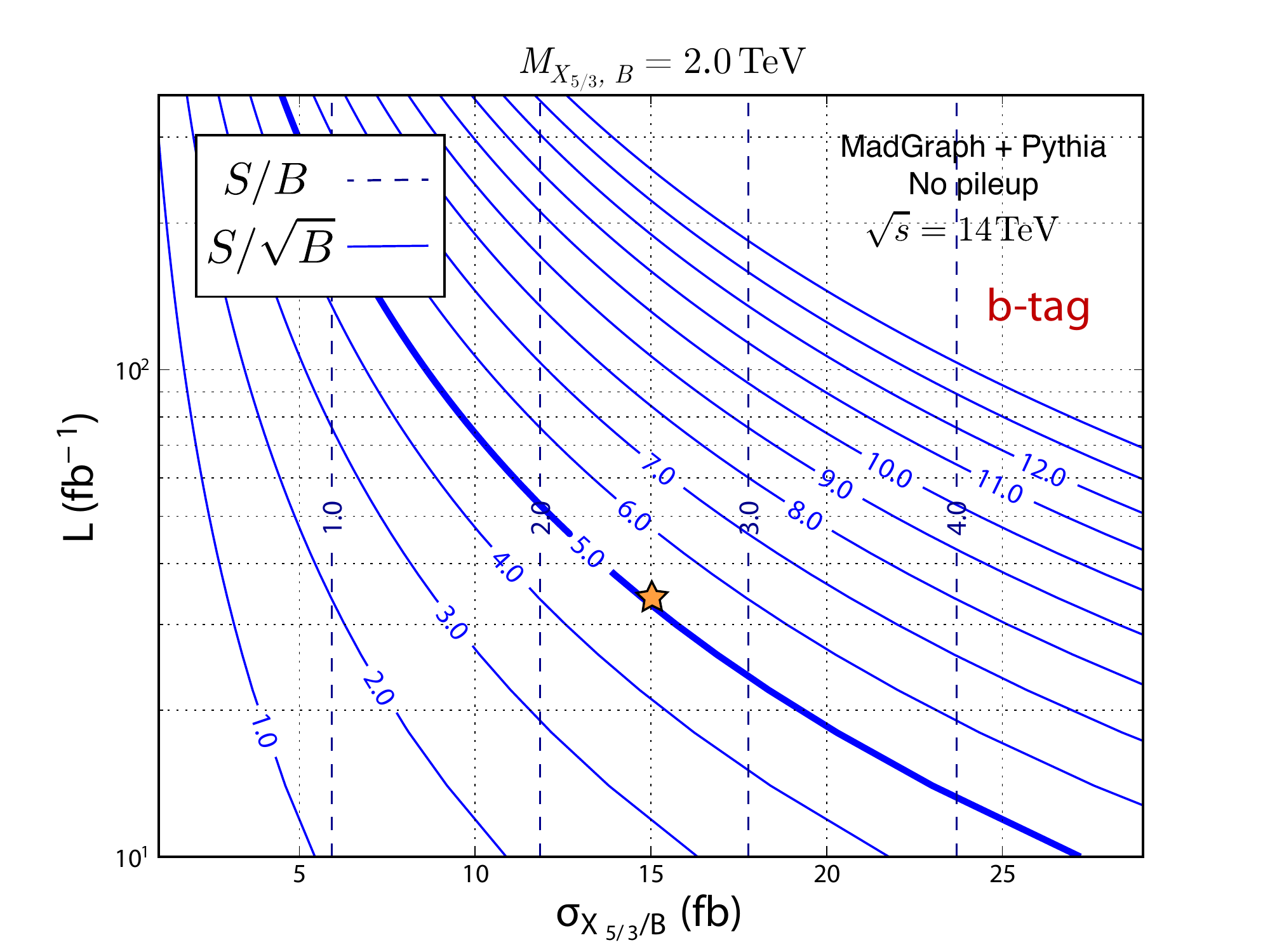}\includegraphics[width=3.5in]{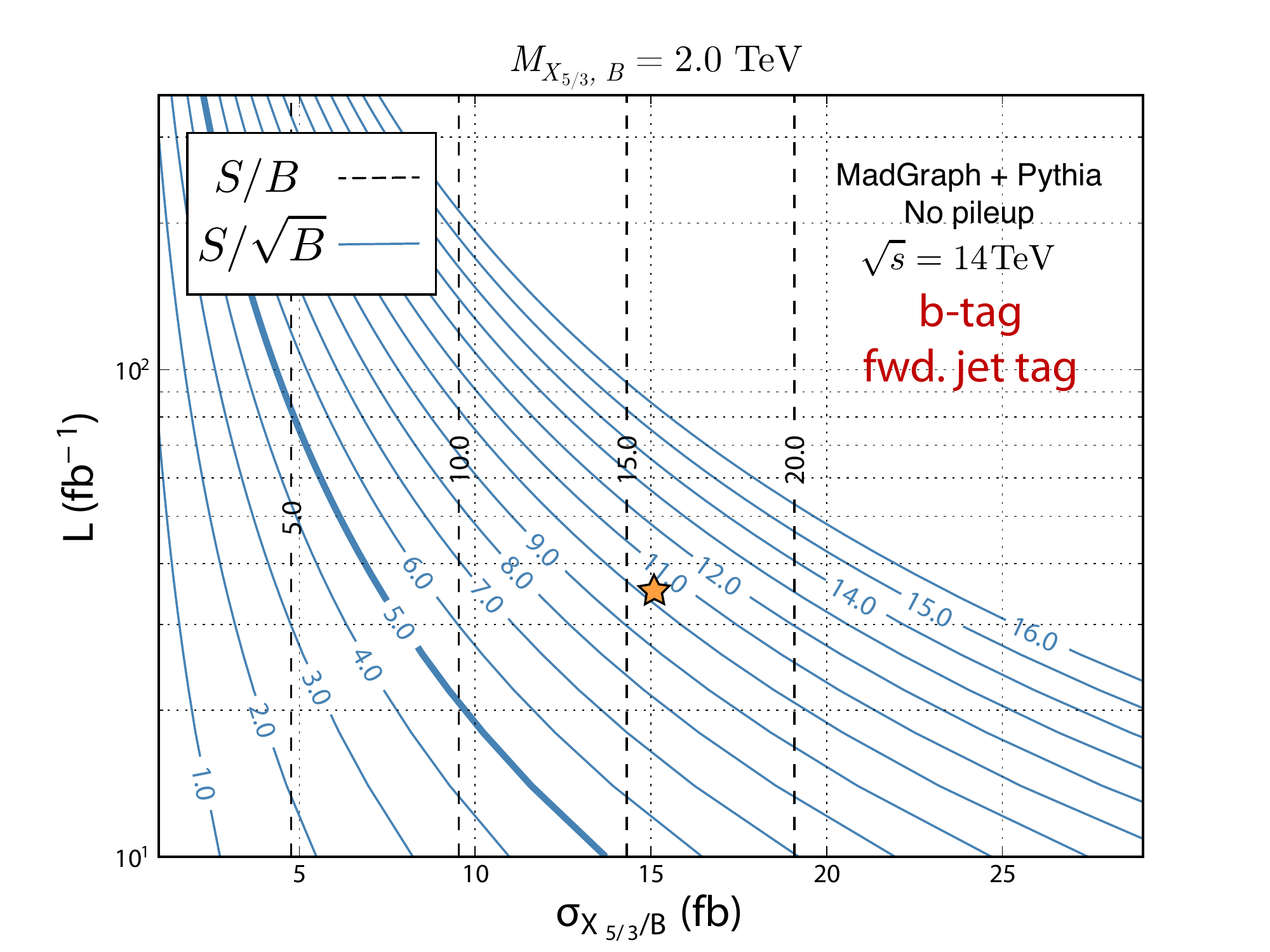}
\caption{Sensitivity to models with large $\MX$, where we expect a high degree of mass degeneracy between $X_{5/3}$ and $B$ partners. The plots on the left \textbf{assume $b$-tagging but not forward jet tagging}, while plots on the right \textbf{assume both $b$-tagging and forward jet tagging}. The inclusive signal cross section and integrated luminosity are on the  $x$ and $y$ axes respectively.  We only show the hadronic top candidate events and no pileup contamination. The solid lines represent contours of constant $S/\sqrt{B}$ at $\mathcal L = 35\, \rm{fb}^{-1}$, while the dashed lines are $S/B$. The selection cuts for each $\MX$ reflect the ones in Table~\ref{tab:cutflow1a}, where we mark the point presented in the table by a star. }
\label{fig:SBresults2}
\end{center}
\end{figure}

We show a more complete representation of our main results with no pileup contamination in Fig.~\ref{fig:SBresults2} for $\MX = 2.0 \TeV$ ,  where we assume that the $X_{5/3}$ and $B$ states are mass degenerate, while Fig.~\ref{fig:SBresults} shows the results for $\MX = 1.0 - 1.75 \TeV$. The plots show contours of constant $S/\sqrt{B}$ (solid lines) for various $\MX$ as a function of the inclusive signal cross section and integrated luminosity. For completeness, we give $S/B$ as dashed lines. The left panels assume the $b$-tagging requirement, but no forward jet tag while the right panels include both $b$-tagging and the forward jet tag. We find that in all considered cases, the future LHC runs have excellent potential for discovery of singly produced top partners, even in the early stages of the experiment and with as low as $20 \,{\rm fb}^{-1}$ of data. In addition, the $14 \TeV$ run of the LHC should be able to achieve a $2 \sigma$ sensitivity, enough to rule out major parts of the parameter space even with $10\, {\rm fb}^{-1}$.

\begin{table}[!]
\begin{center}
 \begin{tabular}{|c|c|c|c|c|c|c|c|c|c|c|c|c|c|c|c|c|c|c|c|c|c|}
 \multicolumn{17}{c}{$\MX = 2.0 \TeV$, $\sigma_{X_{5/3} + B} = 15 \, {\rm fb},$ $L = 35\, {\rm fb^{-1}}$}\\
\hline
\multicolumn{1}{|c|}
   {\redc{$X_{5/3} + B$} }& \multicolumn{2}{|c|}{$\sigma_{s}$ [fb]} & \multicolumn{2}{|c|}{$\sigma_{t\bar{t}}$ [fb]} & \multicolumn{2}{|c|}{$\sigma_{W+{\rm jets}}$ [fb]} &\multicolumn{2}{|c|}{$\epsilon_{s}$ }&\multicolumn{2}{|c|}{$\epsilon_{t\bar{t}}$}& \multicolumn{2}{|c|}{$\epsilon_{W+ {\rm jets}}$}& \multicolumn{2}{|c|}{$S/B$}  & \multicolumn{2}{|c|}{$S/\sqrt{B}$}  \\
 \hline
\hline
  Fat jet candidate & $t$ & $W $ & $t$ & $W$& $t$ & $W$& $t$ & $W$& $t$ & $W$& $t$& $W$& $t$ & $W$  &$t$ & $W$  \\
\hline
Basic Cuts &1.7&1.9&144.0 &487.0 &3807.0 &2301.0&0.38 &0.42 &0.12 &0.41 &0.12&0.08&\tiny{$4\times10^{-4}$ }&\tiny{$6\times10^{-4}$}&0.2&0.2\\
$p_T > 600 \GeV$&1.4&1.6&117.0&430.0&1045.0&747.0&0.31&0.37&0.10&0.36&0.035&0.02&0.001&0.001&0.2&0.2\\
$p_T^{l} > 100 \GeV$&1.3&1.5&61.0&300.0&715.0&502.0&0.30&0.35&0.05&0.25&0.02&0.02&0.002&0.002&0.3&0.3\\
$Ov > 0.5$&1.0&1.1&25.0&150.0&131.0&172.0&0.22&0.22&0.02&0.13&0.004&0.006&0.006&0.003&0.5&0.3\\
$\mX > 1.5 \TeV$&0.9&1.0&2.4&91.0&55.0&118.0&0.19&0.22&0.002&0.08&0.002&0.004&0.01&0.004&0.7&0.4\\
$m_{j^\prime l} > 200 \GeV$&0.8&0.3&0.9&11.0&45.0&37.0&0.18&0.07&$8\times 10^{-4}$&0.009&0.001&0.001&0.02&0.02&0.7&0.7\\
\hline
$b$-tag \& no fwd. tag&\bluebold{0.3}&0.1&\bluebold{0.04}&2.0&\bluebold{0.08}&0.6&0.07&0.03&$4\times 10^{-5}$&0.002&$2 \times 10^{-6}$&$2\times 10^{-5}$&\bluebold{2.5}&0.1&\bluebold{5.2}&1.0\\
\hline
fwd. tag \& no $b$-tag&\bluebold{0.5}&0.2&\bluebold{0.2}&2.5&8.0&5.0&0.11&0.05&$2 \times 10^{-4}$&0.002&$3\times 10^{-4}$&$2 \times 10^{-4}$&\bluebold{0.06}&0.07&\bluebold{1.0}&1.0\\
\hline
$b$-tag and fwd. tag&\bluebold{0.2}&0.1&\bluebold{0.01}&0.5&\bluebold{ $<$ 0.01}&0.07&0.04&0.02&$1\times 10^{-5}$&$4\times 10^{-4}$&$<10^{-6}$&$2 \times 10^{-6}$&\bluebold{15.7}&0.3&\bluebold{10.2}&1.5\\
\hline
\end{tabular}

\caption{ Example cutflow for signal and background events for $\MX = 2.0 \TeV$ and inclusive cross sections $\sigma_{X_{5/3} + B}$.  $\sigma_{s, t\bar{t}, W+{\rm jets}}$ are the signal/background cross sections including all branching ratios, whereas $\epsilon$ are the efficiencies of the cuts relative to the generator level cross sections.  The results assume no pileup contamination. The signal cross section assumes both $X_{5/3}$ and $B$ production. \label{tab:cutflow1a}}
\end{center}
\end{table}

\begin{table}[!]
\begin{center}

 \begin{tabular}{|c|c|c|c|c|c|c|c|c|c|c|c|c|c|c|c|c|c|c|c|c|c|}
 \multicolumn{17}{c}{$\MX = 1.75 \TeV$, $\sigma_{\Xx} = 50\, {\rm fb},$ $L = 20\, {\rm fb^{-1}}$} \\
\hline
\multicolumn{1}{|c|}
   {$X_{5/3} / B$} & \multicolumn{2}{|c|}{$\sigma_{s}$ [fb]} & \multicolumn{2}{|c|}{$\sigma_{t\bar{t}}$ [fb]} & \multicolumn{2}{|c|}{$\sigma_{W+{\rm jets}}$ [fb]} &\multicolumn{2}{|c|}{$\epsilon_{s}$ }&\multicolumn{2}{|c|}{$\epsilon_{t\bar{t}}$}& \multicolumn{2}{|c|}{$\epsilon_{W+ {\rm jets}}$}& \multicolumn{2}{|c|}{$S/B$}  & \multicolumn{2}{|c|}{$S/\sqrt{B}$}  \\
 \hline
\hline
  Fat jet candidate & $t$ & $W $ & $t$ & $W$& $t$ & $W$& $t$ & $W$& $t$ & $W$& $t$& $W$& $t$ & $W$  &$t$ & $W$  \\
\hline
Basic Cuts& 5.6& 5.8&780.0&1975.0&3807.0&2302.0&0.39&0.39&0.05&0.14&0.12&0.08&0.001&0.001&0.4&0.4\\
$p_T > 500 \GeV$&5.1&5.5&314.0&1003.0&2062.0&1339.0&0.34&0.37&0.02&0.07&0.07&0.04&0.002&0.002&0.5&0.5\\
$p_T^{l} > 100 \GeV$&4.8&5.0&180.0&719.0&1353.0&853.0&0.32&0.34&0.01&0.05&0.04&0.03&0.003&0.003&0.5&0.5\\
$Ov > 0.5$&3.4&3.8&108.0&320.0&237.0&259.0&0.23&0.25&0.008&0.02&0.008&0.008&0.01&0.006&0.8&0.6\\
$\mX > 1.5\TeV$&2.5&2.6&7.0&87.0&72.0&141.0&0.17&0.18&$5\times 10^{-4}$&0.006&0.002&0.005&0.03&0.01&1.2&0.8\\
$m_{j^\prime l} > 200 \GeV$&2.3&0.8&3.0&11.0&59.0&42.0&0.15&0.05&$2\times 10^{-4}$&$8 \times 10^{-4}$&0.002&0.001&0.04&0.04&1.3&1.4\\
\hline
$b$-tag \& no fwd. tag&\bluebold{0.8}&0.3&\bluebold{0.3}&2.0&\bluebold{0.1}&0.7&0.06&0.02&$2\times 10^{-5}$&$2\times 10^{-4}$&$3 \times 10^{-6}$&$2\times 10^{-5}$&\bluebold{2.1}&0.3 &\bluebold{5.9}&2.2\\
\hline
fwd. tag \& no $b$-tag&\bluebold{1.5}&0.5&\bluebold{0.7}&3.0&\bluebold{10.0}&6.0&0.1 &0.03 &$5\times 10^{-5}$&$2\times 10^{-4}$&$3\times 10^{-4}$&$2\times 10^{-4}$&\bluebold{0.1}&0.2&\bluebold{2.0}&2.2\\
\hline
$b$-tag and fwd. tag&\bluebold{0.5}&0.2&\bluebold{0.06}&0.6&\bluebold{$<$ 0.01}&0.1&0.04&0.01&$3\times 10^{-6}$&$4\times 10^{-5}$& $< 10^{-6}$&$3 \times 10^{-6}$&\bluebold{9.2}&0.8&\bluebold{9.9}&3.0\\
\hline
\end{tabular}

\vspace{3 mm}
 \begin{tabular}{|c|c|c|c|c|c|c|c|c|c|c|c|c|c|c|c|c|c|c|c|c|c|}
 \multicolumn{17}{c}{$\MX = 1.5 \TeV$, $\sigma_{\Xx} = 100\, {\rm fb},$ $L = 20\, {\rm fb^{-1}}$} \\
\hline
\multicolumn{1}{|c|}
   {$X_{5/3} / B$} & \multicolumn{2}{|c|}{$\sigma_{s}$ [fb]} & \multicolumn{2}{|c|}{$\sigma_{t\bar{t}}$ [fb]} & \multicolumn{2}{|c|}{$\sigma_{W+{\rm jets}}$ [fb]} &\multicolumn{2}{|c|}{$\epsilon_{s}$ }&\multicolumn{2}{|c|}{$\epsilon_{t\bar{t}}$}& \multicolumn{2}{|c|}{$\epsilon_{W+ {\rm jets}}$}& \multicolumn{2}{|c|}{$S/B$}  & \multicolumn{2}{|c|}{$S/\sqrt{B}$}  \\
 \hline
\hline
 Fat jet candidate & $t$ & $W $ & $t$ & $W$& $t$ & $W$& $t$ & $W$& $t$ & $W$& $t$& $W$& $t$& $W$&  $t$& $W$  \\
\hline
Basic Cuts&11.0&10.6&780.0&1975.0&3807.0&2301.0&0.37&0.36&0.05&0.14&0.12&0.08&0.002&0.003&0.7&0.8\\
$p_T > 500 \GeV$&9.3&9.6&314.0&1003.0&2062.0&1339.0&0.31&0.32&0.02&0.07&0.07&0.04&0.004&0.004&0.9&0.9\\
$p_T^{l} > 100 \GeV$&8.5&8.6&180.0&719.0&1353.0&852.0&0.29&0.29&0.01&0.05&0.04&0.03&0.006&0.005&1.0&1.0\\
$Ov > 0.5$&5.8&6.2&108.0&320.0&237.0&259.0&0.20&0.21&0.008&0.02&0.008&0.008&0.02&0.01&1.4&1.1\\
$\mX > 1.3 \TeV$&4.2&4.4&14.0&143.0&106.0&183.0&0.14&0.15&0.001 &0.01 &0.003&0.006&0.04&0.01&1.7&1.1\\
$m_{j^\prime l} > 200 \GeV$&3.8&1.4&5.9&19.0&85.0&50.0&0.13&0.05&$4\times 10^{-4}$&0.001&0.003&0.002&0.04&0.05&1.8&2.0\\
\hline
$b$-tag \& no fwd. tag&\bluebold{1.4}&0.5&\bluebold{0.7}&3.5&\bluebold{0.2}&0.9&0.05&0.02&$5 \times 10^{-5}$&$2 \times 10^{-4}$&$1\times 10^{-5}$&$3\times 10^{-5}$&\bluebold{1.5}&0.3&\bluebold{6.5}&2.9\\
\hline
fwd. tag \& no $b$-tag&\bluebold{2.4}&0.9&\bluebold{1.3}&4.5&\bluebold{14.6}&7.3&0.08&0.03&$9\times 10^{-5}$&$3\times 10^{-4}$&$5\times 10^{-4}$&$2\times 10^{-4}$&\bluebold{0.2}&0.2&\bluebold{2.7}&3.1\\
\hline
$b$-tag and fwd. tag&\bluebold{0.9}&0.3&\bluebold{0.1}&0.9&\bluebold{0.02}&0.1&0.03&0.01&$1\times 10^{-5}$&$6\times 10^{-5}$&$1\times 10^{-6}$&$3\times 10^{-6}$&\bluebold{5.4}&0.8&\bluebold{9.7}&3.8\\
\hline
\end{tabular}

\vspace{3 mm}
 \begin{tabular}{|c|c|c|c|c|c|c|c|c|c|c|c|c|c|c|c|c|c|c|c|c|c|}
 \multicolumn{17}{c}{$\MX = 1.25 \TeV$, $\sigma_{\Xx} = 150\, {\rm fb},$ $L = 20\, {\rm fb^{-1}}$}\\
\hline
\multicolumn{1}{|c|}
   {$X_{5/3} / B$} & \multicolumn{2}{|c|}{$\sigma_{s}$ [fb]} & \multicolumn{2}{|c|}{$\sigma_{t\bar{t}}$ [fb]} & \multicolumn{2}{|c|}{$\sigma_{W+{\rm jets}}$ [fb]} &\multicolumn{2}{|c|}{$\epsilon_{s}$ }&\multicolumn{2}{|c|}{$\epsilon_{t\bar{t}}$}& \multicolumn{2}{|c|}{$\epsilon_{W+ {\rm jets}}$}& \multicolumn{2}{|c|}{$S/B$}  & \multicolumn{2}{|c|}{$S/\sqrt{B}$}  \\
 \hline
\hline
 Fat jet candidate & $t$ & $W $ & $t$ & $W$& $t$ & $W$& $t$ & $W$& $t$ & $W$& $t$& $W$& $t$& $W$&  $t$& $W$  \\
\hline
Basic Cuts&15.1&13.7&780.0&1975.0&3807.0&2301.0&0.34&0.31&0.05&0.14&0.12&0.08&0.003&0.004&1.0&1.0\\
$p_T^{l} > 100 \GeV$&13.4&11.8&389.0&1306.0&2279.0&1330.0&0.31&0.27&0.03&0.09&0.07&0.04&0.005&0.005&1.2&1.2\\
$Ov > 0.5$&8.7&7.8&230.0&468.0&386.0&343.0&0.20&0.18&0.02&0.03&0.01&0.01&0.01&0.01&1.6&1.4\\
$\mX > 1.0\TeV$&7.5&6.5&66.0&333.0&231.0&294.0&0.17&0.15&0.005&0.02&0.008&0.01&0.03&0.01&2.0&1.4\\
$m_{j^\prime l} > 200 \GeV$&6.5&2.3&27.0&51.0 &181.0&72.0 &0.15&0.05&0.002&0.004&0.006&0.002&0.03&0.05&2.0&2.6\\
\hline
$b$-tag \& no fwd. tag&\bluebold{2.3}&0.9&\bluebold{4.7}&9.0&\bluebold{0.6}&1.3&0.05&0.02&$3 \times 10^{-4}$&$6\times 10^{-4}$&$2\times 10^{-5}$&$4 \times 10^{-5}$&\bluebold{0.4}&0.2&\bluebold{4.5}&3.2\\
\hline
fwd. tag \& no $b$-tag&\bluebold{4.0}&1.4&\bluebold{6.7}&12.0&\bluebold{32.0}&11.0&0.09&0.03&$5\times 10^{-4}$&$9 \times 10^{-4}$&0.001&$4\times 10^{-4}$&\bluebold{0.1}&0.2&\bluebold{2.9}&3.7\\
\hline
$b$-tag and fwd. tag&\bluebold{1.4}&0.5&\bluebold{1.0}&2.3&\bluebold{0.07}&0.2 &0.03&0.01&$7\times 10^{-5}$&$1\times 10^{-4}$&$1 \times 10^{-6}$&$1\times 10^{-5}$&\bluebold{1.4}&0.6&\bluebold{6.4}&4.0\\
\hline
\end{tabular}

\vspace{3 mm}
 \begin{tabular}{|c|c|c|c|c|c|c|c|c|c|c|c|c|c|c|c|c|c|c|c|c|c|}
 \multicolumn{17}{c}{$\MX = 1.0 \TeV$, $\sigma_{\Xx} = 300\, {\rm fb},$ $L = 20\, {\rm fb^{-1}}$}\\
\hline
\multicolumn{1}{|c|}
   {$X_{5/3}/B$} & \multicolumn{2}{|c|}{$\sigma_{s}$ [fb]} & \multicolumn{2}{|c|}{$\sigma_{t\bar{t}}$ [fb]} & \multicolumn{2}{|c|}{$\sigma_{W+{\rm jets}}$ [fb]} &\multicolumn{2}{|c|}{$\epsilon_{s}$ }&\multicolumn{2}{|c|}{$\epsilon_{t\bar{t}}$}& \multicolumn{2}{|c|}{$\epsilon_{W+ {\rm jets}}$}& \multicolumn{2}{|c|}{$S/B$}  & \multicolumn{2}{|c|}{$S/\sqrt{B}$}  \\
 \hline
\hline
 Fat jet candidate & $t$ & $W $ & $t$ & $W$& $t$ & $W$& $t$ & $W$& $t$ & $W$& $t$& $W$& $t$& $W$&  $t$& $W$  \\
\hline
Basic Cuts&23.4&19.6&780.0&1975.0&3807.0&2302.0&0.26&0.22&0.05&0.14&0.12&0.08&0.005&0.006&1.6&1.6\\
$p_T^{l} > 100 \GeV$&19.4&15.9&389.0&1306.0&2279.0&1330.0&0.22&0.18&0.03&0.09&0.07&0.04&0.007&0.007&1.7&1.7\\
$Ov > 0.5$&12.0&9.8&230.0&468.0&386.0&343.0&0.14&0.11&0.02&0.03&0.01&0.01&0.02&0.01&2.2&1.9\\
$\mX > 0.8 \TeV$&11.3&9.8&144.0&467.0&316.0&343.0&0.13&0.1&0.01&0.03&0.01&0.01&0.02&0.01&2.4&1.8\\
$m_{j^\prime l} > 200 \GeV$&9.1&3.9&55.0&80.0&247.0&84.0&0.10 &0.04&0.004&0.006&0.008&0.003&0.03&0.06&2.4&3.2\\
\hline
$b$-tag \& no fwd. tag&\bluebold{3.3}&1.5&\bluebold{11.0}&13.0&\bluebold{0.8}&1.5&0.04&0.02&$8 \times 10^{-4}$&$9 \times 10^{-4}$&$3\times 10^{-5}$&$5\times 10^{-5}$&\bluebold{0.2}&0.2&\bluebold{4.2}&3.7\\
\hline
fwd. tag \& no $b$-tag&\bluebold{5.6}&2.3&\bluebold{14.0}&19.0&\bluebold{44.0}&13.0&0.06&0.03&0.001&0.001&0.001&$4 \times 10^{-4}$&\bluebold{0.1}&0.2&\bluebold{3.3}&4.5\\
\hline
$b$-tag and fwd. tag&\bluebold{2.0}&0.9&\bluebold{2.6}&3.5&\bluebold{0.09}&0.21&0.02&0.01&$2\times 10^{-4}$&$2 \times 10^{-4}$&0.0&$1\times10^{-5}$&\bluebold{0.8}&0.5&\bluebold{5.5}&4.7\\
\hline
\end{tabular}

\caption{ Example cutflow for signal and background events for various $\MX$ and inclusive cross sections $\sigma_{\Xx}$.  $\sigma_{s, t\bar{t}, W+{\rm jets}}$ are the signal/background cross sections including all branching ratios, whereas $\epsilon$ are the efficiencies of the cuts relative to the generator level cross sections.  The results assume no pileup contamination. The cut on the fat jet $p_T$ is not shown for the case of $\MX = 1.0, 1.25 \TeV$ since it is included in the Basic Cuts.\label{tab:cutflow1} }
\end{center}
\end{table}

\newpage
\subsection{Effect of Pileup on $\MX$ Sensitivity}

As we pointed out in the previous sections, our event selection criteria contain several observables which are weakly affected by pileup ($i.e.$ $Ov,$ $\MX$,  forward jet tag). However, some of the other selection criteria ($i.e.$ $m_{jl}, p_T^{\rm fj}$) are somewhat pileup sensitive. The lower $p_T$ cut on the fat jet allows for some low fat jet $p_T$ events to migrate into the sample which passes the Basic Cuts due to the fact that we use a large $R=1.0$ cone for fat jet clustering \footnote{Note that, in principle, the effects of pileup can further be suppressed by lowering the size of the fat jet cone without increasing the lower $p_T$ cut on the fat jet}. Furthermore, the effects of pileup on any observable constructed out of the $r=0.4$ jets are limited (compared to the fat jet) by the small jet cone size, but can still be non-negligible at 50 average pileup events. 

\begin{figure}[t]
\begin{center}
\includegraphics[width=3.5in]{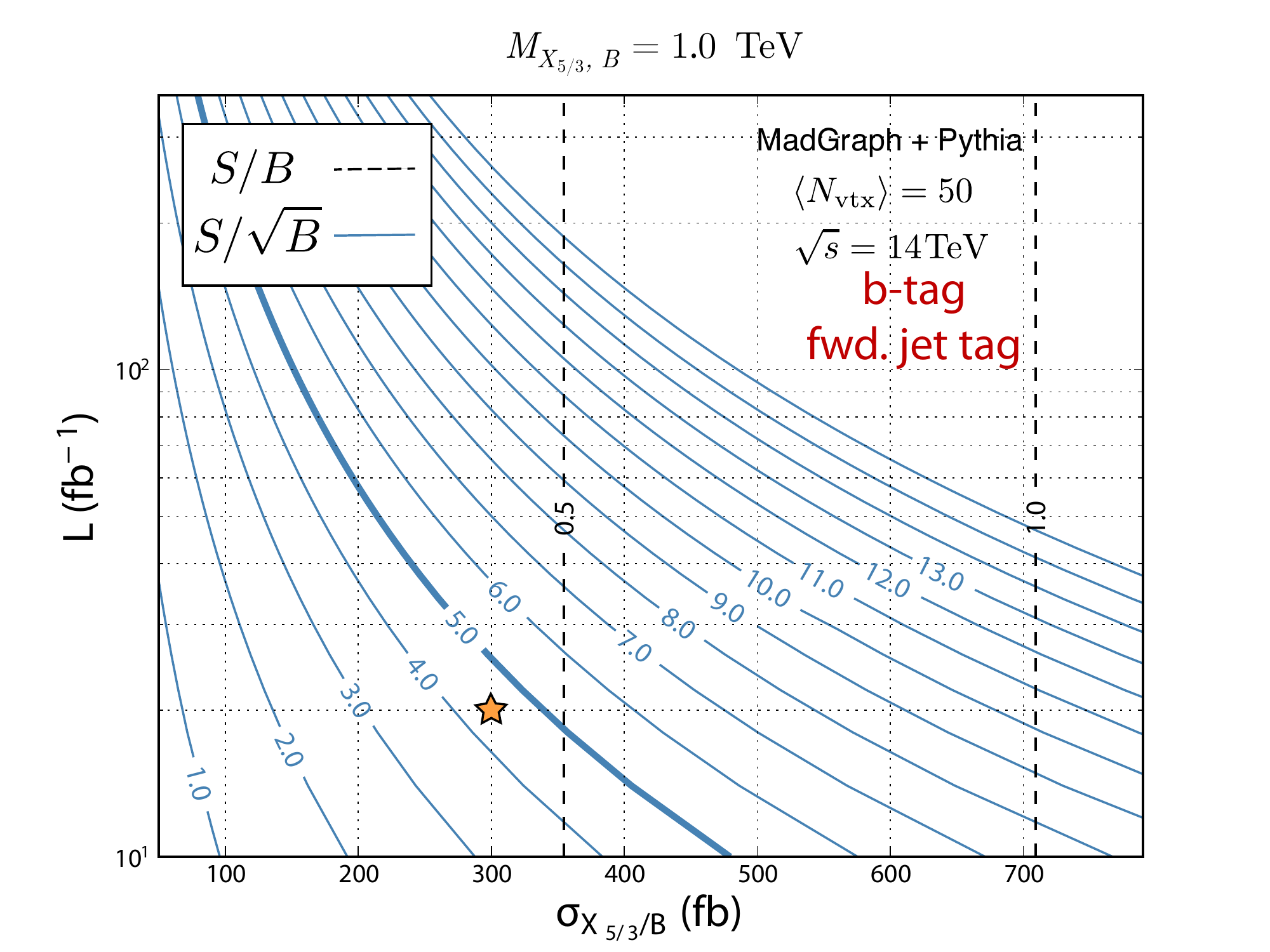}\includegraphics[width=3.5in]{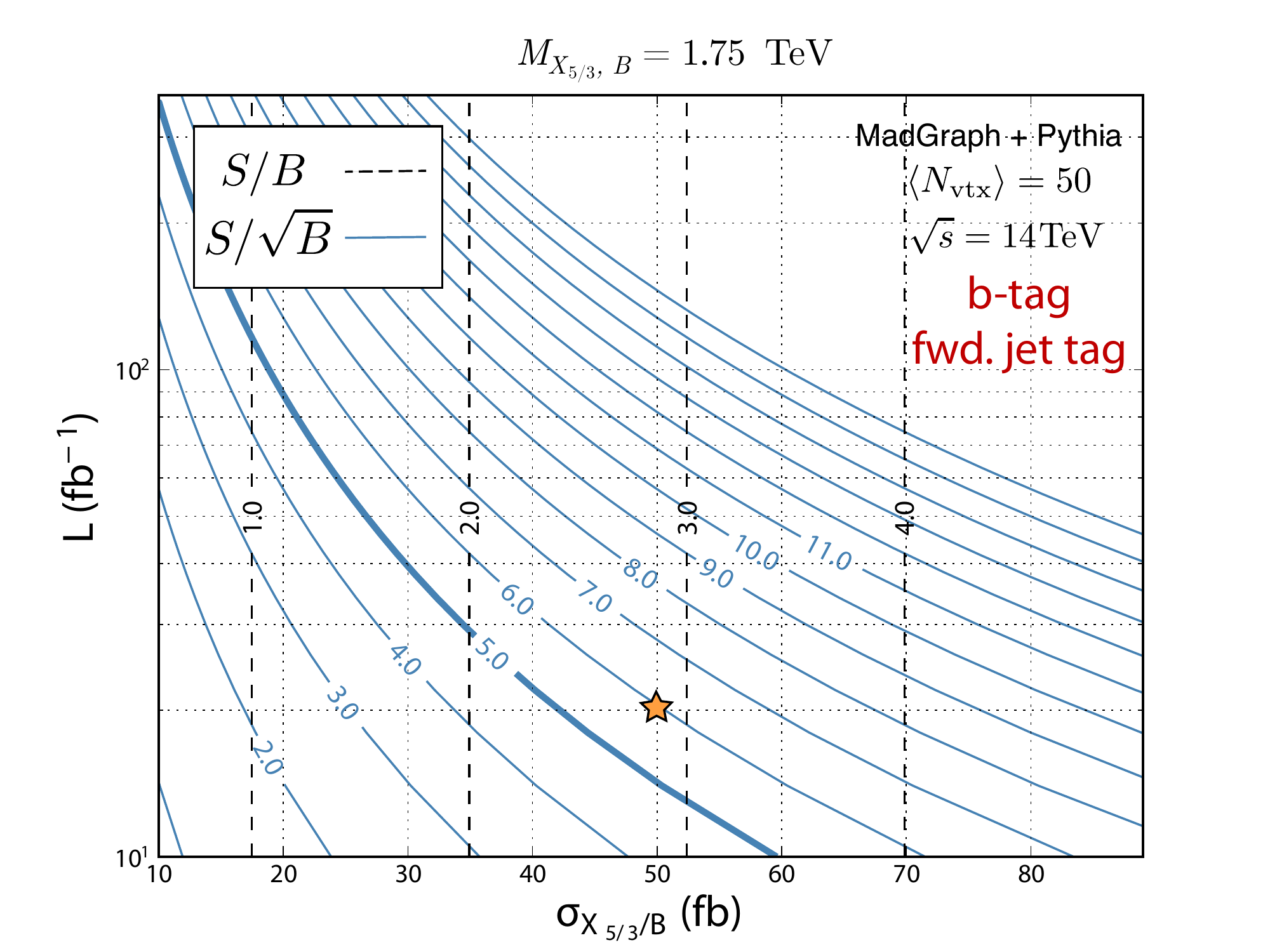} \\
\includegraphics[width=3.5in]{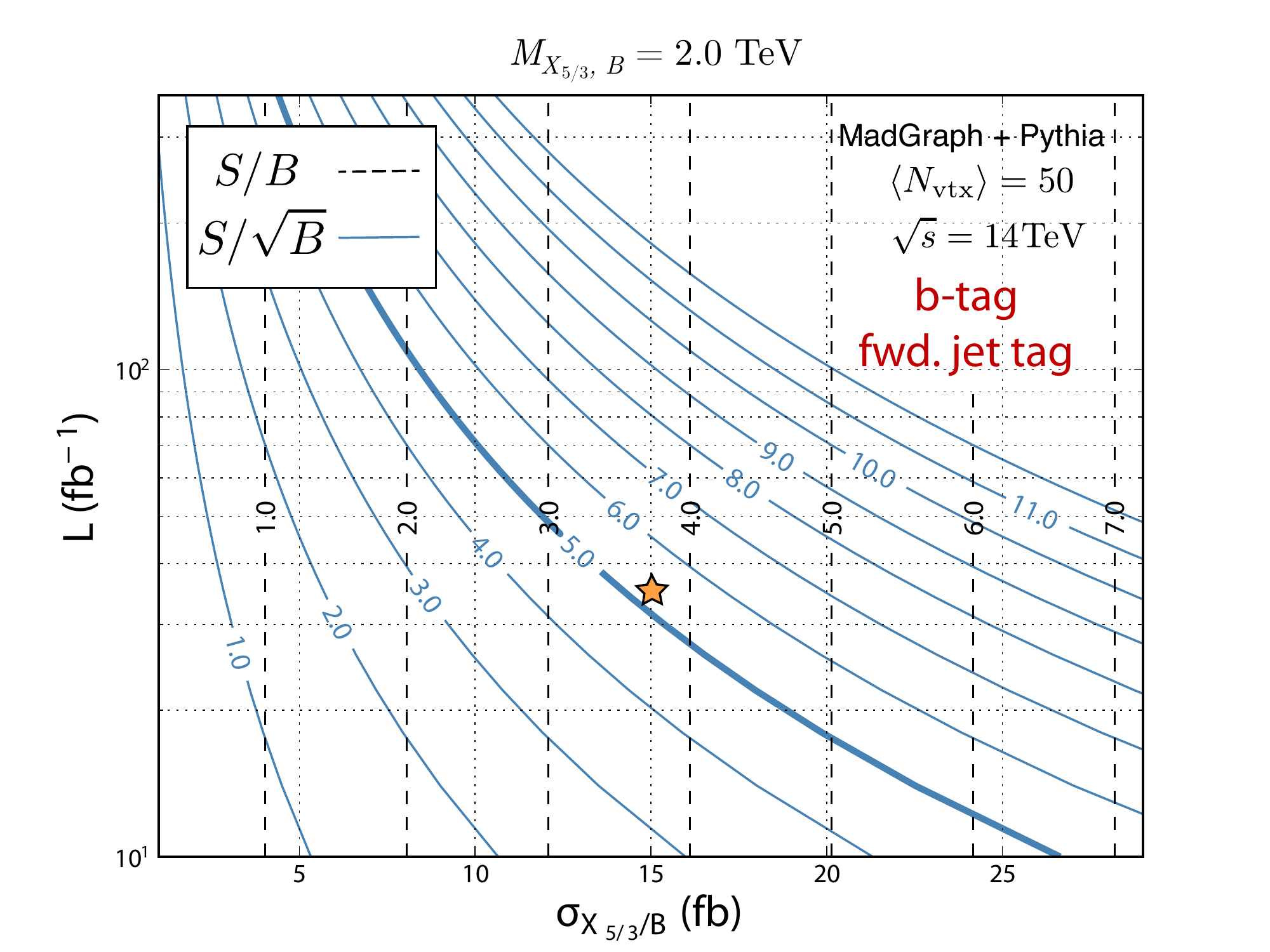}
\caption{Sensitivity to various $\MX$ in presence of \textbf{50 average interactions per bunch crossing}.  The inclusive signal cross section and integrated luminosity are on the $x$ and $y$ axes respectively.  We only show the hadronic top candidate events. The blue solid lines represent contours of constant $S/\sqrt{B}$. The dashed lines are $S/B$. The selection cuts for each $\MX$ reflect the ones in Table~\ref{tab:cutflow2}, where we mark the point presented in the table by a star. }
\label{fig:SBresultsPup}
\end{center}
\end{figure}

Overall effects of pileup on our results are fairly mild and can be mitigated by slight modifications of the cuts on pileup sensitive observables. For illustration, we analyzed three samples of signal events with masses $\MX~=~1.0, 1.75, 2.0 \TeV$ and the relevant backgrounds in the presence of average $\langle N_{\rm vtx}\rangle = 50$ interactions per bunch crossing. In order to reduce the effects of ``pileup induced migration'', we increase the transverse momentum threshold on the fat jet to $p_T > 600 \GeV$ for $\MX = 1.75 \TeV$ and $p_T > 500 \GeV $ for $\MX = 1.0 \TeV$, as well as shift the cut on the $m_{jl} > 300 \GeV$ in both cases. We do not modify the cuts on pileup insensitive observables. The increase in lower $p_T, m_{jl}$ cuts is most certainly dependent on the amount of pileup contamination and requires further consideration at $ \langle N_{\rm vtx} \rangle > 50 $ pileup events. 

\begin{table}[t]

 \begin{tabular}{|c|c|c|c|c|c|c|c|c|c|c|c|c|c|c|c|c|c|c|c|c|c|}
 \multicolumn{17}{c}{$\MX = 2.0 \TeV$, $\sigma_{X_{5/3} + B} = 15\, {\rm fb},$ $L = 35\, {\rm fb^{-1}},$ $\langle N_{\rm vtx} \rangle = 50$}\\
\hline
\multicolumn{1}{|c|}
  {$\redc{X_{5/3} + B}$} & \multicolumn{2}{|c|}{$\sigma_{s}$ [fb]} & \multicolumn{2}{|c|}{$\sigma_{t\bar{t}}$ [fb]} & \multicolumn{2}{|c|}{$\sigma_{W+{\rm jets}}$ [fb]} &\multicolumn{2}{|c|}{$\epsilon_{s}$ }&\multicolumn{2}{|c|}{$\epsilon_{t\bar{t}}$}& \multicolumn{2}{|c|}{$\epsilon_{W+ {\rm jets}}$}& \multicolumn{2}{|c|}{$S/B$}  & \multicolumn{2}{|c|}{$S/\sqrt{B}$}  \\
 \hline
\hline
  Fat jet candidate & $t$ & $W $ & $t$ & $W$& $t$ & $W$& $t$ & $W$& $t$ & $W$& $t$& $W$& $t$ & $W$  &$t$ & $W$  \\
\hline
Basic Cuts&1.6&2.3&76.0&556.0&5921.0&3879.0&0.36&0.51&0.06&0.46&0.19&0.12&$3\times 10^{-4}$&$4\times 10^{-4}$&0.1&0.1\\
$p_T > 700 \GeV$&1.3&2.0&60.0&506.0&1322.0&1082.0&0.28&0.45&0.05&0.42&0.04&0.04&$9 \times 10^{-4}$&$8 \times 10^{-4}$&0.2&0.2\\
$p_T^{l} > 100 \GeV$&1.2&1.9&23.0&349.0&912.0&733.0&0.27&0.41&0.02&0.29&0.03&0.02&0.001&0.001&0.2&0.2\\
$Ov > 0.5$&1.0&1.3&12.0&170.0&354.0&254.0&0.23&0.30&0.01 &0.14&0.01&0.008&0.003&0.002&0.3&0.3\\
$\MX > 1.5\TeV$&0.9&1.2&0.7&106.0&168.0&160.0&0.20&0.26&$6 \times 10^{-4}$&0.09&0.006&0.005&0.005&0.003&0.4&0.3\\
$m_{jl} > 300 \GeV$&0.8&0.4&0.5&12.0&111.0&27.0&0.17&0.08&$4 \times 10^{-4}$&0.01&0.004&$9 \times 10^{-4}$&0.007&0.02&0.4&0.7\\
\hline
$b$-tag \& no fwd. tag&\bluebold{0.3}&0.1&\bluebold{0.08}&2.7&\bluebold{0.2}&0.5&0.07&0.03&$7 \times 10^{-5}$&0.002&$5\times 10^{-6}$&$2 \times 10^{-5}$&\bluebold{1.3}&0.09&\bluebold{3.7}&1.0\\
\hline
fwd. tag \& no $b$-tag&\bluebold{0.5}&0.3&\bluebold{0.2}&3.7&\bluebold{32.0}&7.8&0.1 &0.06&$2 \times 10^{-4}$&0.003&0.001&$3 \times 10^{-4}$&\bluebold{0.02}&0.05&\bluebold{0.6}&0.9\\
\hline
$b$-tag and fwd. tag&\bluebold{0.2}&0.1&\bluebold{0.03}&0.9&\bluebold{0.03}&0.1&0.05&0.02&$2 \times 10^{-5}$&$7 \times 10^{-4}$&$1\times 10^{-6}$&$4\times 10^{-6}$&\bluebold{3.7}&0.2&\bluebold{5.3}&1.3\\
\hline
\end{tabular}

\vspace{5 mm}
 \begin{tabular}{|c|c|c|c|c|c|c|c|c|c|c|c|c|c|c|c|c|c|c|c|c|c|}
 \multicolumn{17}{c}{$\MX = 1.75 \TeV$, $\sigma_{\Xx} = 50\, {\rm fb},$ $L = 20\, {\rm fb^{-1}},$ $\langle N_{\rm vtx} \rangle = 50$}\\
\hline
\multicolumn{1}{|c|}
  {$X_{5/3} / B$} & \multicolumn{2}{|c|}{$\sigma_{s}$ [fb]} & \multicolumn{2}{|c|}{$\sigma_{t\bar{t}}$ [fb]} & \multicolumn{2}{|c|}{$\sigma_{W+{\rm jets}}$ [fb]} &\multicolumn{2}{|c|}{$\epsilon_{s}$ }&\multicolumn{2}{|c|}{$\epsilon_{t\bar{t}}$}& \multicolumn{2}{|c|}{$\epsilon_{W+ {\rm jets}}$}& \multicolumn{2}{|c|}{$S/B$}  & \multicolumn{2}{|c|}{$S/\sqrt{B}$}  \\
 \hline
\hline
  Fat jet candidate & $t$ & $W $ & $t$ & $W$& $t$ & $W$& $t$ & $W$& $t$ & $W$& $t$& $W$& $t$ & $W$  &$t$ & $W$  \\
\hline
Basic Cuts&5.6&7.0&1281.0&3587.0&5921.0&3879.0&0.38&0.47&0.09&0.25&0.19&0.12&\tiny{$8 \times 10^{-4}$}&\tiny{$7\times 10^{-4}$}&0.3&0.3\\
$p_T > 600 \GeV$&4.8&6.5&283.0&1410.0&2458.0&1921.0&0.32&0.44&0.02&0.10 &0.08&0.06 &0.002&0.001&0.4&0.4\\
$p_T^{l} > 100 \GeV$&4.5&5.9&140.0&978.0&1595.0&1214.0&0.30&0.40 &0.01&0.07&0.05&0.04&0.003&0.002&0.5&0.4\\
$Ov > 0.5$&3.7&4.2&97.0&393.0&554.0&378.0&0.25&0.29&0.007&0.03&0.02&0.01&0.006&0.005&0.7&0.6\\
$\MX > 1.5 \TeV$&2.8&3.1&6.2&103.0&200.0&191.0&0.19&0.20&$4\times 10^{-4}$&0.007&0.007&0.006&0.01&0.009&0.9&0.7\\
$m_{jl} > 300 \GeV$&2.2&0.9&2.9&12.0&132.0&32.0&0.15&0.06&$2\times 10^{-4}$&$8 \times 10^{-4}$&0.004&0.001&0.02&0.05&0.9&1.5\\
\hline
$b$-tag \& no fwd. tag&\bluebold{0.9}&0.3&\bluebold{0.7}&2.8&\bluebold{0.2}&0.6&0.06&0.02&$5\times 10^{-5}$&$2\times 10^{-4}$&$1\times 10^{-5}$&$2\times 10^{-5}$&\bluebold{1.0}&0.3&\bluebold{4.1}&2.1\\
\hline
fwd. tag \& no $b$-tag&\bluebold{1.6}&0.6&\bluebold{1.0}&4.2&\bluebold{39.0}&9.4&0.11&0.04&$7\times 10^{-5}$&$3\times 10^{-4}$&0.001&$3\times 10^{-4}$&\bluebold{0.04}&0.1&\bluebold{1.1}&1.9\\
\hline
$b$-tag and fwd. tag&\bluebold{0.6}&0.2&\bluebold{0.2}&1.1&\bluebold{0.03}&0.2&0.04&0.02&$1\times 10^{-5}$&$8\times 10^{-5}$&$1\times10^{-6}$&$1\times 10^{-5}$&\bluebold{2.8}&0.5&\bluebold{5.9}&2.4\\
\hline
\end{tabular}

\vspace{5 mm}
 \begin{tabular}{|c|c|c|c|c|c|c|c|c|c|c|c|c|c|c|c|c|c|c|c|c|c|}
 \multicolumn{17}{c}{$\MX = 1.0 \TeV$, $\sigma_{\Xx} = 300\, {\rm fb},$ $L = 20\, {\rm fb^{-1}},$ $\langle N_{\rm vtx} \rangle = 50$}\\
\hline
\multicolumn{1}{|c|}
  {$X_{5/3} / B$} & \multicolumn{2}{|c|}{$\sigma_{s}$ [fb]} & \multicolumn{2}{|c|}{$\sigma_{t\bar{t}}$ [fb]} & \multicolumn{2}{|c|}{$\sigma_{W+{\rm jets}}$ [fb]} &\multicolumn{2}{|c|}{$\epsilon_{s}$ }&\multicolumn{2}{|c|}{$\epsilon_{t\bar{t}}$}& \multicolumn{2}{|c|}{$\epsilon_{W+ {\rm jets}}$}& \multicolumn{2}{|c|}{$S/B$}  & \multicolumn{2}{|c|}{$S/\sqrt{B}$}  \\
 \hline
\hline
  Fat jet candidate & $t$ & $W $ & $t$ & $W$& $t$ & $W$& $t$ & $W$& $t$ & $W$& $t$& $W$& $t$ & $W$  &$t$ & $W$  \\
\hline
Basic Cuts&29.7&31.9&1281.0&3587.0&5921.0&3879.0&0.33&0.36&0.09&0.25&0.19&0.13&0.004&0.004&1.6&1.5\\
$p_T > 500 \GeV$&22.3&26.9&715.0&2588.0&4220.0&3081.0&0.25&0.30&0.05&0.18&0.14&0.10&0.004&0.004&1.4&1.3\\
$p_T^{l} > 100 \GeV$&18.8&21.6&300.0&1630.0&2461.0&1752.0&0.21&0.24&0.02&0.10&0.08&0.06&0.007&0.006&1.6&1.4\\
$Ov > 0.5$&14.5&12.2&201.0&528.0&789.0&464.0&0.16&0.14&0.01&0.04&0.03&0.02&0.01&0.01&2.0&2.0\\
$\MX > 0.8 \GeV$&13.9&12.2&115.0&527.0&653.0&463.0&0.16&0.14&0.008&0.04&0.02&0.02&0.02&0.01&2.2&2.0\\
$m_{jl} > 300 \GeV$&9.3&4.0&47.0&73.0&416.0&84.0&0.10 &0.04&0.003&0.005&0.01&0.003&0.02 &0.06&1.9&3.3\\
\hline
$b$-tag \& no fwd. tag&\bluebold{3.4}&1.7&\bluebold{15.0}&16.0&\bluebold{1.2}&1.7&0.04&0.02&0.001&0.001 &$4\times10^{-5}$&$6\times10^{-5}$&\bluebold{0.2}&0.2&\bluebold{3.8}&3.7\\
\hline
fwd. tag \& no $b$-tag&\bluebold{6.3}&2.7&\bluebold{18.0}&25.0&\bluebold{123.0}&25.0&0.07&0.03&0.001&0.002&0.004&$8 \times 10^{-4}$&\bluebold{0.04}&0.1&\bluebold{2.4}&4.0\\
\hline
$b$-tag and fwd. tag&\bluebold{2.3}&1.2&\bluebold{5.2}&5.7&\bluebold{0.3}&0.4&0.03&0.01&$4\times 10^{-4}$&$4\times 10^{-4}$&$1\times 10^{-5}$&$1\times 10^{-5}$&\bluebold{0.4}&0.4&\bluebold{4.4}&4.2\\
\hline
\end{tabular}

\caption{Example cutflow for signal and background events \textbf{in the presence of $\langle N_{\rm vtx} \rangle = 50$ interactions per bunch crossing}, for various $\MX$ and inclusive cross sections $\sigma_{\Xx}$. No pileup subtraction/correction techniques have been applied to the samples. $\sigma_{s, t\bar{t}, W+{\rm jets}}$ are the signal/background cross sections including all branching ratios, whereas $\epsilon$ are the efficiencies of the cuts relative to the generator level cross sections.  The results for $\MX = 2.0 \TeV$ assume both $X_{5/3}$ and $B$ production. \label{tab:cutflow2}}

\end{table}

In addition to shifting and broadening kinematic distributions, high pileup is likely to produce uniformly distributed, soft ``pileup jets'' which could mimic leptonic top decays in case they land close enough to the hardest lepton. In order to reduce the effect of fake pileup jets on the event cathegorisation criteria from Section~\ref{sec:tagger} ($i.e.$ whether the event is a hadronic top or hadronic $W$ candidate), we consider only $r =0.4$ jets with $p_T > 50 \GeV$ which are in the vicinity of the hard lepton. And while it is in principle possible to design alternative criteria for cathegorising events into hadronic top and hadronic $W$ candidates,  here we choose to postpone this detail until future studies. 

Table~\ref{tab:cutflow2} and Fig.~\ref{fig:SBresultsPup} show the effects of pileup on our results in more detail.  On a cut-by-cut basis, we find that the signal cross section remains weakly affected by pileup at $\langle N_{\rm vtx}\rangle = 50$ interactions per bunch crossing, with the efficiencies of each cut remaining at a few percent  level compared to our study with no-pileup. The background events are somewhat more pileup sensitive, especially $W$+jets, as multi-jet events are characterized by more soft components and hence more pileup susceptible. 

We find that \textbf{without any pileup correction or subtraction}, we can achieve the same signal cross sections as in our studies without pileup while the amount of background events which survive the event selections is increased by roughly a factor of $2 - 3$. Still, we find that for our benchmark data points,  a $\sim 6\sigma$ sensitivity is achievable for $\MX = 1.75 \TeV$ and $\sim 4 \sigma$ for $\MX = 1.0 \TeV$ with $20 \, {\rm fb^{-1}}$, assuming both $b$-tagging and forward jet tagging. Our results on effects of pileup on signal significance can be interpreted as the most pessimistic scenario and a lower limit on how well the experiments can perform as a function of pileup mitigation efficiency. 

Future LHC experiments are likely to employ advanced pileup subtraction techniques using track information and overtaxing, which could only further improve the performance of event selection in a high pileup environment. However, it is important to note that since we employ a number of already pileup insensitive observables, it is likely that no aggressive pileup subtraction technique will be necessary to recover the full power of our events selection.

\subsection{A Few Remarks on the Complementarity of Top Partner Searches}\label{sec:ifsignal}

In case a top partner is discovered at the LHC, combining results from different channels could greatly improve the significance of the signal. Yet, there is additional information one can obtain from measurements of both same sign di-lepton and other decay channels.

For instance, a possible mass degeneracy between the $X_{5/3}$ and $B$ states could be difficult to untangle with the current mass resolution of the LHC experiments. 
In case a signal is observed, considering only the invariant mass distribution of a $tW$ system or the $H_T$ distribution would likely not be sufficient to determine whether there are one or more resonances observed in the signal events. Complementary information from same sign di-lepton channel could aid in resolving the mass degeneracy. As noted before, same sign di-lepton searches are sensitive only to the production o the $X_{5/3}$ partner and not the $B$ state. A simple cross section measurement (upon unfolding) of both the same sign di-lepton and lepton-jet channels should thus show a difference
\begin{equation}
	\Delta \sigma = \sigma_{X_{5/3} + B}^{l+{\rm fj}} - \sigma_{X_{5/3}}^{2l} \sim \sigma_{B},
\end{equation}
where $l+ {\rm fj}$ refers to lepton-jet channel and $2l$ represents the same sign di-lepton channels. Note that normalising $\Delta \sigma$ with, say, the sum of same sign di-lepton and lepton-jet cross sections can further reduce the systematic uncertainties. 
 
Furthermore,  indirectly deducing the presence of a $B$ in the signal is also possible by considering charge asymmetries. As outlined in Section~\ref{sec:model.prod}, $X_{5/3}$ production dominates over $\bar{X}_{5/3}$ because the former is produced from $g$ and an up-type quark in the initial state while the latter is produced from $g$ and a down-type quark. In the same sign di-lepton search one should thus observe an excess of the $l^+l^+$ signal over the $l^-l^-$ events. Analogously, if the lepton charge would be measured in the lepton-jet events we are investigating here ($e.g.$ top partner decays into $Wt\rightarrow l \nu_l j j b$), one should observe an excess of $l^+$ events in the final state over $l^-$ if the decay results from  $X_{5/3} or \bar{X}_{5/3}$. If however the decay results from a $B$ or $\bar{B}$, there is no charge asymmetry. $\bar{B}$ production dominates over $B$ production, again because of the larger $u$ quark PDF in the initial state, but $B$ and $\bar{B}$ decay into $W^+W^-$ and a b-jet, and for both, the final state lepton can arise from either the $W^+$ or the $W^-$ decay with equal probabilities. 

In conclusion, if the $X\rightarrow Wt\rightarrow l \nu_l j j$ signal arises from both, $X_{5/3}$ and $B$ (and their antiparticles), the charge asymmetry is partially washed out and does not match the charge asymmetry of the di-lepton signal, and hence indirectly pointing towards the presence of the $B$ state. A possible advantage of the lepton asymmetry measurement would also be a reduced sensitivity to experimental systematics, although it will be susceptible to the effects of charge symmetric backgrounds. However, given our results from previous sections and a an $S/B > 1$, it is likely that the background effects on the charge asymmetry will be manageably low.

\section{Conclusions}\label{sec:conclusions} 
In this paper we study the potential of the early run-II of the LHC to discover and measure heavy fermionic top partners. 
So far, most experimental studies have been focusing on pair production relying on same-sign di-lepton signals as a main feature for distinguishing the top partner signals from the SM background. However, as pointed out in Ref.~\cite{Azatov:2013hya},  single production has an advantage of utilizing an efficient boosted tagging strategy without loosing signal efficiency from requiring two leptonic decays. In addition, the single production cross section becomes larger than that of pair production in the higher mass region ({\it e.g.} somewhere between $1\TeV$ and $1.5\TeV$ depending on models), which makes the single production process more relevant for the upcoming run of the LHC. 
In conjunction with the our usage of jet substructure physics and $b$-tagging,  we also propose a new method to tag forward jets that characterise our signal events. We demonstrate that both our substructure and forward jet handles are robust against contamination from pileup.

For the purpose of illustration, we focused on partial composite scenarios for the top sector, where both top quark chiralities 
consist of an elementary fermion field which has a sizable mixing with the strong dynamics sector. We use the Minimally Composite Higgs Model,  based on the coset space $SO(5)/SO(4)$ as the benchmark model for signal events, where we kept the signal cross section a free parameter in order to reduce the model dependence of our results.
Our analysis considered the most significant signal which comes from the singly produced charge $5/3$ and $-1/3$ partners ($X_{5/3}$ and $B$),  and their conjugates.  $X_{5/3}$ is typically the lightest top partner, with the mass splitting of $B$ and $X_{5/3}$ becoming small at high $M_4$. 
In addition, the decay topology of $X_{5/3}$ and $B$ is effectively identical when the semi-leptonic final states are considered, such that the combined signal typically has the largest cross section.

The singly produced $X_{5/3}$ and $B$ partners appear in a final state with an additional top and a light jet, so that the signal has a $t\bar{t}Wj$ event topology. For our search strategy, we require that only one of the daughter products of the top partners  (top or $W$) decays leptonically, but not simultaneously. For jet substructure analysis we employ the \verb|TemplateTagger v.1.0| implementation of the Template Overlap Method, which is relatively robust against a large pile-up contamination. The presence of two highly boosted objects allows for a straight-forward reconstruction of the top partner mass, despite the missing energy component and high pileup.

Since our signal has an additional high energy forward jet, we propose a new approach to forward jet tagging in order to limit the pileup contamination in the forward region. We propose to cluster the jets in the forward region with a cone smaller than the standard $r = 0.4$ (i.e. $r=0.1, 0.2$), which does not require an elaborate re-calibration of jet observables, since all we require is to tag the forward jet as opposed to measure it. In addition, we include a semi-realistic $b$-tagging algorithm into our analysis, as multiple $b$-jets appear in our signal events. As our forward jet tagging proposal is new, we presented the result of our analysis both with and without forward jet tagging, while  we found that we can achieve the best result when both $b$-tagging and our forward jet tagging are employed.

The main results of our analysis can be summarized as follows: 
\begin{itemize}
\item We showed that Run-II of the LHC at 14 TeV can detect and measure $2\TeV$ top partners in a lepton-jet final state,  with almost $5 \sigma$ signal significance and $S/B > 1$ at $35 \fb^{-1}$\,. The results assume a total production cross section of $15 \fb$, an average 50 interactions per bunch crossing and no pileup subtraction. In a no-pileup environment, the significance is approximately twice as high. 
\item A sizeable part of the model parameter space parts which result in a 2 TeV top partner  can be ruled at $2\sigma$ with as little as $10 \fb^{-1}$. 
\item High levels of pileup ($i.e.$ 50 interactions per bunch crossing) present a challenge for the lepton-jet final states. However, even with no pileup correction/subtraction lepton-jet channels provide sufficient sensitivity to major parts of the fermonic top partner parameter space, whereby the use of several pileup-insensitive observables greatly reduces the effects of pileup contamination.
\item The searches for singly produced fermionic top partners will greatly benefit from the introduction of a forward jet tag, with the additional factor of $\sim 2$ in the overall rejection power at $60 \%$ signal efficiency. We proposed a simple new procedure of how to mitigate effect of high pileup levels on forward jet multiplicity.
\item We find that the sensitivity the experiments can achieve in the hadronic $W$-leptonic top channel is comparable to the hadronic top-leptonic $W$ channel for $\MX \sim 1 \TeV$, while the sensitivity of hadronic top channel is superior for higher masses.
 \end{itemize}
 
Note that it will be straightforward to combine our current analysis with the conventional same-sign lepton searches in the single production of charge $5/3$ and $-1/3$, as well as pair production channels. Furthermore, our method can be easily adapted in other top partners searches, including charge $2/3$ partners, and other models of top partners beyond the minimal composite Higgs models. We also want to emphasize that our analysis is done independent of the underlying physics model, by keeping the signal cross section a free parameter, such that any new physics searches with a $t\bar{t}Wj$ event topologies can use our result directly. 

Finally, in case a signal is observed at the future LHC runs, a combination of lepton-jet channels and same sign di-lepton channels offers valuable information beyond the simple improvement in signal significance. A possible mass degeneracy between the heavy partner states can be disentangled by comparing results of same sign di-lepton measurements and signals from lepton-jet events, as the former is sensitive only to $5/3$ charge states, while additional states might appear in the latter.

\bigskip
\emph{Acknowledgements:} \\
The authors would like to thank the CERN theory group for the hospitality during the initial stages of this project. The heavy numerical calculations required for this project could not be possible without the support and understanding of Lorne Levinson and Pierre Choukroun of the Weizmann Institute. This work was supported by the National Research Foundation of Korea(NRF) grant funded by the Korea government(MEST) (No. 2012R1A2A2A01045722),
and also supported by Basic Science Research Program through the National Research Foundation of Korea(NRF) funded by the ministry of Education,
Science and Technology (No. 2013R1A1A1062597). GP is supported by the IRG, by the Gruber award, and ERC-2013-CoG grant (TOPCHARM \# 614794). Furthermore, a significant part of this work was done when GP held a Staff position at CERN. 
SL and TF are also supported by Korea-ERC researcher visiting program through the National Research Foundation of Korea(NRF) (No. 2014K2a7B044399 and No. 2014K2a7A1044408).
\newpage
\appendix
\section{$SO(5)/SO(4)$ Essentials}\label{sec:app1}
We define here notation used in the main text and collect some useful
expressions for the ${\rm SO}(5)/{\rm SO}(4)$ coset. These relations are included for completeness and the convenience of the reader. They have been given before in Ref.~\cite{Delaunay:2013pwa}, mostly following the notation of Ref.~\cite{DeSimone:2012fs}.

The 10 generators of ${\rm SO}(5)$ generators in the fundamental representation are written as
\beq
(T^\alpha_{L})_{IJ} = -\frac{i}{2}\left[\frac{1}{2}\varepsilon^{\alpha\beta\gamma}
\left(\delta_I^\beta \delta_J^\gamma - \delta_J^\beta \delta_I^\gamma\right) +
\left(\delta_I^\alpha \delta_J^4 - \delta_J^\alpha \delta_I^4\right)\right]\,,\nonumber
\eeq
\beq
(T^\alpha_{R})_{IJ} = -\frac{i}{2}\left[\frac{1}{2}\varepsilon^{\alpha\beta\gamma}
\left(\delta_I^\beta \delta_J^\gamma - \delta_J^\beta \delta_I^\gamma\right) -
\left(\delta_I^\alpha \delta_J^4 - \delta_J^\alpha \delta_I^4\right)\right]\,,
\label{eq:SO4_gen}
\eeq
\begin{equation}
T^{i}_{IJ} = -\frac{i}{\sqrt{2}}\left(\delta_I^{i} \delta_J^5 - \delta_J^{i} \delta_I^5\right)\,,
\label{eq:SO5/SO4_gen}
\end{equation}
where $I,J=1,\ldots ,5$.
The above basis is convenient because it explicitly isolates the 6 unbroken generators $T^{\alpha}_{L,R}$ ($\alpha=1,2,3$) of the $\textrm{SO}(4) \simeq \textrm{SU}(2)_L \times \textrm{SU}(2)_R$ subgroup from the broken ones $T^{i}$ ($i=1,\ldots, 4$), associated with the
coset $\textrm{SO}(5)/\textrm{SO}(4)$.
The generators in eqs.~(\ref{eq:SO4_gen}) and (\ref{eq:SO5/SO4_gen}) are normalized
such that ${\rm Tr}[T^AT^B]=\delta^{AB}$. It is convenient to collectively denote $T^\alpha_{L,R}$ as $T^a$ ($a=1,\ldots,6$), where $T^{1,2,3}=T^{1,2,3}_L$ and $T^{4,5,6}=T^{1,2,3}_R$. In the basis of Eq.~\eqref{eq:SO4_gen}, $T^a$ are bock-diagonal
\begin{equation}
T^a=\left(\begin{array}{cc}t^a &0 \\ 0 &0 \end{array}\right)\,,
\end{equation}
where $t^a$ are the 6 ${\rm SO}(4)$ generators in the fundamental representation of ${\rm SO}(4)$.

The explicit form of the Goldstone matrix as a function of the Goldstone
fields $\Pi_i$ is
\beq
U_{gs}= U_{gs}(\Pi) = \exp\left[i \frac{\sqrt{2}}{f} \Pi_{i} T^{i}\right]=\left(\begin{matrix}
\displaystyle \textbf{1}_{4\times 4}-{\vec\Pi\vec\Pi^T\over \Pi^2} \left(1-\cos{\Pi\over f}\right)\hspace{1em}&
\displaystyle {\vec\Pi\over \Pi}\sin{\Pi\over f}\\
\rule{0pt}{2.em}\displaystyle -{\vec\Pi^T\over \Pi}\sin{\Pi\over f}&\displaystyle \cos{\Pi\over f}\\
\end{matrix}\right)
\label{gmatr}
= \left(\begin{matrix}
1&0&0&0&0\\
0&1&0&0&0\\
0&0&1&0&0\\
0&0&0&\cos{\frac{\hat{h}}{f}}&\sin{\frac{\hat{h}}{f}}\\
0&0&0&-\sin{\frac{\hat{h}}{f}}&\cos{\frac{\hat{h}}{f}}\\
\end{matrix}\right)\,,
\eeq
where $\vec{\Pi}\equiv (\Pi_1,\Pi_2,\Pi_3,\Pi_4)^T$ and $\Pi\equiv\sqrt{\vec{\Pi}\cdot\vec{\Pi}}$, and where the last equation holds in unitary gauge, where the  Goldstone multiplet reduces to
\beq
\vec{\Pi}=\left(\begin{matrix} 0 \\ 0\\ 0\\ \bar{h} \end{matrix}\right)\,,
\label{unigauge}
\eeq
with $\bar{h}=v + h$.
The components of the CCWZ $d_\mu$ and $e_\mu\equiv e_\mu^at^a$ symbols are 
\begin{eqnarray}
d_\mu^{\,i}&&=\sqrt{2}\left(\frac{1}{f}-\frac{\sin{\frac{\Pi}{f}}}\Pi\right)\frac{\left(\vec{\Pi}\cdot \nabla_\mu\vec{\Pi}\right)}{\Pi^2}\Pi^{i}+\sqrt{2}\,\frac{\sin{\frac{\Pi}{f}}}\Pi\nabla_\mu\Pi^{i}\,,\nonumber\\
e_{\mu}^a&&=-A_\mu^a+4i\sin^2\left({\frac{\Pi}{2f}}\right)\ \frac{\vec{\Pi}^T t^a\nabla_\mu\vec{\Pi}}{\Pi^2}\,.
\label{dande}
\end{eqnarray}
$\nabla_\mu\Pi$ is the derivative of the Goldstone fields $\Pi$ ``covariant'' under the EW gauge group,
\begin{equation}
\nabla_\mu\Pi^{i}=\partial_\mu\Pi^{i}-i A_\mu^a\left(t^a\right)^{i}_{\  j}\Pi^{ j}\,,
\end{equation}
where $A^a_\mu$ contains the  elementary SM gauge fields written in an ${\rm SO}(5)$ notation
that is
\bea
A_\mu^a T^a &=& \frac{g}{\sqrt{2}}W^+_\mu\left(T_L^1+i T_L^2\right)+\frac{g}{\sqrt{2}}W^-_\mu\left(T_L^1-i T_L^2\right)\nonumber\\
&&+g \left(c_w Z_\mu+s_w A_\mu \right)T_L^3+g' \left(c_w A_\mu-s_w Z_\mu \right)T_R^3\,,
\label{gfd}
\eea
where $s_w$ and $c_w$ are respectively the sine and cosine of the weak mixing angle.
Note that the $d_\mu$ and $e_\mu$ symbols transform under the unbroken ${\rm SO}(4)$ symmetry as a fourplet and an adjoint, respectively. 
In unitary gauge, the $e_\mu$ symbol components reduce to
\beq\label{esymb}
e^{1,2}_\mu= -\cos^2\left(\frac{\bar{h}}{2f}\right) g W_\mu^{1,2}\,,\quad  e^3_{\mu}=-\cos^2\left(\frac{\bar{h}}{2f}\right) g W^{3}_\mu-\sin^2\left(\frac{\bar{h}}{2f}\right) g' B_\mu\,,
\eeq
\beq
e^{4,5}_\mu=-\sin^2\left(\frac{\bar{h}}{2f}\right) g W_\mu^{1,2}\,,\quad e^6_{\mu}=  -\cos^2\left(\frac{\bar{h}}{2f}\right) g' B_\mu-\sin^2\left(\frac{\bar{h}}{2f}\right) g W^{3}_\mu\,,
\eeq
with $W^1_\mu=(W^+_\mu+W^-_\mu)/\sqrt{2}$, $W_\mu^2=i(W_\mu^+-W_\mu^-)/\sqrt{2}$, $W_\mu^3=c_w Z_\mu+s_w A_\mu$ and $B_\mu=c_wA_\mu -s_w Z_\mu$,  while the $d_\mu$ components read
\beq\label{dsymb}
d^{1,2}_\mu  =  -\sin(\bar{h}/f) \frac{g W^{1,2}_{\mu}}{\sqrt{2}}\,,\quad
d^{3}_\mu  = \displaystyle \sin(\bar{h}/f) \frac{g' B_{\mu}- g W^{3}_{\mu}}{\sqrt{2}}\,,\quad
d^{4}_\mu  = \displaystyle \frac{\sqrt{2}}{f} \partial_\mu h\,.
\eeq

\section{Details of Composite Higgs Models with Partially Composite Top}
The model used in this article in order to illustrate the potential of boosted top searches in discovering composite quarks in composite Higgs models is the MCHM$_5$, which is based on the breaking of $SO(5)\times U(1)_X\rightarrow SO(4)\times U(1)_X\simeq SU(2)_{R}\times SU(2)_{L}\times U(1)_X$ of a strongly coupled theory. The $SU(2)_{L}$ and  a $U(1)$ subgroup of $SU(2)_R\times U(1)_X$ are gauged in order to provide the electroweak gauge bosons. The non-linearly realized Higgs is parameterized  by the Goldstone boson matrix which in unitary given in Eq.\eqref{gmatr}.

Beyond the (pseudo-) Goldstone boson Higgs, the low energy description the strongly coupled sector is expected to contain scalar, fermionic and vector resonances, typically at or below a scale $4\pi f$. Here, we use a bottom-up approach and only include a minimal set of light fermionic resonances. The symmetry structure of the strong dynamics does not fix the embedding of the fermionic resonances. For simplicity we assume that the top partners live in a single $\pmb5$ multiplet (transforming non-linearly under $SO(5)$) with a $U(1)_{X}$ charge of $2/3$, while the elementary third generation quarks are embedded as incomplete $\pmb5$ multiplets (transforming linearly under $SO(5)$) 
\begin{align}
\tilde{\psi}=\frac{1}{\sqrt{2}}\left[\begin{matrix}
iB'-iX_{5/3}\\
B'+X_{5/3}\\
iT+iX_{2/3}\\
-T+X_{2/3}\\
\sqrt{2}\tilde{T}\\
\end{matrix}\right]_{\frac{2}{3}}=\left[\begin{matrix}
\tilde{\psi}_{4}\\
\tilde{\psi}_{1}\\
\end{matrix}\right]_{\frac{2}{3}}, \;\;\;\;
q_{L}^{t5}=\frac{1}{\sqrt{2}}\left[\begin{matrix}
ib'_{L}\\
b'_{L}\\
it'_{L}\\
-t'_{L}\\
0\\
\end{matrix}\right],\;\;\;\;t_{R}^{5}=\left[\begin{matrix}
0\\
0\\
0\\
0\\
t'_{R}\\
\end{matrix}\right].
\end{align}
The 3rd family (partner) particle content along with its quantum numbers is summarized in Table~\ref{tab:particles}. The states given above are the gauge eigenstates of the model, which mix due to EWSB as discussed below. The resulting mass eigenstates are two states $b,B$ with (electromagnetic) charge $-1/3$, four states $t,T_{f1},T_{f2},T_{s}$ with charge $2/3$, and the state $X_{5/3}$ with charge $5/3$. 

\begin{table}[t]
\begin{center}
\begin{tabular}{|c|c|c|c|c|c|c|}
\hline
& $\left(\begin{array}{c} t'_L \\ b'_L\end{array}\right)$  & $t'_R$ & $b'_R$ & $\left(\begin{array}{c} X_{5/3} \\ X_{2/3}\end{array}\right)$ & $\left(\begin{array}{c} T' \\ B' \end{array}\right)$& $\tilde{T}$ \\ \hline
$SU(3)_c$ & $\pmb3$ & $\pmb3$ & $\pmb3$ & $\pmb3$ & $\pmb3$ & $\pmb3$ \\\hline
$SO(5)$ & $\pmb5^\star$  & $\pmb5^\star$  & $\pmb5^\star$  & \multicolumn{3} {|c|} {$\pmb5$}  \\\hline
$SO(4)$ & $\pmb4^\star$ & $\pmb1$ & $\pmb1$ &  \multicolumn{2} {|c|} {$\pmb4$}  & $\pmb1$ \\\hline
$SU(2)_L$ & $\pmb2$ & $\pmb1$ & $\pmb1$ & $\pmb2$ & $\pmb2$ & $\pmb1$ \\\hline
$U(1)_X$ & $2/3$ & $2/3$ & $2/3$ & $2/3$ & $2/3$ & $2/3$ \\\hline
$U(1)_Y$   &  $1/6$ & $2/3$ & $-1/3$ & $7/6$ & $1/6$ & $2/3$ \\\hline
\end{tabular} 
\end{center}
\caption{Quantum numbers of the 3rd family (partner) quark gauge eigenstates. ``$\star$'' indicates incomplete representations.}
\label{tab:particles}
\end{table}

In what follows, we adopt the Callan-Coleman-Wess-Zumino prescription in order to write down the effective Lagrangian in a non-linearly invariant way under $SO(5)$.  The Lagrangian of the model is
\begin{equation}
\begin{aligned}
\mathcal{L} =&+i\bar{q}'_{L}\slashed{D}q'_{L}+i\bar{t}'_{R}\slashed{D}t'_{R}+i\bar{b}'_{R}\slashed{D}b'_{R}\\ &+i\bar{\tilde{\psi}}_{4}\slashed{D}\tilde{\psi}_{4}+i\bar{\tilde{\psi}}_{1}\slashed{D}\tilde{\psi}_{1}-M_{4}\bar{\tilde{\psi}}_{4}\tilde{\psi}_{4}-M_{1}e^{i\phi}\bar{\tilde{\psi}}_{1}\tilde{\psi}_{1}\\
&+(ic_{L}\bar{\tilde{\psi}}^{i}_{L4}\gamma^{\mu}d_{\mu i}\tilde{\psi}_{L1}+ic_{R}\bar{\tilde{\psi}}^{i}_{R4}\gamma^{\mu}d_{\mu i}\tilde{\psi}_{R1}+h.c.)\\
&-(y_{L}f\bar{q}^{t5}_{L}U\tilde{\psi}_{R}+y_{R}f\bar{t}^{5}_{R}U\tilde{\psi}_{L}+h.c.)\,.\\
\end{aligned}
\label{eq:5model}
\end{equation}
The first line denotes the kinetic terms for the elementary fermions with $\bar{q}'_L=(\bar{t}'_L,\bar{b}'_L)$, and the Standard model covariant derivatives. The second line contains the composite quark mass terms with a fourplet mass $M_4$ and a singlet mass $M_1$ as well as the kinetic terms, where the covariant derivatives for the singlet and four-plet are given by $D_{\mu}\tilde{\psi}_{1}=(\partial_{\mu}-ig'XB_{\mu}-ig_{s}G_{\mu})\tilde{\psi}_{1}$ and $D_{\mu}\tilde{\psi}_{4}=(\partial_{\mu}+ie_{\mu}-ig'XB_{\mu}-ig_{s}G_{\mu})\tilde{\psi}_{4}$ respectively. The explicit form of the Cartan-Maurer forms $d_\mu$ and $e_\mu$ in unitary gauge have been given in Appendix \ref{sec:app1}.  The third line of the Lagrangian describes electroweak gauge boson and Higgs interactions with composite quarks which arise purely in the strong sector. The structure of the non-linear couplings is dictated by the Cartan-Maurer one-form $d_\mu$ which contains combinations of electroweak gauge and higgs bosons. The size of the interactions depends on the parameters $c_{L,R}$. Finally, last line of Eq.~(\ref{eq:5model}) shows the coupling terms between the elementary and the composite quark sector whose structure is dictated by the Goldstone boson matrix, with the strength being controlled by $\lambda_{L,R}$. These terms induce mass mixing such that the lightest mass eigenstates (which are identified  with the Standard Model $b$ and $t$ quark) are ``partially composite'', {\it i.e.} they are linear combinations of the elementary and the composite quarks.

\subsection{Partial compositeness: masses and mixing}

Entering the Goldstone matrix into the effective Lagrangian and expanding around the vacuum expectation value, we obtain the quark mass terms
\beq
\mathcal{L}_{m,h} = - \bar{\psi}^t_L M^t\psi^t_R - \bar{\psi}^b_L M^b\psi^b_R - M_4 \bar{X}_{5/3 \, L} X_{5/3 \, R} +\mbox{ h. c.}\,,
\eeq
where  $\bar{\psi} ^t_{L,R}\equiv(\bar{t}',\bar{T},\bar{X}_{2/3},\bar{\tilde{T}})_{L,R}$, $\bar{\psi} ^b_{L,R}\equiv(\bar{b}',\bar{B}')_{L,R}$\,, 
\beq
M^t=\left(\begin{matrix}
0 & y_{L}f\cos^{2}{\frac{\epsilon}{2}} & y_{L}f\sin^2{\frac{\epsilon}{2}} & -\frac{y_{L}f}{\sqrt{2}}\sin{\epsilon} \\
\frac{y_{R}f}{\sqrt{2}}\sin{\epsilon}&M_{4}&0&0\\
-\frac{y_{R}f}{\sqrt{2}}\sin{\epsilon}&0&M_{4}&0\\
y_{R} f \cos{\epsilon} & 0 & 0 & M_1 e^{i\phi}
\end{matrix}\right)\,\,\, \mbox{ and } \,\,\,
M^b =  \left(\begin{matrix}
0&y_{L}f\\
0&M_{4}\\
\end{matrix}\right).
\label{massmat}
\eeq
The mass matrices depend on the fourplet and singlet mass scales $M_4$ and $M_1$ and the left- and right-handed pre-Yukawa couplings $y_{L,R}$. A priory all these parameters are complex. However, all but one phase can be absorbed by field redefinitions of the quarks and quark partners. We choose the phase remaining phase $\phi$ to be on the singlet mass term (as indicated in the Lagrangian Eq.\eqref{eq:5model}) and $y_{L,R}$ and $M_{1,4}$ to be real in what follows, while $c_{L,R}$ are complex parameters.

$X_{5/3}$ is the only state with electric charge $5/3$ and as such must be a mass eigenstate with mass $M_4$. The charge $-1/3$ mass eigenstate are
\beq
\left(\begin{matrix} b_{L/R} \\ B_{L/R} \end{matrix}\right)
\equiv\psi^b_{m \,L/R}=U^b_{L/R}\psi^b \,,
\label{Ubtrafo}
\eeq
where
\beq
U^b_{L/R}=\left(
\begin{matrix} 
\cos{\theta^b_{L/R}} & \sin{\theta^b_{L/R}}\\
- \sin{\theta^b_{L/R}}& \cos{\theta^b_{L/R}} 
\end{matrix}
\right)\,, \,\,
\,\, \tan{\theta^b_R}=0  \,  ,\, \, \,  \tan{\theta^b_L}= - \frac{y_L f}{M_4} \,,
\label{Ubdef}
\eeq
with masses $m_b=0$ and $M_B=\sqrt{M_4^2+ y_L^2 f^2}$, where $b$ is identified with the SM-like bottom quark, while $B$ is a heavy partner state.\footnote{In this article, we treat the bottom quark as massless. In order to induce a non-zero bottom mass, additional bottom partner quarks need to be introduced which however typically mix weakly with the partners of the top-partner multiplet such that we ignore them, here.}

\bigskip

In the charge $2/3$ quark sector, the elementary top mixes with the two fourplet states $T, X_{2/3}$ as well as with the singlet state $\tilde{T}$. For our phenomenological studies, we perform the diagonalization numerically. To provide a qualitative discussion, here, we provide some approximate results by expanding the mass matrix in  $\epsilon\equiv v/f$. The charge $2/3$ mass eigenstates are 
\beq
\left(\begin{matrix} t_{L/R} \\ T_{f1,L/R} \\ T_{f2,L/R} \\ T_{s,L/R} \end{matrix}\right)\equiv \psi^t_{m\, L/R} = U^{t,\tilde{\phi}}_{L/R}U^t_{L/R} \psi^t,
\label{Uttrafo}
\eeq
where $U^{t,\tilde{\phi}}_L= \mathbbm{1}$, $U^{t,\tilde{\phi}}_R =diag(e^{i\tilde{\phi}},1,1,1)$, with $\tilde{\phi}$ being the phase of $1-\frac{M_4}{ M_1 }e^{-i\phi}$,
{\tiny
\beq
U_L^t=\left(
\begin{array}{cccc}
 \frac{M_4}{ M_{Tf2} } & -\frac{y_L f}{ M_{Tf2} } & 0 & \frac{\epsilon}{\sqrt{2}} \frac{ y_L f
   \left(y_R^2 f^2+e^{-i \phi }  M_1  M_4\right)}{  M_{Tf2}  M^2_{Ts} } \\\\
 0 & 0 & -1 & -\frac{\epsilon}{\sqrt{2}} \frac{ y_R^2 f^2}{ \left(M^2_{Ts}-M_4^2\right)} \\\\
 -\frac{y_L f}{ M_{Tf2} } & -\frac{M_4}{ M_{Tf2} } & 0 & 
  \frac{\epsilon}{\sqrt{2}}  \frac{ M_4 y_R^2 f^2-e^{-i \phi }  M_1  y_L^2 f^2}{  M_{Tf2} \left(M^2_{Ts}-M^2_{Tf2}\right)} \\\\
  \frac{\epsilon}{\sqrt{2}}\frac{ y_L f \left(e^{i \phi }  M_1  \left(M_{Ts}^2-M_4^2\right)-y_R^2 f^2 M_4\right)}{  M^2_{Ts} \left(M^2_{Ts}-M^2_{Tf2}\right)} &
  - \frac{\epsilon}{\sqrt{2}}\frac{  \left(y_R^2 f^2\left(M_{Ts}^2-y_L^2 f^2\right)-e^{i \phi }  M_1  M_4 y_L^2 f^2\right)}{ M^2_{Ts}
   \left(M^2_{Ts}-M^2_{Tf2}\right)} & \frac{\epsilon}{\sqrt{2}}\frac{ y_R^2 f^2}{
   \left(M_{Ts}^2-M_4^2\right)} & -1 \\
\end{array}
\right)\,, \nn
\label{UtLdef}
\eeq
}

{\tiny
\beq
U_R^t=\left(
\begin{array}{cccc}
 -\frac{ M_1 }{M_{Ts}  } & \frac{\epsilon}{\sqrt{2}} \frac{ y_R f \left( M_1  M_4+e^{i \phi}  y_L^2  f^2 \right)}{ M^2_{Tf2} M_{Ts} } & 
 -\frac{\epsilon}{\sqrt{2}} \frac{ y_R f  M_1 }{M_4 M_{Ts}} & \frac{e^{i \phi } y_R f}{ M_{Ts} } \\\\
  \frac{\epsilon}{\sqrt{2}}\frac{  y_R f \left( M_1 ^2-M_4^2\right)}{ M_4 \left(M^2_{Ts}-M^2_4\right)} & 0 & -1 & 
- \frac{\epsilon}{\sqrt{2}} \frac{e^{i \phi } y_R^2 f^2  M_1 }{ M_4\left(M^2_{Ts}-M_4^2\right)} \\\\
 - \frac{\epsilon}{\sqrt{2}}  \frac{ y_R f \left(M_4\left(M^2_{1}-M_{Tf2}^2\right)+e^{-i \phi } y_L^2 f^2  M_1  \right) }{ M^2_{Tf2} \left(M^2_{Ts}-M^2_{Tf2}\right)} &
   -1 & 0 &- \frac{\epsilon}{\sqrt{2}}\frac{ \left(\left(M^2_{Tf2}-y_R^2 f^2\right) y_L^2 f^2 - e^{i \phi }  M_1  M_4 y_R^2 f^2\right)}{ M^2_{Tf2} \left(M^2_{Ts}-M^2_{Tf2}\right)} \\\\
 -\frac{y_R f}{ M_{Ts} } & - \frac{\epsilon}{\sqrt{2}}\frac{  \left(M_4 y_R^2 f^2-e^{i \phi }  M_1  y_L^2 f^2\right)}{  M_{Ts}   \left(M^2_{Ts}-M^2_{Tf2}\right)} &
   \frac{\epsilon}{\sqrt{2}}\frac{ M_4 y_R^2f^2}{  M_{Ts}  \left(M^2_{Ts}-M_4^2 \right)} &
   -\frac{e^{i \phi }  M_1 }{ M_{Ts} } \\\\
\end{array}
\right)\,, \nn
\label{UtRdef}
\eeq
}
and the masses are given by 
\bea
m_t&=&\frac{v}{\sqrt{2}}\frac{|  M_1 -e^{-i\phi} M_4|}{f} \frac{y_L f}{ \sqrt{M_4+y^2_L f^2}}\frac{y_R f }{ \sqrt{ M_1 ^2+y^2_R f^2}}+\mathcal{O}(\epsilon^3), \label{eq:topmass}\\
M_{Tf1}&=& M_4 +\mathcal{O}(\epsilon^2)\,,\\
M_{Tf2}&=& \sqrt{M^2_4+y^2_L f^2} +\mathcal{O}(\epsilon^2)\,,\\
M_{Ts}&=& \sqrt{ M_1 ^2+y^2_R f^2} +\mathcal{O}(\epsilon^2)\,,
\eea

The structure of the mixing matrices and masses can easily be understood from a mass insertion picture: At leading order (ignoring electroweak symmetry breaking), $t_L$ can only mix with states in an $SU(2)$ doublet ({\it  i.e.} the charge $2/3$ members of the fourplet: $T$ and $X_{2/3}$) while $t_R$ can only mix with the $SU(2)$ (and thereby $SO(4)$) singlet state $\tilde{T}$. This mixing induces mass corrections for the singlet state and (one linear combination of) the fourplet states, while the lightest eigenstate does not obtain a mass at this order. Generating a mass for this state requires mixing of $SU(2)$ doublet and singlet states and therefore at least one insertion of $v/f$.  Therefore, $m_t$ as well as all matrix elements of $U_{L/R}$ between $SU(2)$ doublet and singlet components  are (at most) of $\mathcal{O}(\epsilon)$.

\bigskip
For our later phenomenological studies, let us discuss typical parameter ranges and mass scales. In order to avoid too large fine-tuning, the compositeness scale $f$ should be close to the electroweak scale. On the other hand, electroweak precision constraints imply $f\gtrsim 800$ TeV \cite{Grojean:2013qca,Ciuchini:2013pca} so that we assume $f$ to lie at the TeV scale. The composite mass scales $M_1$ and $M_4$ arise from the condensation of the strongly coupled theory and therefore have a natural value between $f$ and $\sim 4 \pi f$. Searches for top partners in the 8 TeV LHC run impose a bound of $M_{4,1}\gtrsim 800$ GeV already, and in this article, we aim to explore prospects for LHC at 13 TeV to explore top partner masses around 2 TeV, {\it i.e.} above the scale $f$. Finally, requiring the top mass Eq.\ref{eq:topmass} to take its measured value requires $y_L$ and $y_R$ to be $\mathcal{O}(1)$. Therefore, the typical partner we consider contains the $SO(4)$ singlet partner $T_s$ whose mass scale is set by $ M_1 $, a almost degenerate $SU(2)$ doublet $(X_{5/3}, T_{f1}$ with mass $M_4$ and a second almost degenerate $SU(2)$ doublet $(T_{f2}, B)$ which for $f < M_4$ and $y_L \sim 1$ is also close to degenerate with the former $SU(2)$ doublet.  

\subsection{Interactions of quarks with quark partners in the gauge eigenbasis}

The interaction terms of the model are derived by writing out the Goldstone matrix, the $d_\mu$ and the $e_\mu$ symbols in the effective Lagrangian Eq.~(\ref{eq:5model}) and  expanding in $\epsilon\equiv v/f$. We first calculate the couplings in the gauge eigenbasis  $\bar{\psi} ^t_{L,R}\equiv(\bar{t}',\bar{T},\bar{X}_{2/3},\bar{\tilde{T}})_{L,R}$, $\bar{\psi} ^b_{L,R}\equiv(\bar{b}',\bar{B}')_{L,R}$, $\bar{X}_{5/3\, L,R}$. 

The pre-Yukawa terms yield a contribution to Higgs-quark couplings
\beq
\mathcal{L}_{h,yuk} =  - h\bar{\psi}^t_L G^h_{yuk} \psi^t_R +\mathcal{O}(\epsilon^2)+\mbox{ h. c. },
\eeq
where 
\beq
G^h_{yuk}=\left(\begin{matrix}
0 &- \frac{y_{L}}{2}\sin{\epsilon} & \frac{y_{L}}{2}\sin{\epsilon} & -\frac{y_{L}}{\sqrt{2}}\cos{\epsilon} \\
\frac{y_{R}}{\sqrt{2}}\cos{\epsilon}&0&0&0\\
-\frac{y_{R}}{\sqrt{2}}\cos{\epsilon}&0&0&0\\
- y_{R}  \sin{\epsilon} & 0 & 0 & 0\\
\end{matrix}\right).
\label{res1}
\eeq

\bigskip

The kinetic terms include an $e$-term contribution which yields
\bea
\mathcal{L}_{\rm q, gauge}= \sum_{\alpha=L,R}&&\bar{\psi}^{b}_\alpha\,\slashed{W}^-G^{B,g}_\alpha\psi^{t}_\alpha+\bar{X}_{5/3\,\alpha}\,\slashed{W}^+G^{X,g}_\alpha\psi^{t}_\alpha+\bar{\psi}^{t}_\alpha\,\slashed{Z}G^{Zt,g}_\alpha\psi^{t}_\alpha+\bar{\psi}^{b}_\alpha\,\slashed{Z}G^{Zb,g}_\alpha\psi^{b}_\alpha+ \mbox{ h.c. }\nonumber\\
&&+ \mbox{ canonical EM and QCD interactions }
\label{eq:Lgeb}
\eea
with
\bea
 G^{B,g}_\alpha  & = &
\frac{g}{\sqrt{2}}
\left(
\begin{array}{cccc}
\delta_\alpha^L&0&0&0\\
0&\cos^{2}{\frac{\epsilon}{2}}&\sin^{2}{\frac{\epsilon}{2}}&0
\end{array}
\right) \, ,\label{res2} \\
G^{X,g}_\alpha & = &\frac{g}{\sqrt{2}}
\left(
\begin{array}{cccc}
0&\sin^{2}{\frac{\epsilon}{2}}&\cos^{2}{\frac{\epsilon}{2}}& 0 
\end{array}
\right) \,,\\
G^{Zt,g}_\alpha & =& \frac{g}{2c_w}
\left(
\begin{array}{cccc}
\delta_\alpha^L&0&0&0\\
0&\cos{\epsilon}&0&0\\
0&0&-\cos{\epsilon}&0\\
0&0&0&0
\end{array}
\right) -\frac{2g}{3}\frac{s_w^2}{c_w}\, \cdot \,\,  {\bf 1}\, ,\label{GZmat}\\
G^{Zb,g}_\alpha &=&-\frac{g}{2 c_w}\left(
\begin{matrix}
 \delta_\alpha^L & 0\\
 0 & 1
 \end{matrix}\right)
 +\frac{g}{3}\frac{s_w^2}{c_w}\, \cdot \,\,  {\bf 1}\,,\label{res3}
\eea
where $\delta_\alpha^L$ is 1 for $\alpha=L$ and 0 for $\alpha=R$.

\bigskip

The $d_\mu$ term interactions in Eq.~(\ref{eq:5model}) yield further contributions to the quark interactions with gauge bosons and the Higgs which read
\bea
\mathcal{L}_c &=&\sum_{\alpha=L,R}\left[-\frac{i c_\alpha}{f}\left(\bar{T}_\alpha-\bar{X}_{2/3 \, \alpha}\right) \gamma^\mu\left(\partial_\mu h \right)\tilde{T}_\alpha\right.\label{c1term}\\
&&\left. \,\,\,-\frac{g}{\sqrt{2}}c_\alpha \sin\frac{h+v}{f}\left(\frac{1}{\sqrt{2}c_w}\left(\bar{T}_\alpha+\bar{X}_{2/3\, \alpha}\right)\,\slashed{Z} \tilde{T}_\alpha+\bar{B}'_\alpha\,\slashed{W}^- \tilde{T}_\alpha-\bar{X}_{5/3\, \alpha}\,\slashed{W}^+ \tilde{T}_\alpha\right) \right]+\mbox{h.c.}. \nonumber
\eea 
The contribution of the $d$-terms to quark - gauge boson interactions can easily be read off from the last line, leading to contributions analogous to Eq.~(\ref{eq:Lgeb}) with 
\bea
 G^{B,c}_\alpha  & = &
\frac{g}{\sqrt{2}}
\left(
\begin{array}{cccc}
0&0&0&0\\
0&0&0&-c_\alpha \sin{\epsilon}
\end{array}
\right) \, ,\label{res4} \\
G^{X,c}_\alpha & = &\frac{g}{\sqrt{2}}
\left(
\begin{array}{cccc}
0&0&0&  c_\alpha \sin{\epsilon}
\end{array}
\right) \,,\\
G^{Zt,c}_\alpha & =& \frac{g}{2c_w}
\left(
\begin{array}{cccc}
0&0&0&0\\
0&0&0&c_\alpha \sin{\epsilon}\\
0&0&0&c_\alpha \sin{\epsilon}\\
0&c_\alpha \sin{\epsilon}&c_\alpha \sin{\epsilon}&0
\end{array}
\right)\,,\\
G^{Zb,c}_\alpha &=&0,\label{res5}
\eea
To rewrite the first term of Eq.~(\ref{c1term}) we partially integrate it and make use of the quark equations of motion
\bea
i\slashed{\partial}\psi^t_L&=& M^t\psi^t_R\,,\\
i\slashed{\partial}\psi^t_R&=& (M^t)^\dagger\psi^t_L\,,\\
i\bar{\psi}^t_L\overleftarrow{\slashed{\partial}}&=& -\bar{\psi}^t_R (M^t)^\dagger\,,\\
i\bar{\psi}^t_R\overleftarrow{\slashed{\partial}}&=& -\bar{\psi}^t_L (M^t)\,,
\eea
to obtain
\bea
\mathcal{L}_c &\subset& \frac{c_L h}{f}\left[-\bar{\tilde{T}}_L\left(M^t_{2i}-M^t_{3i}\right)\psi^t_{R \,i}+\left(\bar{T}_L-\bar{X}_{2/3 \, L}\right)M^t_{4 \,i}\psi^t_{R \, i}  \right]\,,\nonumber\\
&&+\frac{c_R h}{f}\left[- \bar{\psi}^t_{L\, i} \left(M^t_{i\,2}-M^t_{i\,3}\right)\tilde{T}_R + \bar{\psi}^t_{L\, i}M^t_{i \, 4} \left(T_R-X_{2/3\, R}\right)\right] + \mbox{ h.c. }\,,\nn\\
&=& -\bar{\psi}^t_L h G^h_c \psi^t_R +\mbox{ h.c. },
\eea
where, using Eq.~(\ref{massmat})
\bea
G^h_{c}&=&\left(\begin{matrix}
0 &-\frac{c_R M^t_{1 4}}{f} & \frac{c_R M^t_{1 4}}{f}  &\frac{c_R \left(M^t_{12}-M^t_{13}\right)}{f} \\
-\frac{c_L M^t_{41}}{f} &0&0& \frac{c_R M^t_{22}}{f}-  \frac{c_L M^t_{44}}{f}\\
\frac{c_L M^t_{41}}{f} &0&0& - \frac{c_R M^t_{33}}{f}+  \frac{c_L M^t_{44}}{f}\\
\frac{c_L \left(M^t_{21}-M^t_{31}\right)}{f} & -\frac{c_R M^t_{44}}{f}  +\frac{c_L M^t_{22}}{f} & \frac{c_R M^t_{44}}{f}- \frac{c_L M^t_{33}}{f}  & 0\\
\end{matrix}\right)\nn\\
&=& \left(\begin{matrix}
0 &\frac{c_R y_L}{\sqrt{2}} \sin{\epsilon} & -\frac{c_R y_L}{\sqrt{2}} \sin{\epsilon} &c_R y_L \cos{\epsilon} \\
- c_L y_R \cos{\epsilon}&0&0&-\frac{c_LM_1-c_R M_4}{f} \\
c_L y_R \cos{\epsilon}&0&0&\frac{c_LM_1-c_R M_4}{f} \\
\sqrt{2} c_L y_R \sin{\epsilon} & \frac{c_LM_4-c_R M_1}{f} &- \frac{c_LM_4-c_R M_1}{f} & 0\\
\end{matrix}\right).\label{res6}
\eea

\bigskip

Collecting all interaction terms then yields the interaction Lagrangian in the gauge eigenbasis

\bea
\mathcal{L}_{\rm q, int}&=& \sum_{\alpha=L,R}\left[\bar{\psi}^{b}_\alpha\,\slashed{W}^-G^{B}_\alpha\psi^{t}_\alpha+\bar{X}_{5/3\,\alpha}\,\slashed{W}^+G^{X}_\alpha\psi^{t}_\alpha+\bar{\psi}^{t}_\alpha\,\slashed{Z}G^{Zt}_\alpha\psi^{t}_\alpha+\bar{\psi}^{b}_\alpha\,\slashed{Z}G^{Zb}_\alpha\psi^{b}_\alpha\right]\nonumber\\
&& - \bar{\psi}^t_L h G^h \psi^t_R  + \mbox{ h.c. }+ \mbox{ canonical EM and QCD interactions } + \mbox{ higher order in } \frac{v+h}{f}\,,
\label{eq:Lint}
\eea
with
\bea
 G^{B}_\alpha  & = &
\frac{g}{\sqrt{2}}
\left(
\begin{array}{cccc}
\delta_\alpha^L&0&0&0\\
0&\cos^{2}{\frac{\epsilon}{2}}&\sin^{2}{\frac{\epsilon}{2}}&-c_\alpha \sin{\epsilon}
\end{array}
\right) \, ,\label{res2} \\
G^{X}_\alpha & = &\frac{g}{\sqrt{2}}
\left(
\begin{array}{cccc}
0&\sin^{2}{\frac{\epsilon}{2}}&\cos^{2}{\frac{\epsilon}{2}}& c_\alpha \sin{\epsilon} 
\end{array}
\right) \,,\\
G^{Zt}_\alpha & =& \frac{g}{2c_w}
\left(
\begin{array}{cccc}
\delta_\alpha^L&0&0&0\\
0&\cos{\epsilon}&0&c_\alpha \sin{\epsilon}\\
0&0&-\cos{\epsilon}&c_\alpha \sin{\epsilon}\\
0&c_\alpha \sin{\epsilon}&c_\alpha \sin{\epsilon}&0
\end{array}
\right) -\frac{2g}{3}\frac{s_w^2}{c_w}\, \cdot \,\,  {\bf 1}\, ,\\
G^{Zb}_\alpha &=& -\frac{g}{2 c_w}\left(
\begin{matrix}
 \delta_\alpha^L & 0\\
 0 & 1
 \end{matrix}\right)
 + \frac{g}{3}\frac{s_w^2}{c_w}\, \cdot \,\,  {\bf 1}\,,\label{BZbcoupl}\\
G^h&=& \left(\begin{matrix}
0 &-\frac{y_L}{2}\left(1-\sqrt{2} c_R\right) \sin{\epsilon} &\frac{y_L}{2}\left(1-\sqrt{2} c_R\right) \sin{\epsilon} &- \frac{y_L}{\sqrt{2}}\left(1-\sqrt{2} c_R\right) \cos{\epsilon} \\
\frac{y_R}{\sqrt{2}}(1- \sqrt{2}c_L) \cos{\epsilon}&0&0&-\frac{c_LM_1-c_R M_4}{f} \\
-\frac{y_R}{\sqrt{2}}(1- \sqrt{2}c_L) \cos{\epsilon}&0&0&\frac{c_LM_1-c_R M_4}{f} \\
-y_R(1- \sqrt{2}c_L) \sin{\epsilon}& \frac{c_LM_4-c_R M_1}{f} &- \frac{c_LM_4-c_R M_1}{f} & 0\\
\end{matrix}\right).\nonumber\\ \label{eq:Gh}
\eea

Again, the coupling structure is easily understood in terms of $SU(2)$ multiplets in the $\epsilon$ expansion. Concerning the gauge couplings, at leading order, the elementary states couple SM-like, and the fourplet and singlet composite quarks have canonical couplings determined by their charge. At $O(\epsilon)$, the $d$-terms lead to interactions between EW gauge bosons, fourplet and singlet states. Furthermore, there are no gauge interactions with one elementary and one composite quark; these are solely induced due to the mixing or the mass eigenstates. The higgs - quark interactions obtain contributions from the pre-yukawa terms which where also responsible for the mass mixing. In addition, the $d$-terms contain derivative interactions of the Higgs to singlet and fourplet quarks which can be rewritten as Yukawa couplings via the quark equations of motion.

\subsection{Derivation and discussion of the  interactions of quarks with quark partners in the mass eigenbasis}\label{sec:app33}

From Eq.~(\ref{eq:Lint}), the interactions of the physical states are obtained by rotating into the mass eigenbasis via the transformations $U^{t/b}_{L/R}$ given in Eqs.~(\ref{Ubdef},\ref{UtLdef},\ref{UtRdef}). For our simulations we implemented the full set of interactions and diagonalized the mass matrix numerically, but the main phenomenological features can be readily understood from the dominant couplings of the lightest quark partner states to SM gauge bosons and SM-like quarks which are relevant for the single-production of the quark partner as well as its decay channels.
\bigskip

\begin{figure}
\begin{center}
\includegraphics[width=\textwidth]{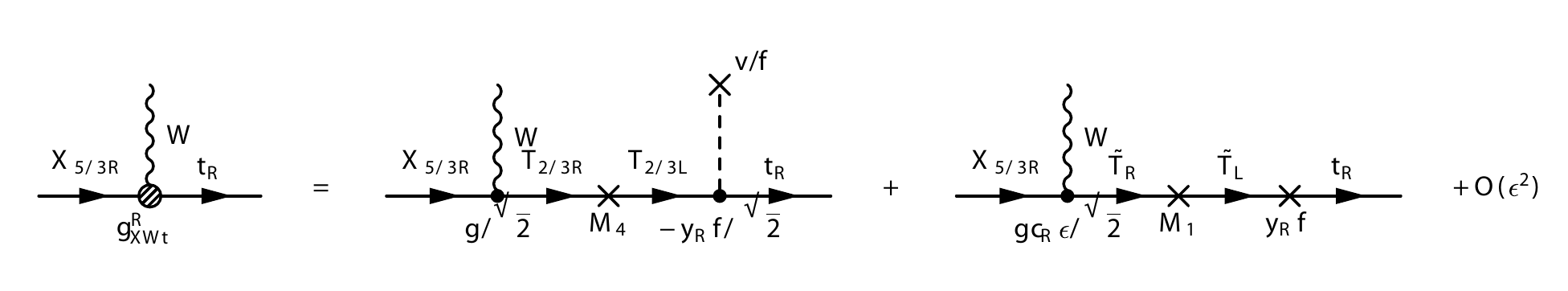}
\end{center}
\caption{Contributions to $g^{R}_{XWt}$ at $\mathcal{O}(\epsilon)$ from the mass insertion picture.\\ At $\mathcal{O}(1)$, the gauge eigenstate $X_{5/3\, R}$ only couples to $W^+$ and  $X_{3/2\, R}$ via the $e$-term. The $X_{5/3\, R}$ then mixes via mass and VEV insertions with $t'_R$ and $\tilde{T}_R$, which make up the $\mathcal{O}(1)$ components of the mass eigenstate  $t_R$. From the mass matrix in Eq.~(\ref{massmat}) it can be seen that the only mass insertion combination at $\mathcal{O}(\epsilon)$  goes from $X_{2/3\,R}$ through $ X_{2/3\,L}$ to $t'_R$. Combining the couplings and mass insertions and taking into account that the $t'_R$ component of $t_R$ has a coefficient $M_1/M_{Ts}$ yields the first contribution to the coupling $g^{R}_{XWt}$ in Eq.~(\ref{eq:XWtcoupl}).\\
At $\mathcal{O}(\epsilon)$, the gauge eigenstate $X_{5/3\, R}$ couples to $W^+$ and  $\tilde{T}_R$ via the $d$-term. $T_R$ mixes via $\tilde{T}_L$ to $t'_R$ at $\mathcal{O}(1)$. Projecting $t'_R$ on $t_R$ and assembling the couplings and insertions yields the second term of  $g^{R}_{XWt}$ in Eq.~(\ref{eq:XWtcoupl}). \\ The analogous analysis for $g^{L}_{XWt}$ results in couplings of $\mathcal{O}(\epsilon^2)$ because the mixing of $X_(2/3\, L)$ to $t'_L$ is of $\mathcal{O}(\epsilon)$ while the mixing of $\tilde{T}_L$ to $t'_L$ is of $\mathcal{O}(\epsilon^2)$. 
Couplings of other heavy quark partners to SM quarks and EW gauge bosons or the Higgs can be understood analogously. 
 }
 \label{fig:gRMI}
 \end{figure}

\noindent
\textbf{$X_{5/3}$:}\\
The exotically charged $X_{5/3}$ has mass $M_4$ and is thus the lightest fourplet quark partner. Its couplings to only SM particles are
\bea
g^{L}_{XWt}&=&G^X_{L\, i} \left(U^t_L \right)^\dagger_{\,\, i 1} = \mathcal{O}(\epsilon^2)\,,\\
g^{R}_{XWt}&=&G^X_{R\, i} \left(U^t_R \right)^\dagger_{\,\, i 1} =\frac{g}{\sqrt{2}}\left( U^{*t}_{R\, 13} +c_R \epsilon  U^{*t}_{R\, 14} \right)+\mathcal{O}(\epsilon^2)\,,\nonumber\\
&=& - \frac{g e^{-i\tilde{\phi}}}{\sqrt{2}}\frac{\epsilon}{\sqrt{2}}\left(\frac{ y_R f  M_1 }{M_4 M_{Ts}}-\sqrt{2} c_R \frac{e^{-i \phi } y_R f}{M_{Ts}}\right)+\mathcal{O}(\epsilon^2)\,.
\label{eq:XWtcoupl}
\eea
Other couplings to two SM particles are forbidden due to (electric) charge conservation. The structure of the dominant right-handed coupling can be understood from the mass insertion picture as shown in Fig.~\ref{fig:gRMI}.

\bigskip

\noindent
\textbf{$B$:}\\
The  $B$ has charge $-1/3$ and can a priori couple to $Wt$, $Zb$, or $hb$. However, the $b$ does not have any pre-yukawa couplings within this model so that a $\bar{B}hb$ coupling term is absent. A $\bar{B}Zb$ coupling is absent as well. In the gauge eigenbasis, no $\bar{B}'Zb'$ couplings are present. In the right-handed sector, $b'_R$ and $B'_R$ are already mass eigenstates. The left-handed coupling in Eq.(\ref{BZbcoupl}) is universal for $b'_L$ and $B'_L$, and rotation into the mass eigenbasis does not induce a ``mixed'' $\bar{B}Zb$ interaction. $\bar{B}Wt$ are present and given by  
\bea
g^{L}_{BWt}&=&U^b_{L\, 2i}G^B_{L\, i j} \left(U^t_L \right)^\dagger_{\,\, j 1}  =\mathcal{O}(\epsilon^2)\,.\\
g^{R}_{BWt}&=&U^b_{R\, 2i}G^B_{R\, i j} \left(U^t_R \right)^\dagger_{\,\, j 1} =\frac{g}{\sqrt{2}}\left( U^{*t}_{R\, 12} -c_R \epsilon  U^{*t}_{R\, 14} \right)+\mathcal{O}(\epsilon^2)\nn\\
&=&  \frac{g e^{-i\tilde{\phi}}}{\sqrt{2}}\frac{\epsilon}{\sqrt{2}}\left(\frac{ y_R f \left( M_1  M_4+e^{-i \phi}  y_L^2  f^2 \right)}{ M^2_{Tf2} M_{Ts} }-\sqrt{2} c_R  \frac{e^{-i \phi } y_R f}{ M_{Ts} }\right)+\mathcal{O}(\epsilon^2)\,.\label{BcouplR}
\eea

\bigskip

\noindent
\textbf{$T_{f1}$ and $T_{f2}$:}\\
The states $T_{f1}$ and $T_{f2}$ have charge $-1/3$ and can a priori couple to $Wb$, $Zb$, or $hb$. However, both left- and right-handed couplings to $Wb$ are of order $\epsilon^2$, and so are the left-handed couplings to $Zt$. The leading order right-handed couplings to $Zt$ are
\bea
g^{R}_{Tf1Zt}&=& - \frac{g e^{-i\tilde{\phi}}}{2 c_w}\frac{\epsilon}{\sqrt{2}}\left(\frac{ y_R f M_1}{ M_{Ts} M_4 }
+\sqrt{2} c_R  \frac{e^{-i \phi } y_R f}{ M_{Tf2} }\right)+\mathcal{O}(\epsilon^2)\,,\\
g^{R}_{Tf2Zt}&=& - \frac{g e^{-i\tilde{\phi}}}{2 c_w}\frac{\epsilon}{\sqrt{2}}\left(\frac{ y_R f \left( M_1  M_4+e^{-i \phi}  y_L^2  f^2 \right)}{ M^2_{Tf2} M_{Ts} }
+\sqrt{2} c_R  \frac{e^{-i \phi } y_R f}{ M_{Ts} }\right)+\mathcal{O}(\epsilon^2)\,,
\eea
while the leading couplings to the Higgs are
\bea
\lambda_{Tf1L\,h\,tR}&=& - \frac{e^{-i\tilde{\phi}}y_R}{\sqrt{2}}\left(\frac{ M_1}{ M_{Ts}}
-\sqrt{2} \frac{c_L \left(1-e^{-i\phi}\right)M_1+ c_R e^{-i\phi}M_4}{M_{Ts}}\right), \\
\lambda_{Tf2L\,h\,tR}&=& \frac{e^{-i\tilde{\phi}}y_R}{\sqrt{2}}\left(\frac{ M_1  M_4+e^{-i \phi}  y_L^2  f^2 }{M_{Tf2} M_{Ts}} - \sqrt{2} \frac{c_L\left(1-e^{-i\phi}\right)M_4^2+ c_R e^{-i\phi} M^2_{f2}}{M_{Tf2} M_{Ts}}\right)\,.
\eea

\noindent
\textbf{$T_{s}$:}\\
The state $T_{s}$ also has charge $-1/3$. Its  dominant couplings to $Wb$, $Zb$, and $hb$ are
\bea
g^{L}_{TsWb}&=&\frac{g}{\sqrt{2}}\frac{\epsilon}{\sqrt{2}}\left(\frac{y_L f\left(e^{-i\phi} M_1 M_4+y^2_Rf^2\right)}{M_{Tf2}M^2_{Ts}} -\frac{\sqrt{2}c_L y_L f}{M_{Tf2}}\right)\,,\\
g^{L}_{TsZt}&=& \frac{g}{2 c_w}\frac{\epsilon}{\sqrt{2}}\left(\frac{y_L f\left(e^{i\phi} M_1 M_4+y^2_Rf^2\right)}{M_{Tf2}M^2_{Ts}} +\frac{\sqrt{2}c_L y_L f}{M_{Tf2}}\right)\,,\\
\lambda_{TsR\,h\,tL}&=&\frac{y_L}{\sqrt{2}}\frac{\left(e^{-i\phi} M_1 M_4+y^2_Rf^2\right)-\sqrt{2}c_L\left(e^{-i\phi}M_1^2+y^2_R f^2 \right)}{M_{Tf2}M_{Ts}}\,.\label{eq:Tshtcoupl}
\eea

\subsection{Decays of top partners}\label{sec:app34}
Note from the coupling expressions:  For all top-partners, the dominant couplings to $W,Z,h$ and an SM quark are chiral (either left- or right-handed coupling dominates). In this case, the partial widths for a decay of a fermion $F$ into a fermion $f$ and a gauge boson or Higgs are
\bea
\Gamma(F\rightarrow W f) &=&M_F\frac{M^2_F}{m^2_W}\frac{|g|^2_{eff}}{32 \pi}\Gamma_W\,,\\
\Gamma(F\rightarrow Z f)  &=&M_F\frac{M^2_F}{m^2_W}\frac{|g|^2_{eff}}{32 \pi}\Gamma_Z\,,\\
\Gamma(F\rightarrow h f)  &=&M_F\frac{|\lambda|^2_{eff}}{32 \pi}\Gamma_h\,,
\eea
where $\Gamma_{W,Z,h}= 1+\mathcal{O}(\frac{m^2_{W/Z/h}}{M^2_F})$ are kinematic functions. Using these relations we can estimate the partial widths and BRs of the different top partners, using the effective couplings Eqs.~(\ref{eq:XWtcoupl} - \ref{eq:Tshtcoupl}).

\bigskip
$X_{5/3}$:\\
\bea
\Gamma(X_{5/3}\rightarrow W t) &\approx&M_4 \frac{M^2_4}{m^2_W}\frac{|g|^2_{eff}}{32 \pi}\Gamma_W\nonumber\\
&\approx&\frac{M_4}{32 \pi} \frac{M_4^2}{g^2 v^2 /4}\frac{g^2 \epsilon^2 f^2}{4 M^2_4} y^2_R\left|\frac{ M_1 -\sqrt{2}c_R e^{-i\phi} M_4}{M_{Ts}}\right|^2\nonumber\\
&=& M_4 \, \frac{y^2_R}{32 \pi} \left|\frac{ M_1 -\sqrt{2}c_R e^{-i\phi} M_4}{M_{Ts}}\right|^2\,.
\eea
Note:
\begin{itemize}
\item Although the effective coupling is $\mathcal{O}(\epsilon)$, the partial width is not $\epsilon$ suppressed.
\item The partial width is proportional to $|y_R^2 c^2_R|$ (for large $c_R$).
\item For $y_R c_R\lesssim 5$ the resonances are still  narrow width $(\Gamma/M\lesssim 25\%)$. 
\end{itemize}

$X_{5/3}$ is the lightest fourplet state, so the BR of this channel is 100\% unless $M_{Ts} < M_4$. The ``cascade'' decay $X_{5/3}\rightarrow W T_s$ is kinematically suppressed by a factor $\sim(1-(M_{Ts}/M_4)^2)^2$, so that this decay only plays a role when the is a substantial mass splitting. In this case, the partner states in the $\bf 4$ are merely decoupled, and direct searches for the $T_s$ partner are more promising. 

\bigskip

$B$:\\
The discussion of $B$ decays is analogous to the above discussion of $X_{5/3}$ decays. The dominant decay channel is $B\rightarrow W^- t$ which occurs through the right-handed coupling given in Eq.\eqref{BcouplR}:
\bea
\Gamma(B\rightarrow W t) &\approx&M_4 \frac{M^2_4}{m^2_W}\frac{|g|^2_{eff}}{32 \pi}\Gamma_,\nonumber\\
&\approx&\frac{M_4}{32 \pi} \frac{M_4^2}{g^2 v^2 /4}\frac{g^2 \epsilon^2 f^2}{4 M^2_4} y^2_R\left|\frac{M_1}{M_{Ts}}\frac{M^2_4}{M^2_{Tf2}}+e^{-i\phi}\frac{M_4}{M_{Tf2}}\frac{y^2_L f^2}{M_{Tf2}M_{Ts}}-\sqrt{2}c_R e^{-i\phi}\frac{M_4}{M_{Ts}}\right|^2 \nn\\
&=& M_4 \frac{y^2_R}{32 \pi}\left|\frac{M_1}{M_{Ts}}\frac{M^2_4}{M^2_{Tf2}}+e^{-i\phi}\frac{M_4}{M_{Tf2}}\frac{y^2_L f^2}{M_{Tf2}M_{Ts}}-\sqrt{2}c_R e^{-i\phi}\frac{M_4}{M_{Ts}}\right|^2\,.
\eea
The decays $B\rightarrow Z b$ and $B\rightarrow h b$ which are a priory allowed by the quantum numbers only occur at higher order in $\epsilon$ and are therefore suppressed.   Cascade decays to $T_{f1,2}$ are either kinematically forbidden or suppressed by a factor $\sim(1-(M_{Tf1,2}/M_B)^2)^2$. As the $B$ is mass degenerate with the $T_{f2}$ and only marginally mass split from the $M_{Tf1}$, these decays are negligible. Finally again, the ``cascade'' decay $B\rightarrow W^- T_s$ is kinematically suppressed by a factor $\sim(1-(M_{Ts}/M_4)^2)^2$, so that this decay only plays a role when the is a substantial mass splitting.

\bigskip

$T_{f1}$ and $T_{f2}$:\\
The total widths of $T_{f1,2}$ are of the same order as for $X_{5/3}$, but they have two dominating decay channels into $ht$ or $Zt$, while $Wb$ is suppressed. Concerning the BR for $T_{f1}$:
\bea
\frac{\Gamma(T_{f1}\rightarrow Z t)}{\Gamma(T_{f1}\rightarrow h t)}&\approx& \frac{M_{Tf1}\frac{y^2_R}{64 \pi}\left|\frac{M_1}{M_4 }+\sqrt{2} c_R  e^{-i \phi } \right|^2}{M_{Tf1}\frac{y^2_R}{64 \pi}  \left|\frac{ M_1}{ M_{Ts}}
-\sqrt{2} \frac{c_L \left(1-e^{-i\phi}\right)M_1+ c_R e^{-i\phi}M_4}{M_{Ts}}\right|^2 } = \frac{\left|\frac{M_1}{M_4 }+\sqrt{2} c_R  e^{-i \phi } \right|^2}{\left|\frac{ M_1}{ M_{Ts}}
-\sqrt{2} \frac{c_L \left(1-e^{-i\phi}\right)M_1+ c_R e^{-i\phi}M_4}{M_{Ts}}\right|^2 }\,.
\eea
In the limit of no $d$-term ($c_{L,R}=0$) as well as in the limit in which $c_R$ dominates, this yields BRs of $\sim M^2_{Ts}/(M^2_{Ts}+M^2_4)$ and $\sim M^2_4/(M^2_{Ts}+M^2_4)$ for the decays into $Zt$ and $ht$ , while for  $d$- and $e$-term contributions of similar size, the terms can enhance each of the BRs, depending on the size and phase of $c_{L,R}$ and $\phi$.\\
For $T_{f2}$ we obtain analogously:
\bea
\frac{\Gamma(T_{f2}\rightarrow Z t)}{\Gamma(T_{f2}\rightarrow h t)}&\approx&  \frac{\left|\frac{ M_1}{M_4 }
+\sqrt{2} c_R  e^{-i \phi } \right|^2 }
{\left|\frac{ M_1  M_4+e^{-i \phi}  y_L^2  f^2 }{M_{Tf2} M_{Ts}} - \sqrt{2} \frac{c_L\left(1-e^{-i\phi}\right)M_4^2+ c_R e^{-i\phi} M^2_{Tf2}}{M_{Tf2} M_{Ts}}\right|^2}\,,
\eea 
which in the large $c_R$ limit yields BRs of $\sim M^2_{Ts}/(M^2_{Ts}+M^2_{Tf2})$ and $\sim M^2_{Tf2}/(M^2_{Ts}+M^2_{Tf2})$, but for $c_R\sim 1$, again, the BRs can change either way.\\

\bigskip

$T_{s}$:\\
For $T_s$ we get decays into $Wb$, $Zt$ and $ht$. In addition, for $M_{Ts} >> M_4$, decays into fourplet partners can play a role, while for $M_{Ts}\sim M_4$ they are kinematically forbidden or at least suppressed. The ratios of decay rates are
\bea
\frac{\Gamma(T_{s}\rightarrow Z t)}{\Gamma(T_{s}\rightarrow h t)}&\approx&  \frac{\left|\frac{\left(e^{i\phi} M_1 M_4+y^2_Rf^2\right)}{M_{Tf2}M_{Ts}} +\sqrt{2}c_L\frac{M_{Ts} }{M_{Tf2}}\right|^2 }{\left|\frac{\left(e^{-i\phi} M_1 M_4+y^2_Rf^2\right)-\sqrt{2}c_L\left(e^{-i\phi}M_1^2+y^2_R f^2 \right)}{M_{Tf2}M_{Ts}}\right|^2 }\,,\\
\frac{\Gamma(T_{s}\rightarrow W b)}{\Gamma(T_{s}\rightarrow Z t)}&\approx&  2 \frac{\left|\frac{\left(e^{-i\phi} M_1 M_4+y^2_Rf^2\right)}{M_{Tf2}M_{Ts}} -\sqrt{2}c_L\frac{M_{Ts} }{M_{Tf2}}\right|^2 }
{\left|\frac{\left(e^{i\phi} M_1 M_4+y^2_Rf^2\right)}{M_{Tf2}M_{Ts}} +\sqrt{2}c_L\frac{M_{Ts} }{M_{Tf2}}\right|^2 }.
\eea 
In the limit $c_L\rightarrow 0$, this yields BRs of $2:1:1$ to $Wb$, $Zt$, and $ht$, up to kinematic corrections, while for $c_L\sim 1$ again, BRs vary.

\bigskip

Summarizing: The $X_{5/3}$ and $B$ decay $\sim100\%$ into $Wt$, rather independently of the chosen parameters, while for the charge $2/3$ partners, BRs are strongly parameter dependent.  

\subsection{Concluding remarks}

In this Appendix we derived the Feynman rules for the interactions of SM-like quarks to its composite partners. One main result -- the interactions in the gauge eigenbasis -- is given in Eq.\eqref{eq:Lint}. This Lagrangian describes all interactions of two quarks one gauge boson or a Higgs, only omitting higher dimensional operators in which two quarks couple to a larger number of Higges or gauge fields. For our phenomenological studies in this article, we used this Lagrangian and diagonalized the quark mass matrix Eq.\eqref{massmat} numerically, without using an expansion in $\epsilon$ to obtain the interactions in the mass eigenbasis. The expressions derived in Appendix \ref{sec:app33} and \ref{sec:app34}  for the couplings and decay widths are calculated at $\mathcal{O}(\epsilon)$ and are only given for illustration and in order to cross check our numerical implementation.    

\label{sec:app3}
\newpage
\bibliography{draft_paper}
\end{document}